\documentclass[twocolumn,twocolappendix,trackchanges]{aastex63}

\newcommand{\kms}{\mathrm{km\,s}^{-1}}

\usepackage{amsmath}	
\usepackage{amssymb}	
\usepackage{multirow}

\received{June 1, 2019}
\revised{January 10, 2019}
\accepted{\today}
\submitjournal{ApJS}

\shorttitle{Self-consistent RVs from LAMOST MRS DR7}
\shortauthors{Zhang et al.}
\graphicspath{{./}{figures/}}

\begin{document}


\title{Self-consistent Stellar Radial Velocities from LAMOST Medium-Resolution Survey (MRS) DR7}

\correspondingauthor{Bo Zhang (LAMOST Fellow)}
\email{bozhang@bnu.edu.cn}

\author[0000-0002-6434-7201]{Bo Zhang}
\affiliation{Department of Astronomy, Beijing Normal University, Beijing 100875, People's Republic of China}

\author[0000-0002-2577-1990]{Jiao Li}
\affiliation{Key Laboratory of Space Astronomy and Technology, National Astronomical Observatories, Chinese Academy of Sciences, Beijing 100101, People's Republic of China}




\author[0000-0002-6039-8212]{Fan Yang}
\affiliation{Department of Astronomy, Beijing Normal University, Beijing 100875, People's Republic of China}
\affiliation{Key Laboratory of Optical Astronomy, National Astronomical Observatories, Chinese Academy of Sciences, Beijing 100101,  People's Republic of China}
\affiliation{School of Astronomy and Space Science, University of Chinese Academy of Sciences, Beijing, 100049, People's Republic of China}

\author[0000-0003-4829-6245]{Jian-Ping Xiong}
\affiliation{Key Laboratory of Optical Astronomy, National Astronomical Observatories, Chinese Academy of Sciences, Beijing 100101,  People's Republic of China}
\affiliation{School of Astronomy and Space Science, University of Chinese Academy of Sciences, Beijing, 100049, People's Republic of China}


\author[0000-0001-8241-1740]{Jian-Ning Fu}
\affiliation{Department of Astronomy, Beijing Normal University, Beijing 100875, People's Republic of China}

\author[0000-0002-1802-6917]{Chao Liu}
\affil{Key Laboratory of Space Astronomy and Technology, National Astronomical Observatories, Chinese Academy of Sciences, Beijing 100101, People's Republic of China}


\author[0000-0003-3347-7596]{Hao Tian}
\affil{Key Laboratory of Space Astronomy and Technology, National Astronomical Observatories, Chinese Academy of Sciences, Beijing 100101, People's Republic of China}

\author[0000-0001-7607-2666]{Yin-Bi Li}
\affil{Key Laboratory of Optical Astronomy, National Astronomical Observatories, Chinese Academy of Sciences, Beijing 100101,  People's Republic of China}

\author[0000-0002-6868-6809]{Jia-Xin Wang}
\affil{Department of Astronomy, Beijing Normal University, Beijing 100875, People's Republic of China}

\author[0000-0002-0831-3111]{Cai-Xia Liang}
\affiliation{National Cancer Center, Peking Union Medical College and Chinese Academy of Medical Sciences, Beijing 100021, People's Republic of China}

\author[0000-0002-4391-2822]{Yu-Tao Zhou}
\affil{Department of Astronomy, School of Physics, Peking University, Beijing 100871, People’s Republic of China}
\affil{Kavli Institute for Astronomy and Astrophysics, Peking University, Beijing 100871, People’s Republic of China}

\author[0000-0002-7660-9803]{Wei-kai Zong}
\affiliation{Department of Astronomy, Beijing Normal University, Beijing 100875, People's Republic of China}

\author[0000-0003-1972-0086]{Cheng-Qun Yang}
\affil{Shanghai Astronomical Observatory, Chinese Academy of Sciences, Shanghai 200030, People's Republic of China}

\author[0000-0002-9528-4805]{Nian Liu}
\affil{Physics and Space Science College, China West Normal University, Nanchong 637002, People’s Republic of China}

 \author[0000-0002-3701-6626]{Yong-Hui Hou}
 \affiliation{Nanjing Institute of Astronomical Optics \& Technology, National Astronomical Observatories, Chinese Academy of Sciences, Nanjing 210042, People’s Republic of China}
 \affiliation{School of Astronomy and Space Science, University of Chinese Academy of Sciences, Beijing, 100049, People's Republic of China}











\begin{abstract}
Radial velocity (RV) is among the most fundamental physical quantities obtainable from stellar spectra and is rather important in the analysis of time-domain phenomena.
The LAMOST Medium-Resolution Survey (MRS) DR7 contains 5 million single-exposure stellar spectra at spectral resolution $R\sim7\,500$.
However, the temporal variation of the RV zero-points (RVZPs) of the MRS survey, which makes the RVs from multiple epochs inconsistent, has not been addressed.
In this paper, we measure the RVs of the 3.8 million single-exposure spectra (for 0.6 million stars) with signal-to-noise ratio (SNR) higher than 5 based on cross-correlation function (CCF) method, and propose a robust method to self-consistently determine the RVZPs exposure-by-exposure for each spectrograph with the help of \textit{Gaia} DR2 RVs.
Such RVZPs are estimated for 3.6 million RVs and can reach a mean precision of $\sim 0.38\,\mathrm{km\,s}^{-1}$.
The result of the temporal variation of RVZPs indicates that our algorithm is efficient and necessary before we use the absolute RVs to perform time-domain analysis.
Validating the results with APOGEE DR16 shows that our absolute RVs can reach an overall precision of 0.84/0.80 $\mathrm{km\,s}^{-1}$ in the blue/red arm at $50<\mathrm{SNR}<100$, while 1.26/1.99 $\kms$ at $5<\mathrm{SNR}<10$.
The cumulative distribution function (CDF) of the standard deviations of multiple RVs ($N_\mathrm{obs}\geq 8$) for 678 standard stars reach 0.45/0.54, 1.07/1.39, and 1.45/1.86 $\kms$ in the blue/red arm at 50\%, 90\%, and 95\% levels, respectively.
The catalogs of the RVs, RVZPs, and selected candidate RV standard stars are available at \url{https://github.com/hypergravity/paperdata}.
\end{abstract}




\section{Introduction} \label{sec:intro}
Large spectroscopic surveys, for example, RAVE \citep{2006AJ....132.1645S, 2020AJ....160...83S}, SDSS/SEGUE \citep{2009AJ....137.4377Y}, LAMOST \citep{2012RAA....12.1197C, 2012RAA....12..735D, 2012RAA....12..723Z, 2015RAA....15.1095L}, APOGEE \citep{2017AJ....154...94M}, GALAH \citep{2015MNRAS.449.2604D}, \textit{Gaia}-ESO \citep{2012Msngr.147...25G} and \textit{Gaia}-Radial Velocity Spectrometer \citep[\textit{Gaia}-RVS,][]{2004MNRAS.354.1223K, 2018A&A...616A...5C},  have obtained tens of millions of stellar spectra over the last two decades, aiming at understanding the formation and evolution of the Galaxy.
Among the most fundamental physical quantities derived from stellar spectra is radial velocity (RV) which forms the basis of many studies such as stellar multiplicity \citep[e.g.,][]{2014ApJ...788L..37G,  2017MNRAS.469L..68G, 2018MNRAS.476..528E, 2020ApJS..249...31Y}, stellar kinematics \citep[e.g.,][]{2020ApJ...899..110T, 2020arXiv200505980B}, and Galactic substructures \citep{2019ApJ...886..154Y, 2020ApJ...905....6X}.

The LAMOST, after a five-year low-resolution spectroscopic survey (LRS, $R\sim1\,800$, $3\,800\mathrm{\AA}<\lambda<9\,000\mathrm{\AA}$), has started a five-year medium-resolution spectroscopic survey \citep[MRS, $R\sim7\,500$, $4\,950 \mathrm{\AA}<\lambda<5\,350 \mathrm{\AA}$ and $6\,300 \mathrm{\AA}<\lambda<6\,800 \mathrm{\AA}$, see][]{2020arXiv200507210L} since Sep. 2017.
The DR6\footnote{\url{http://dr6.lamost.org/}}, the first data release of the MRS, contains the data obtained during Sep. 2017 to Jun. 2018 and is already public to the international astronomical community.
The DR7\footnote{\url{http://dr7.lamost.org/}} including the data from Sep. 2017 to Jun. 2019 ($\sim$5 million spectra for over 800\,000 stars) is currently open to the Chinese astronomical community only.


At the beginning Sc lamp was used to calibrate the wavelength of LAMOST MRS spectra, and later switched to ThAr lamp in 2018.
Due to the short wavelength coverage, sky lines are not used in wavelength calibration as in LRS \citep{2010ExA....28..195W}.
The released spectra have been applied the barycentric correction and the wavelength uses vacuum standard.
Both the LAMOST pipeline and \cite{2019ApJS..244...27W} have measured RVs from DR7 spectra and estimated a static RV zero-point (RVZP) by comparing the measured RVs to those in the literature of a sample of RV standard stars \citep{2018AJ....156...90H} for each spectrograph.
However, the temporal variation of the RVZPs (between exposures) is not clear so far despite a few trials.
For example, \cite{2019RAA....19...75L} and \cite{2020ApJS..251...15Z} show the temporal variation of RVZPs based on the data from the \textit{Kepler} field using a sample of roughly selected RV-invariant stars, and \cite{2020arXiv200806236R} shows that the RVZPs of the red arm vary between exposures and the difference can reach 4 $\kms$ in the MRS-N fields \citep[nebula survey,][]{2020arXiv200705240W}.

Physically, the variation of the LAMOST MRS RVZPs can be explained by several reasons. 
\begin{enumerate}
  \item The LAMOST telescope has a long optical path and a large focal plane (1.75m diameter) on which 4\,000 fibers are installed \cite{2012RAA....12.1197C}, temperature variation is unavoidable.
  \item Presently, arc lamp exposures are taken every about 2 hours (typical observing time of a plate) which is not frequent enough. The instrument might change its state between lamp exposures.
  \item Since LAMOST MRS needs $\sim$20 ThAr lamps to illuminate the big focal plane simultaneously, the decay and replacement of the lamps and the adjustment of lamp exposure time are very often. They affect the signal-to-noise ratio of the lamp spectra and thus the wavelength calibration consistency over time.
  \item The MRS spectrographs are mounted and dismounted monthly because the MRS survey is scheduled in the 14 bright/gray nights, while the other nights are for LRS survey. Therefore, the focuses are adjusted monthly. 
\end{enumerate}
All these factors and other potential defects in data reduction pipeline are finally reflected in the RVZPs of the spectra.
Therefore, it is insufficient to use the RVs either from the LAMOST pipeline or \cite{2019ApJS..244...27W} at different epochs and perform a time-domain analysis, such as in studies of pulsating stars \citep[the LAMOST-\textit{Kepler} project,][]{2020ApJS..251...15Z, 2020RAA....20..167F}, spectroscopic binaries \citep{2014ApJ...788L..37G, 2017MNRAS.469L..68G,  2019MNRAS.490..550L} and even searching for black holes \citep{2019Natur.575..618L, 2019ApJ...872L..20G}, which are important scientific goals of the MRS survey \citep{2020arXiv200507210L}.

In this paper, aiming at subsequent time-domain analysis, we measure the RVs for the 3.8 million single-exposure spectra with $\mathrm{SNR}>5$ in the LAMOST MRS DR7 v1.1\footnote{\url{http://dr7.lamost.org/v1.1}}, and propose a robust method to self-consistently determine the absolute RVZPs with the help of the \textit{Gaia} DR2 data.
In Section \ref{sec:mrs}, we briefly describe the observaional rules of the LAMOST MRS survey and the instrumental parameters.
In Section \ref{sec:rv}, we describe how we measure RVs.
In Section \ref{sec:rvzpc}, we show the algorithm that determines the RVZPs self-consistently.
In Section \ref{sec:results}, we present our RV and RVZP measurements and assess the precision and self-consistency, and select a sample of candidate RV standard stars based on our results.
The information and instruction of our data products are described in Section \ref{sec:dataproducts}, and the summary of this work is given in Section \ref{sec:conclusion}.

\begin{figure*}[ht!]
\epsscale{1}
\plotone{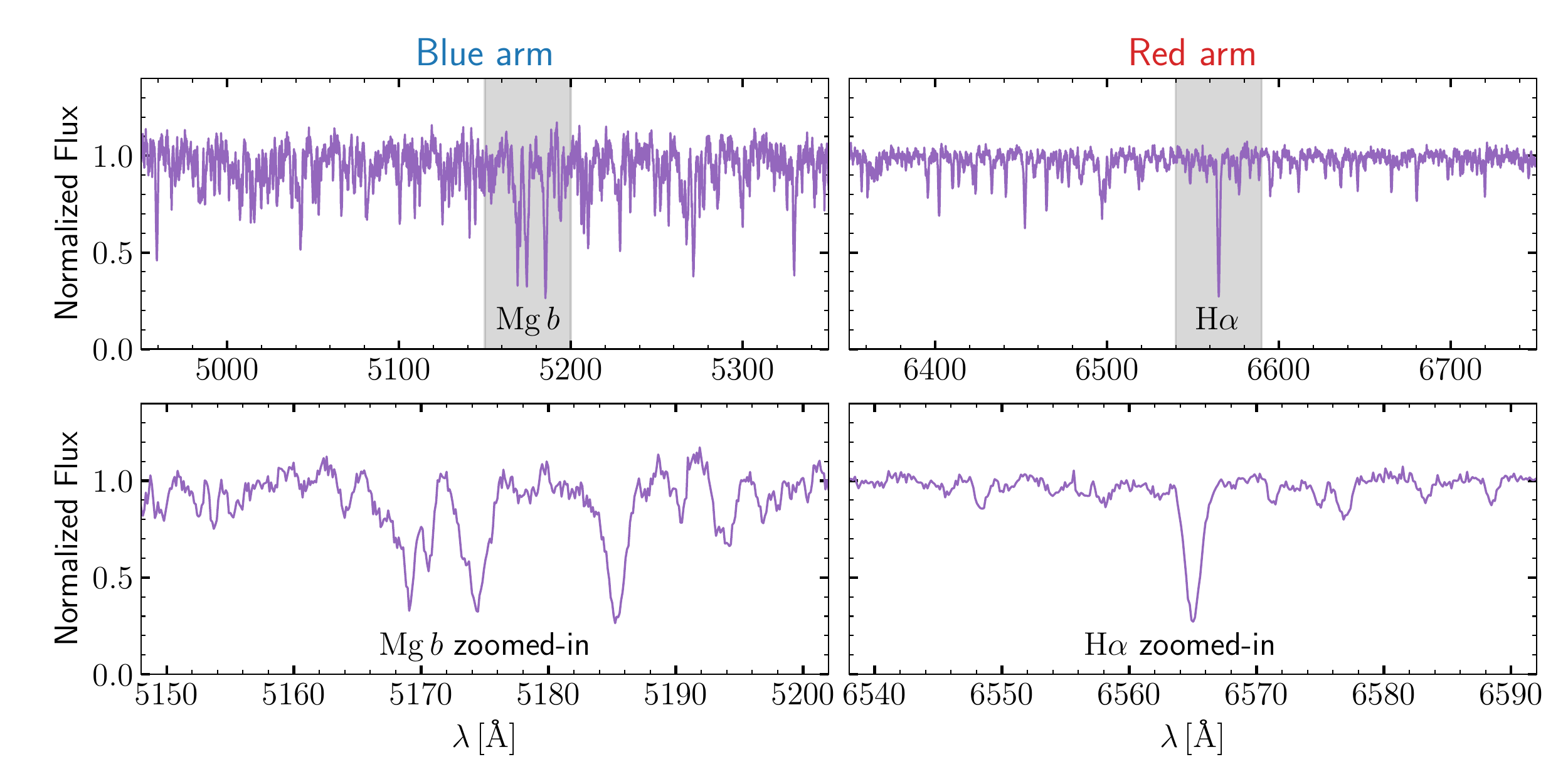}
\caption{An example of a single-exposure spectrum of a K-type giant star observed in the LAMOST DR7 MRS (med-58649-TD193637N444141K01\_sp16-249, $\mathrm{RA}=292.719378^{\circ}$, $\mathrm{Dec}=46.651537^{\circ}$).
The upper panels show the continuum-normalized blue and red arm spectra.
The lower panels show the zoomed-in of the upper panels at the $\mathrm{Mg}\,b$ and $\mathrm{H}\alpha$ region.
The zoomed-in regions are shown with gray blocks.
The SNRs of the blue and red arm are 29 and 61, respectively.}
\label{fig:demospec} 
\end{figure*}

\begin{figure}
\epsscale{1.2}
\plotone{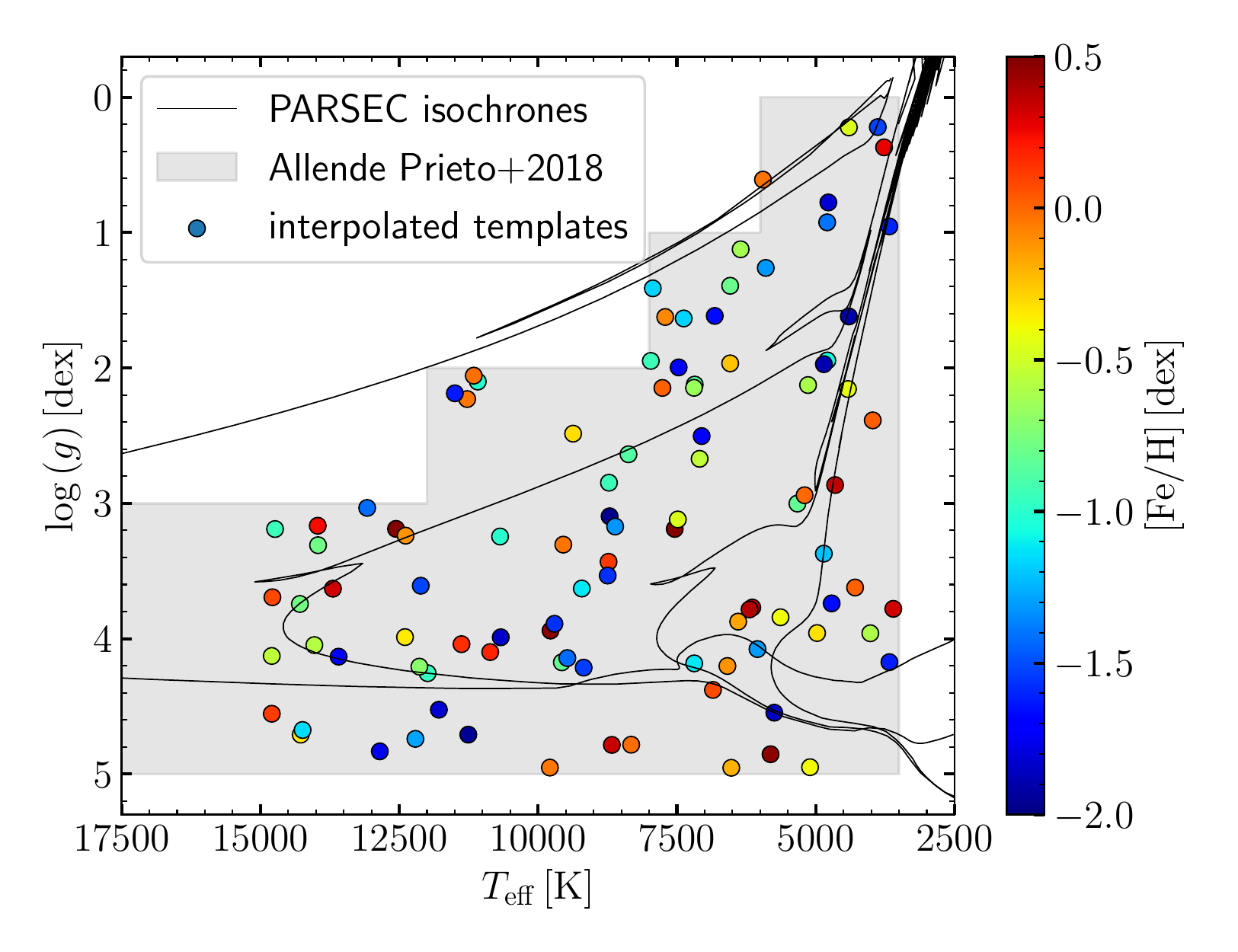}
\caption{The stellar parameters of the 100 randomly drawn templates that are used in RV measurements. The gray area shows the model grid coverage. The black solid lines are the PARSEC isochrones with solar metallicity and $\log_{10}{\tau}=7$, 8, 9 and 10, where $\tau$ is age/yr.}
\label{fig:temp_params}
\end{figure}

\section{The LAMOST MRS Survey} \label{sec:mrs}

\subsection{Targeting and Observational Rules}
The scientific goals of the LAMOST MRS survey mainly include stellar multiplicity, stellar pulsation, star formation, emission nebulae, Galactic archaeology, host stars of exoplanets, and open clusters.
The details of the scientific plan and the survey strategy are described in \citep{2020arXiv200507210L}.
In this section, we summarize them briefly.

Each scientific goal has a PI who is responsible for its targeting.
A {\ttfamily planid} is assigned to each MRS plate (pointing), which has a form of [TD/NT]hhmmss[N/S]ddmmssXnn, where the first two digits denote time-domain (TD, will be repeatedly observed during the five-year survey) or non-time-domain (NT, just observed in a single night), hhmmss[N/S]ddmmss represents for the equatorial coordinate of the plate center and [N/S] means north/south, the digit X is used to denote the scientific goal (see Table~\ref{tab:planid}), and the last two digits represent a serial number.
For example, the {\ttfamily planid} TD062610N184524B01 means it is a time-domain plate dedicated to binarity/multiplicity research.
There also exists some irregular {\ttfamily planid}s which are testing fields, such as HIP8426401, NGC216801, etc.
HIP8426401 means the central star of this plate is HIP84264, and NGC216801 means a plate toward the star cluster NGC2168.

The MRS survey uses the Local Modified Julian Minute (LMJM), an 8-bit integer defined as 1440 $\times$ the Local Modified Julian Date (LMJD) at the beginning of each exposure, as the stamps of each exposure.
Typical exposure time is set to 1200s, while 900s and 600s exposures also exist, depending on the luminosity of targets.
Each NT plate is observed with consecutive 3 exposures while each TD plate is observed until it is unobservable (usually 5-6 1200s exposures are allowed).
An arc lamp exposure is taken at the beginning of each plate, and at the end of an observing night (every $\sim$2 hours).
The targeting is mostly based on \textit{Gaia} DR2 source catalog \citep{2018A&A...616A...1G}.
For a 1200s exposure, the magnitudes corresponding to $\mathrm{S/N}=5$ in blue and red arm spectra are $G\sim$14.5 and 15.2 mag, respectively \citep[see Figure 1 in][]{2020arXiv200507210L}.
Brighter than this magnitude covers the majority of the objects observed by the LAMOST MRS survey.

\begin{deluxetable}{c|c|c}[h]
\tabletypesize{\scriptsize}
\tablenum{1}
\tablecaption{The abbreviation of scientific goals used in {\ttfamily planid}s. \label{tab:planid}}
\tablewidth{0pt}
\tablehead{
\colhead{X} & \colhead{NT/TD} & \colhead{Science}}
\startdata
K & TD & Kepler fields \\
H & TD & high frequency Kepler fields \\
B & TD & binarity / multiplicity\\
T & TD & TESS fields \\
M & NT & Milky Way \\
S & NT & star formation \\ 
C & NT & star cluster \\ 
N & NT & nebula \\
\enddata
\end{deluxetable}

\subsection{The LAMOST MRS Spectra}

\begin{figure}[ht!]
\plotone{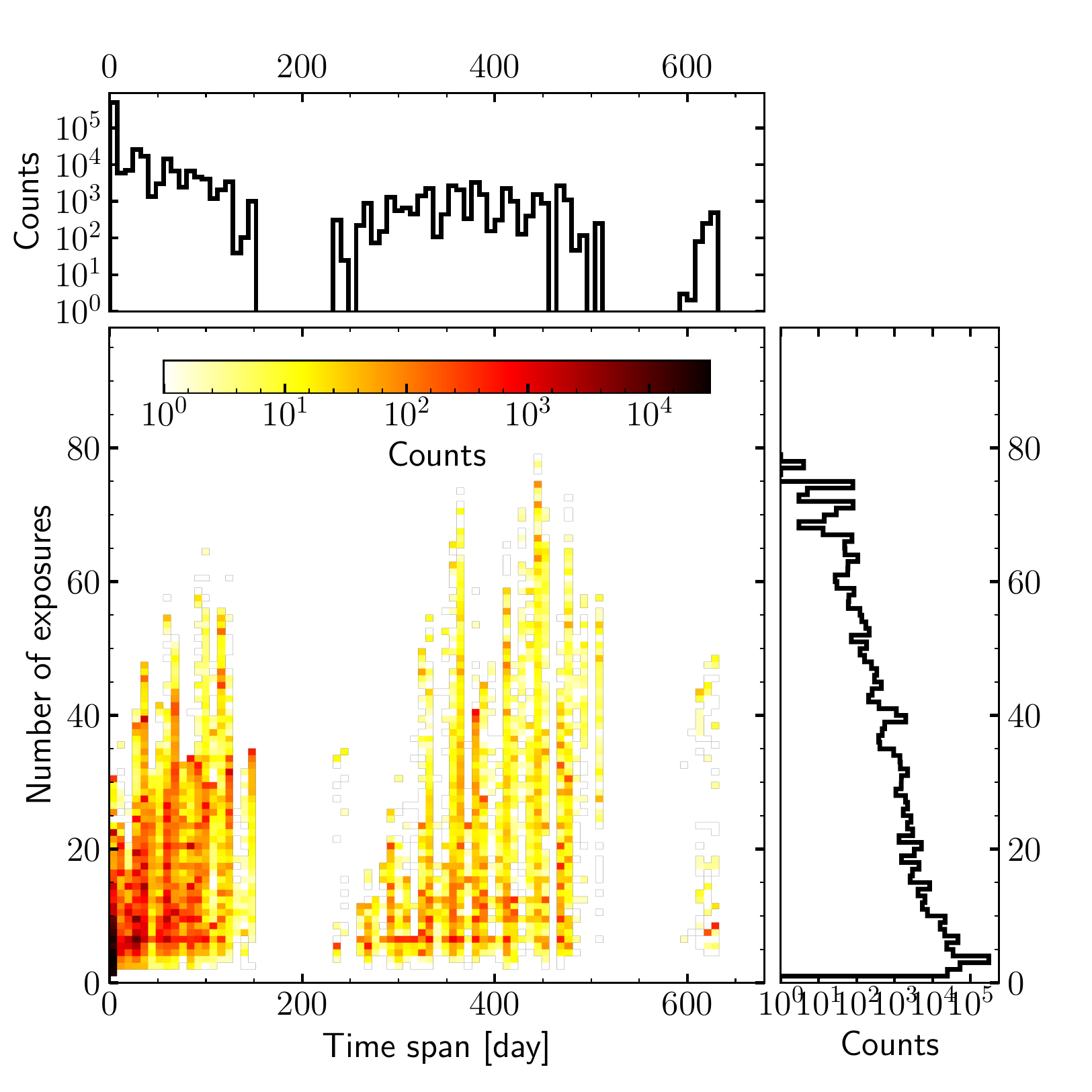}
\caption{The distributions of the number of exposures and the time span of the stars observed in LAMOST MRS DR7.}
\label{fig:mrs_stats}
\end{figure}

The whole LAMOST MRS DR7 ($R\sim7\,500$) contains over 5 million single-exposure spectra of over 800\,000 stars obtained from Sep. 2017 to Jun. 2019. 
In this work, 3\,753\,659 spectra with $\mathrm{SNR}>5$ either in blue arm or red arm for 600\,771 stars are selected.
Their distributions on number of exposures and time span is presented in Figure~\ref{fig:mrs_stats}.
The spectra are oversampled. Typical wavelength steps are 0.11 and 0.14 $\mathrm{\AA}$ at the blue and red arm, respectively.
The sampling rate ($\lambda/\Delta\lambda\sim45\,000$) is as high as 6 times the spectral resolution ($R\sim7\,500$).
In Figure \ref{fig:demospec}, we show a spectrum of a K-type giant star as a demo of the LAMOST MRS survey.
The blue arm and red arm are designed mainly for the Mg I $b$ triplet at around 5\,175$\mathrm{\AA}$ and H$\alpha$ at around 6\,564 $\mathrm{\AA}$.
The released spectra are not corrected with response curves.

\section{Measure Radial Velocities} \label{sec:rv}
\subsection{Preparation for RV Measurements}
\label{sec:preparation}
We noticed that cosmic rays frequently pollute the MRS single-exposure spectra and are neither identified nor removed by the LAMOST pipeline.
Therefore, in our method, the first step is to carefully remove the cosmic rays in spectra.
We smooth spectra with a 21-pixel median filter followed by a 9-pixel Gaussian filter, and remove the original pixels which deviate from the smoothed spectrum by 4 and 8 times the local standard deviation in the upper and lower direction.
The parameters of the filters are set empirically so that the absorption lines in the spectra of A-, F-, G- and K-type stars are not affected.
The removed pixels are replaced by values linearly interpolated using neighboring pixels.
We also note that the two ends of both blue and red arms are sometimes tilted and show unreasonable flux values probably due to extrapolation of the sky flux modeling, so we trim the 50 pixels at both edges of each arm.

The second step is to normalize spectra to pseudo-continuum to place the spectral features (usually absorption lines) on the same scale.
We iteratively fit the spectrum with a smoothing spline function and clip the pixels away from the median values by 3 times the standard deviation in each 100 $\mathrm{\AA}$ window.
The number of iterations is set to 3.
As shown in Figure \ref{fig:demospec}, for a typical K-type star with medium SNR, the normalization is quite adequate for the subsequent RV measurements.

\subsection{Spectral Templates}
We adopt the synthetic grid published by \cite{2018A&A...618A..25A} based on the ATLAS9 stellar atmosphere model \citep{1979ApJS...40....1K, 2012AJ....144..120M} as our spectral library.
We degrade the spectral resolution from 10\,000 to 7\,500 to fit the MRS configuration and converted from air wavelength to vacuum wavelength using the formula proposed by \cite{2000ApJS..130..403M}.
To limit the computational cost of the subsequent RV measurements at a reasonable amount, we interpolate the synthetic library to generate 100 spectral templates with stellar parameters randomly drawn from a uniform distribution in the range $3500<T_\mathrm{eff}/\mathrm{K}<15000$, $0<\log{g}/\mathrm{dex}<5$, $-2<\mathrm{[Fe/H]}/\mathrm{dex}<0.5$ and $-0.5<\mathrm{[\alpha/Fe]}/\mathrm{dex}<0.7$.
Extremely metal-poor and extremely hot templates are not considered because the spectral features are not significant.
Tests on high-SNR MRS spectra show that the sparsity of our templates induces a statistical error $\sim0.10$ and $0.20$ $\kms$ in the blue and red arm, respectively, which are negligible compared to other sources of uncertainties.
Figure \ref{fig:temp_params} shows the distribution of parameters of the 100 spectral templates and a series of PARSEC isochrones \citep{2012MNRAS.427..127B} with solar metallicity and logarithmic age $\log_{10}{\tau}=7$, 8, 9 and 10.
These spectral templates are normalized to pseudo-continuum using the same way as in Section \ref{sec:preparation}.

\begin{figure*}[ht!]
\plotone{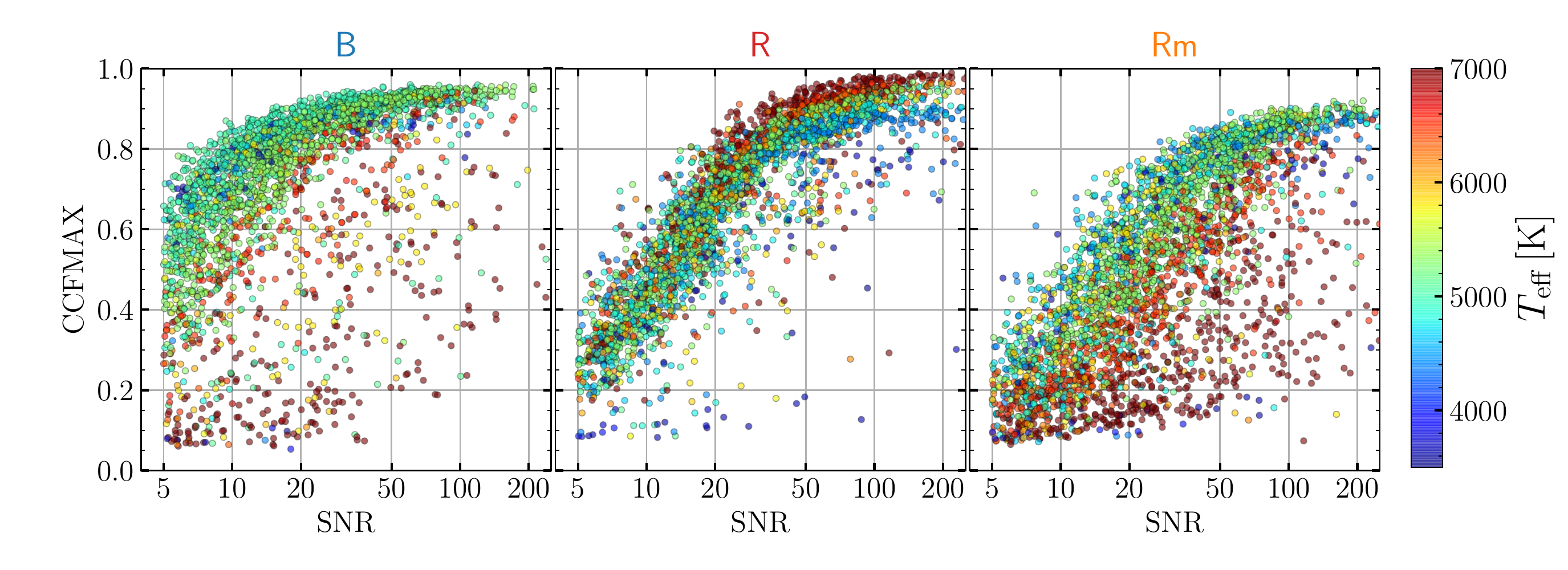}
\caption{The left/middle/right panel shows the SNR--CCFMAX relation for B(blue arm)/R(red arm)/Rm(red arm with $\mathrm{H}\alpha$ masked) CCF results of 5\,000 randomly selected spectra, respectively. Color denotes the $T_\mathrm{eff}$ of the best-match template.}
\label{fig:snr_ccfmax}
\end{figure*}

\begin{figure*}[ht!]
\plotone{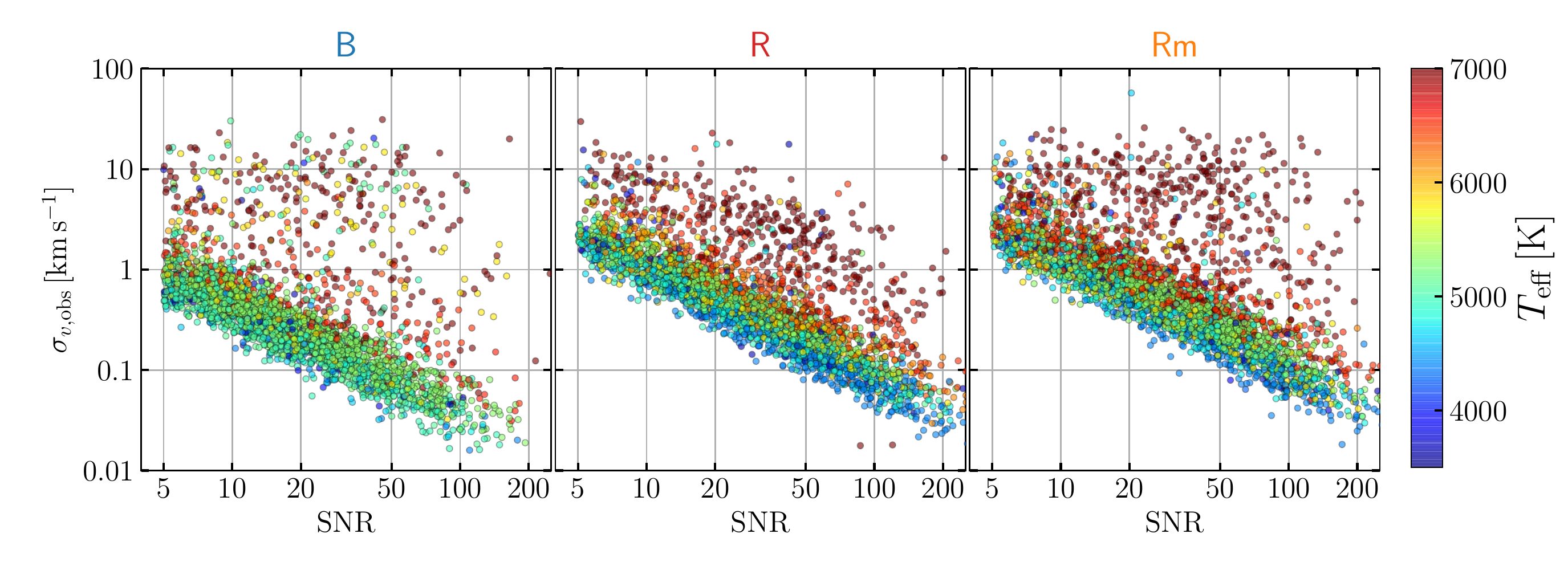}
\caption{The left/middle/right panel shows the SNR--$\sigma_{v,\mathrm{obs}}$ (measurement error) relation for B(blue arm)/R(red arm)/Rm(red arm with $\mathrm{H}\alpha$ masked) CCF results of 5000 randomly selected spectra, respectively. Color denotes the $T_\mathrm{eff}$ of the best-match template.}
\label{fig:snr_rverr}
\end{figure*}

\begin{figure*}[ht!]
\plotone{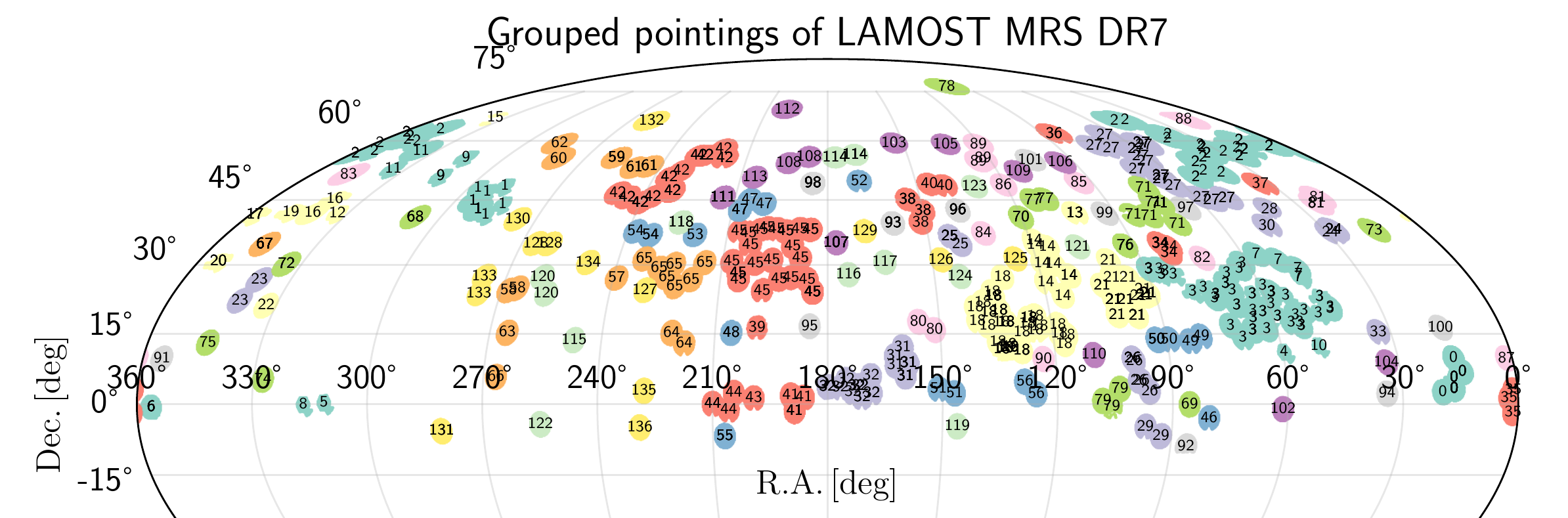}
\caption{Grouped pointings of the LAMOST DR7 MRS observations using a friends-of-friends method with a linking length of 5 degrees. The field-of-view of LAMOST is a circle with 2.5-degree radius if all spectrograph are at work. The figure uses equatorial coordinate system with Mollweide projection, and the pointings in a particular group are shown with the same color.}
\label{fig:groups}
\end{figure*}

\subsection{RV Estimates}
The cross-correlation function \citep[CCF,][]{1979AJ.....84.1511T} method is widely used in spectroscopic surveys to measure stellar RVs \citep[e.g.,][]{2015AJ....150..173N, 2020AJ....160...82S}.
One important advantage is that the CCF can be accelerated using Fast Fourier Transformation (FFT) once the spectrum is continuum-subtracted and re-sampled to a logarithmic wavelength grid.
However, the drawback of such a scheme is that the sampling of the resulting CCF is generally very sparse.
To evaluate the CCF at a smaller RV step, we do not follow the FFT way.
In our implementation, the CCF at an RV of $v$ is evaluated as
\begin{equation}
	\mathrm{CCF}(v|\boldsymbol{F},\boldsymbol{G}) = \frac{\mathrm{Cov}(\boldsymbol{F}, \boldsymbol{G}(v))}{ \sqrt{\mathrm{Var}(\boldsymbol{F})\mathrm{Var}(\boldsymbol{G}(v))}},
\end{equation}
where $\boldsymbol{F}$ is the vector of the normalized observed spectrum, $\boldsymbol{G}(v)$ is the vector of normalized synthetic spectrum shifted by an RV ($v$) and resampled to the wavelength grid of $\boldsymbol{F}$, $\mathrm{Var}$ represents the variance operator, and $\mathrm{Cov}$ represents the covariance operator (see Appendix \ref{app:covariance} for more details).

Deriving the final RV consists of three steps.
\begin{enumerate}
\item The initial estimates are made from an RV grid from $-$1500 to 1500 $\kms$, with a step of 10 $\kms$.
The template with the maximum CCF value is selected as the best-match template, and the corresponding radial velocity is adopted as the initial guess of the final RV of the observed star. And the parameters of the template ($T_\mathrm{eff}$, $\log{g}$, $\mathrm{[Fe/H]}$, $\mathrm{[\alpha/Fe]}$) are recorded, which make a good prior for some following analysis such as stellar atmospheric parameter determination.
\item With the best-match template, we maximize the CCF to determine the final RV ($v_\mathrm{obs}$) using the optimization routine {\tt scipy.optimize.minimize} with the Nelder-Mead algorithm \citep{Nelder1965ASM}. The corresponding CCF value is recorded as CCFMAX to assess the likelihood between the best-match template and the observed spectrum. The SNR--CCFMAX relations are shown in Figure \ref{fig:snr_ccfmax}.
\item To obtain the measurement error $\sigma_{v,\mathrm{obs}}$, we use a Monte Carlo method, namely, we repeat this process 100 times and in each time we add Gaussian random noise to the spectrum according to the flux error. The measurement error is computed using 16th and 84th percentiles, i.e., $\sigma_{v,\mathrm{obs}}=(v_{84}-v_{16})/2$. The SNR--$\sigma_{v,\mathrm{obs}}$ relations are shown in Figure \ref{fig:snr_rverr}.
\end{enumerate}
We avoid the Gaussian fitting to the CCF which is widely used in literature \citep[e.g., ][]{2015AJ....150..173N, 2019ApJS..244...27W}.
The reasons for doing this include that the CCF peak is obviously non-Gaussian and that at low signal-to-noise ratio (SNR) the fitting process is fragile.
The final error of RV includes measurement error and a noise floor term which is due to \replaced{systemtics}{systematics} (e.g., background, detector imperfections, temperature changes, focusing issues, etc) and will be assessed in the following part of this paper.

In this work, one RV is estimated for the blue arm ($v_\mathrm{B}$) and two for the red arm ($v_\mathrm{R}$ and $v_\mathrm{Rm}$).
The $v_\mathrm{Rm}$ is measured using the $\mathrm{H}\alpha$-masked red arm spectrum.
As shown in Figure \ref{fig:snr_ccfmax} and \ref{fig:snr_rverr}, these RV measurements deteriorate rapidly at $\mathrm{SNR}<20$.
For cool stars (e.g., FGK type), at a given SNR, $v_\mathrm{B}$ is more precise than $v_\mathrm{R}$ because of rich spectral features in blue arm \citep[for a detailed discussion of spectral information content in MRS spectra, see][]{2020RAA....20...51Z}.
However, for hot stars, $v_\mathrm{R}$ is more reliable because of the $\mathrm{H}\alpha$ feature and $v_\mathrm{Rm}$ is significantly less precise than $v_\mathrm{R}$  $\mathrm{H}\alpha$. Therefore, we recommend the readers to only consider $v_\mathrm{Rm}$ when the targets have $\mathrm{H}\alpha$ emission.

\begin{figure*}[ht!]
\plotone{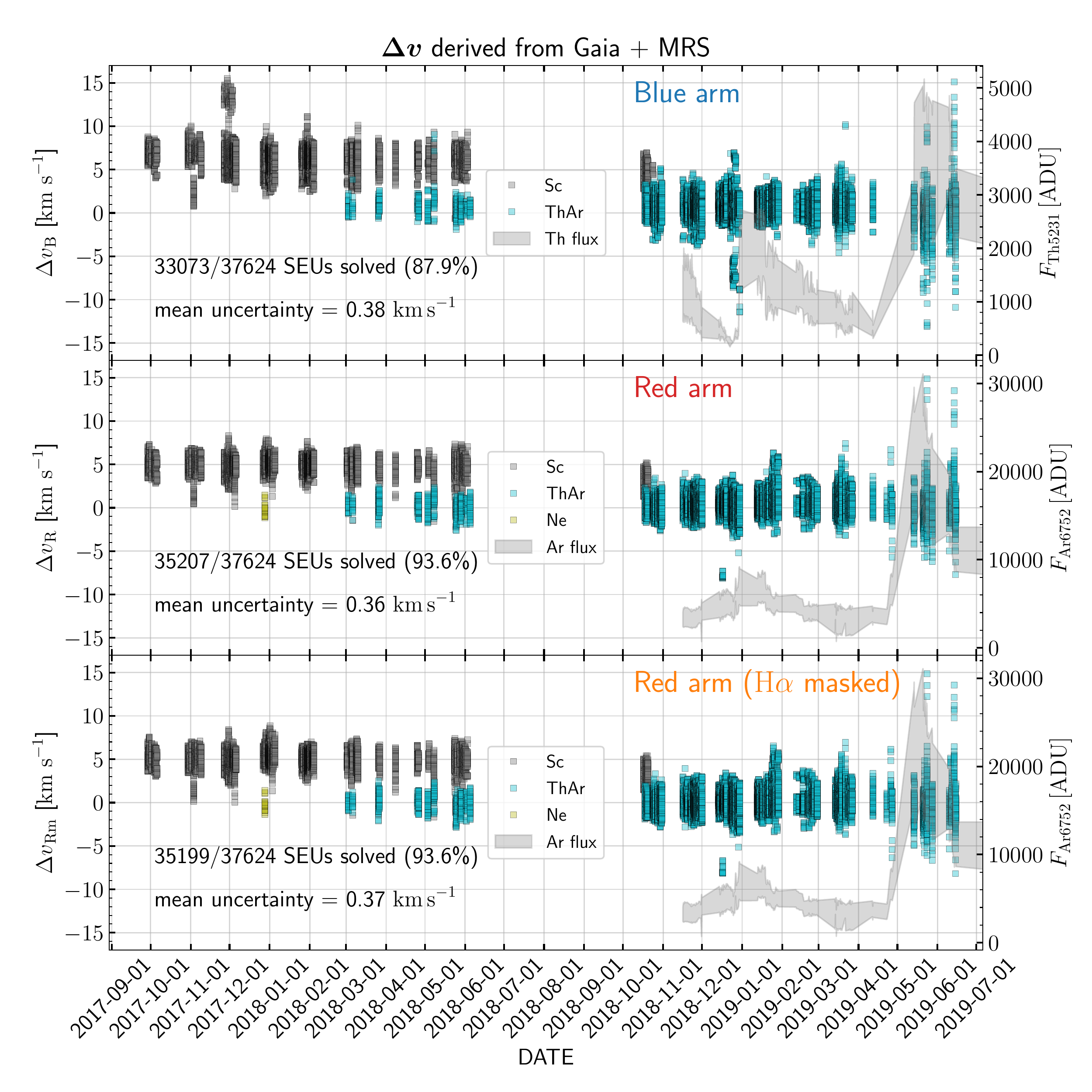}
\caption{The temporal variations of the RVZP correction values for the blue arm ($\Delta v_\mathrm{B}$), red arm ($\Delta v_\mathrm{R}$), and $\mathrm{H}\alpha$-masked red arm ($\Delta v_\mathrm{Rm}$), are shown in the top, middle and bottom panels, respectively.
The gray/cyan/olive square markers show the RVZP correction values of SEUs calibrated with Sc/ThAr/Ne lamp in LAMOST MRS DR7.
The max/min peak flux of Th 5231 $\mathrm{\AA}$ and Ar 6752 $\mathrm{\AA}$ of the 16 spectrographs are shown using gray area.}
\label{fig:rvzpcorr2}
\end{figure*}

\begin{figure*}[ht]
\plotone{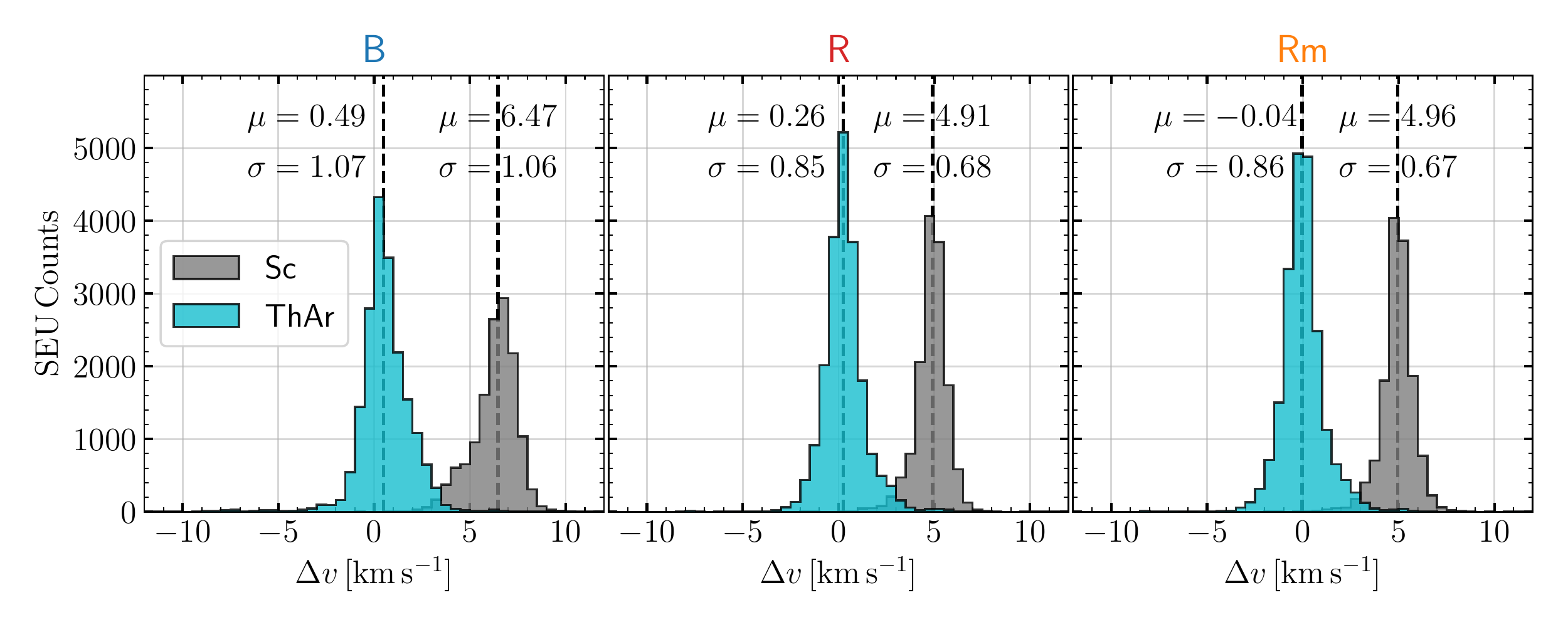}
\caption{The histograms of the $\Delta v_\mathrm{B}$, $\Delta v_\mathrm{R}$ and $\Delta v_\mathrm{Rm}$. Sc and ThAr lamp-calibrated data are shown in gray and cyan, respectively. The $\mu$s and $\sigma$s are calculated using the median and $(q84-q16)/2$, respectively.}
\label{fig:rvzpcorr_hist}
\end{figure*}

\begin{figure*}[ht]
\plotone{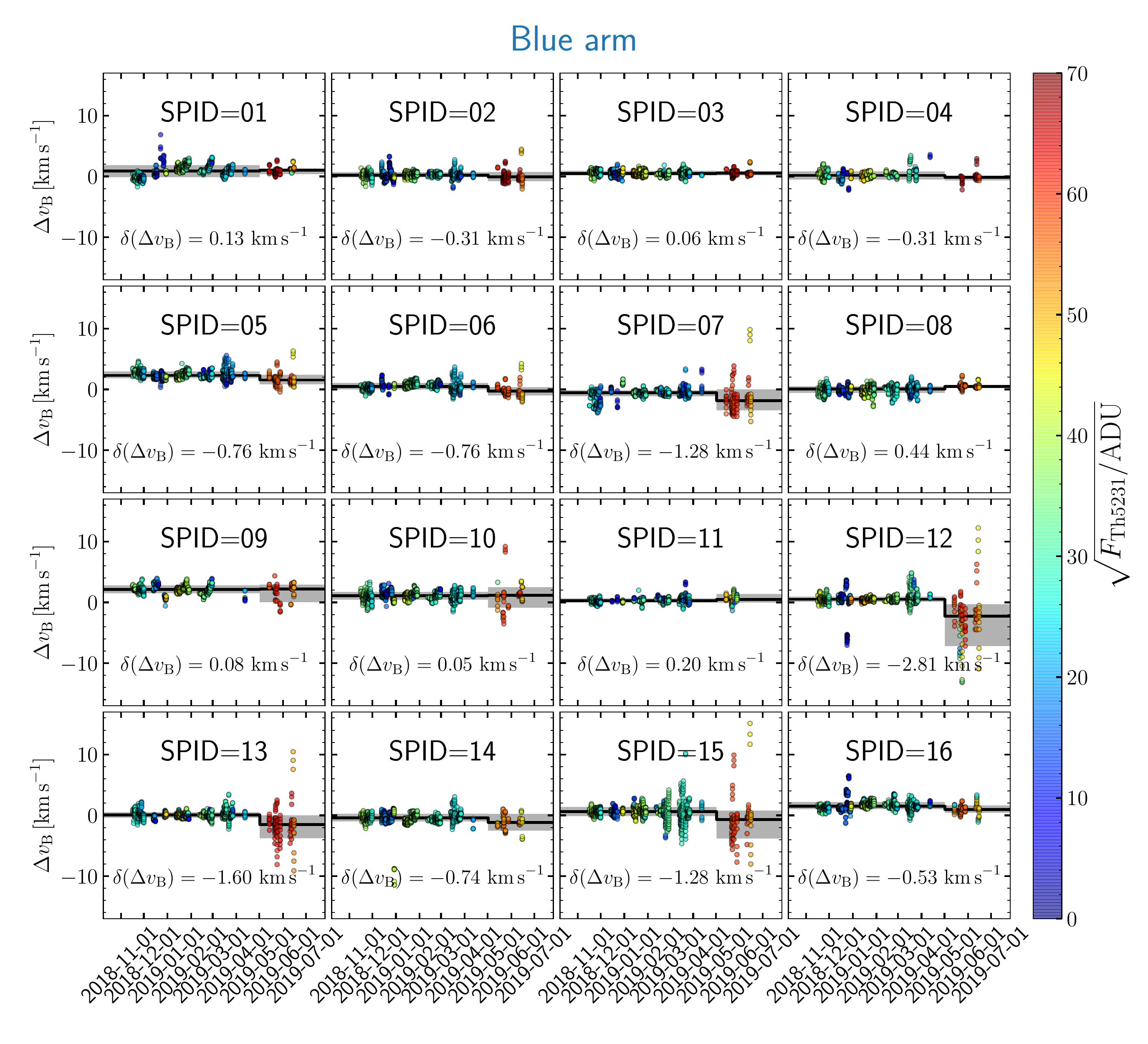}
\caption{The temporal variance of $\Delta v_\mathrm{B}$ before/after 1 May 2019 for the blue arm. The gray filled areas shows the 16 and 84th percentiles of the $\Delta v_\mathrm{B}$ in the two time intervals, and black solid lines show the medians. Color denotes the square root of the peak flux of the Th5231 line, which can be used as an indicator of SNR. Here the new batch of ThAr lamp also introduce RVZP variance.}
\label{fig:temporal_variance_B}
\end{figure*}

\begin{figure*}[ht]
\plotone{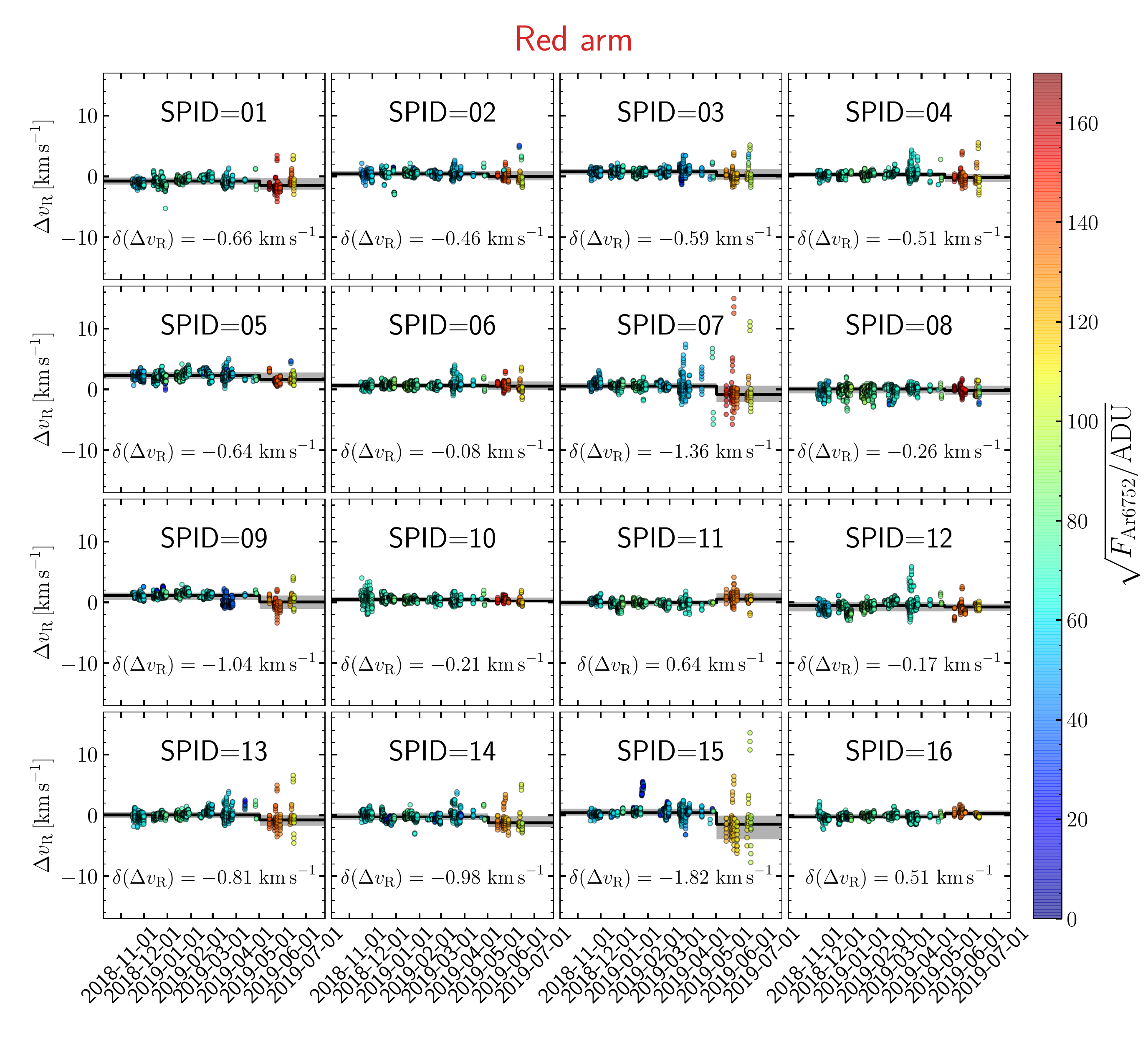}
\caption{The same figure as Figure \ref{fig:temporal_variance_B} but for red arm ($\Delta v_\mathrm{R}$). Color denotes the square root of the peak flux of Ar6752 line.}
\label{fig:temporal_variance_R}
\end{figure*}

\begin{figure*}[ht]
\epsscale{1.2}
\plotone{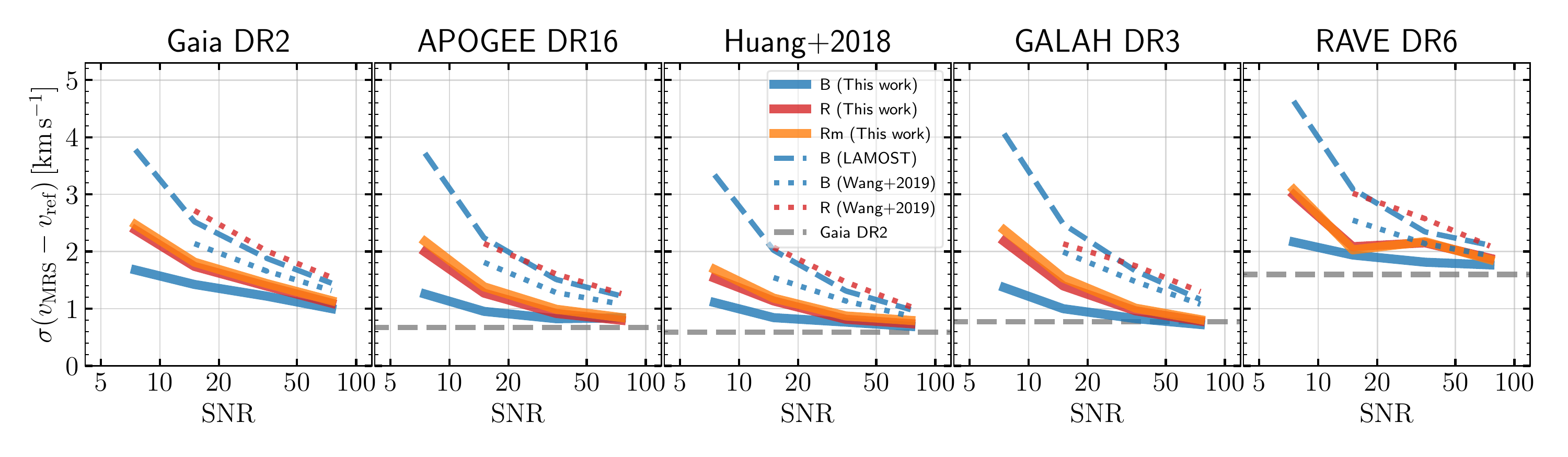}
\caption{The comparisons of our absolute RVs from LAMOST MRS DR7 \citep[including this work, the LAMOST pipeline, and][]{2019ApJS..244...27W} with other datasets, i.e., \textit{Gaia} DR2 \citep{2019A&A...622A.205K}, APOGEE DR16 \citep[Column {\ttfamily VHELIO\_AVG},][]{2020AJ....160..120J}, RV standard stars from \cite{2018AJ....156...90H}, GALAH DR3 \citep[Column {\ttfamily rv\_galah},][]{2020arXiv201102505B}, and RAVE DR6 \citep[Column {\ttfamily hrv\_sparv},][]{2020AJ....160...82S}. The scatters ($\sigma$s) are estimated using Gaussian fitting method.}
\label{fig:compare_rvscale}
\end{figure*}

\begin{figure*}[ht!]
\epsscale{1}
\plotone{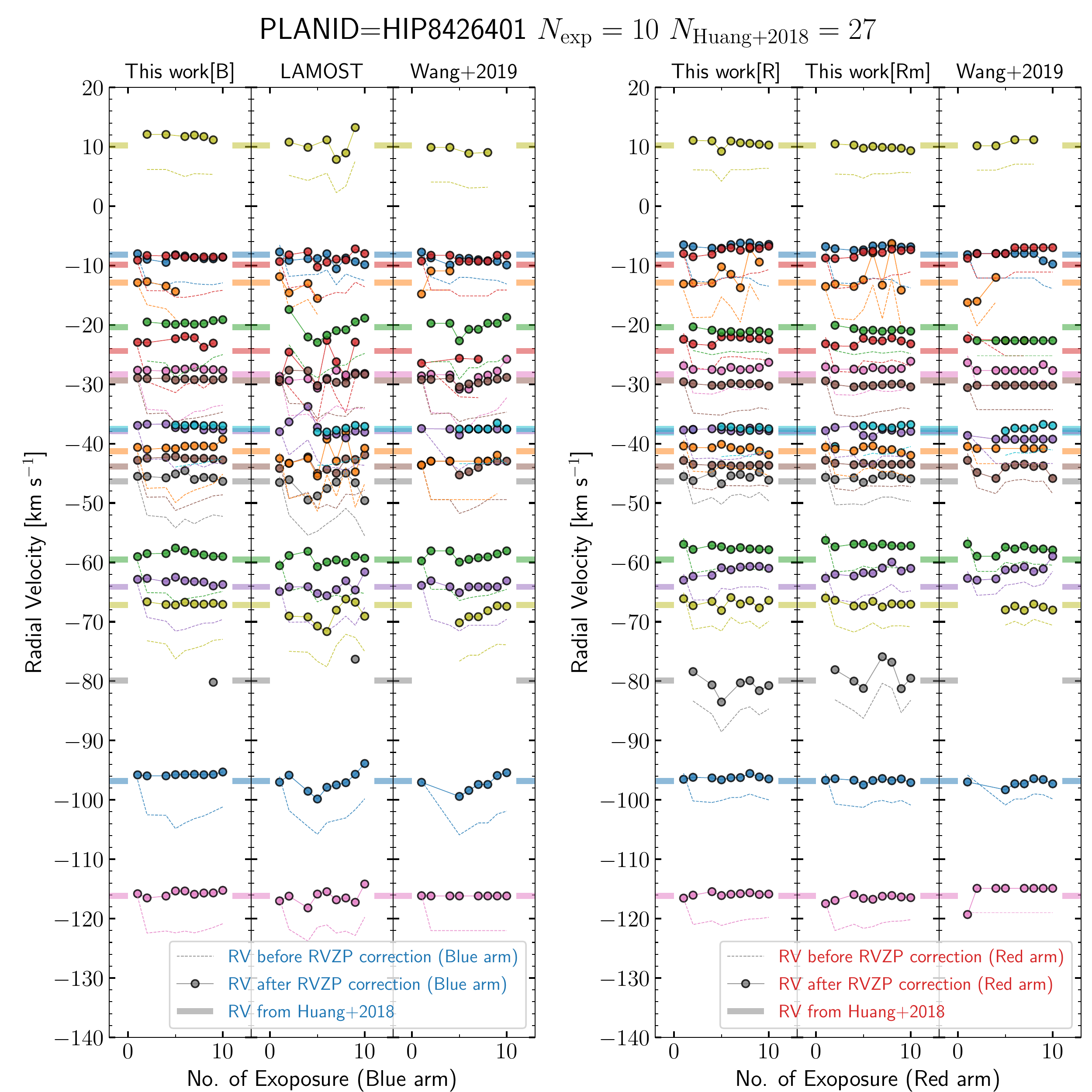}
\caption{A comparison of the RVs of standard stars observed in LAMOST MRS DR7 ({\tt planid}=HIP8426401) determined in this work, the LAMOST and \cite{2019ApJS..244...27W}. 
Each color represents a unique RV standard star.
The thick ticks show their RVs from \cite{2018AJ....156...90H}, the dashed lines show RVs before RVZP corrections, and the circles connected by solid lines show the RVZP-corrected RVs.
The first exposure is on Apr. 2018 and calibrated with Sc lamp, so the correction value is quite different from those of other exposures.}
\label{fig:rv_comparison_HIP8426401}
\end{figure*}

\begin{figure*}[ht!]
\plotone{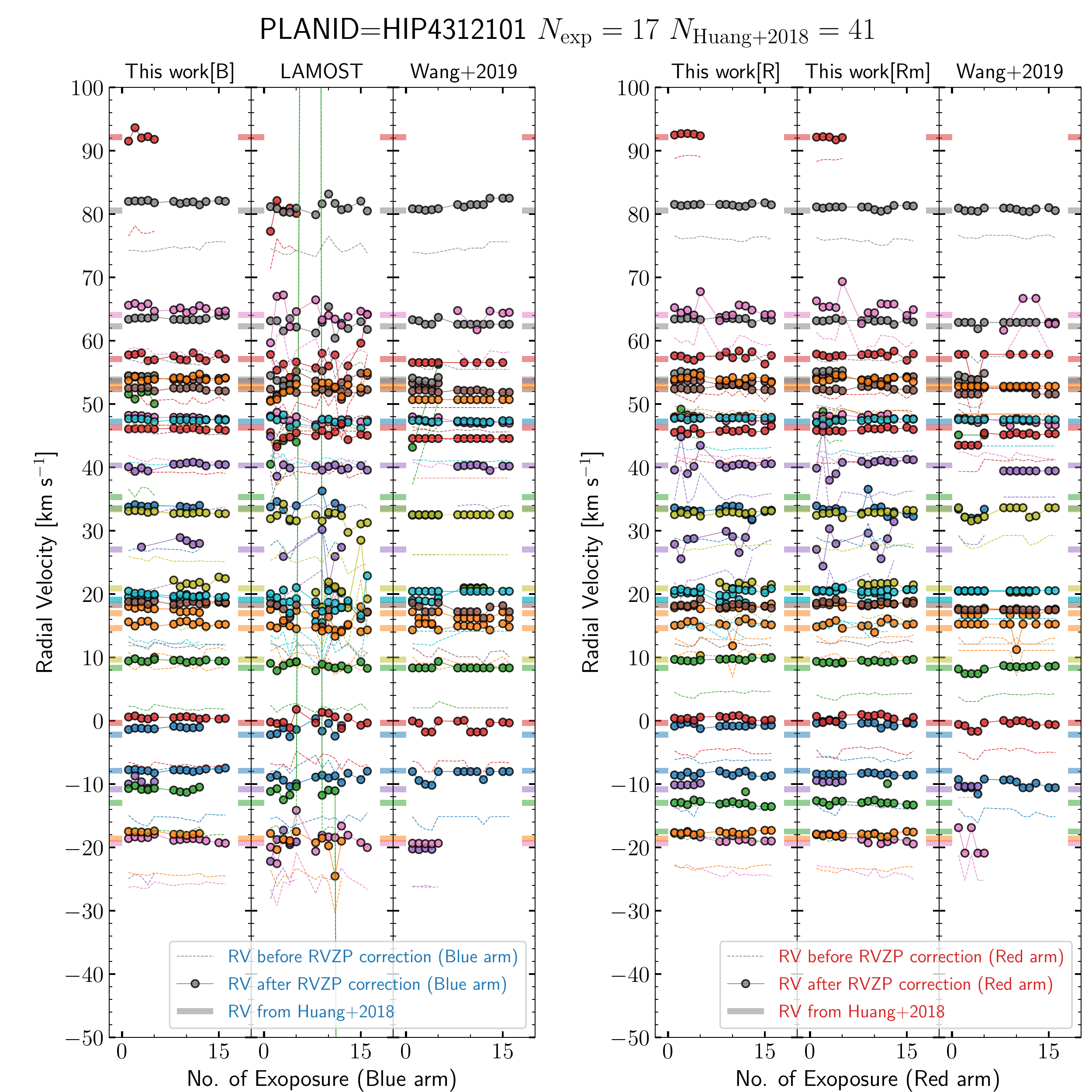}
\caption{The same figure as Figure~\ref{fig:rv_comparison_HIP8426401} but for {\ttfamily planid}=HIP4312101.}
\label{fig:rv_comparison_HIP4312101}
\end{figure*}

\section{Radial velocity zero-points} \label{sec:rvzpc}

In essence, the RV zero-point (RVZP) is the bias of the wavelength solution of a specific spectrum compared to its \textit{true} wavelength solution in terms of radial velocity.
It is affected by many factors, e.g., the condition of the instrument, the quality of the arc lamp exposure, the reduction algorithm, and the non-simultaneous nature of the arc lamp exposure and the object exposure, etc. 
In this paper, we define the RVZP correction value $\Delta v$ by 
\begin{equation} \label{eq:vabs}
	v_\mathrm{abs} = v_\mathrm{obs} + \Delta v,
\end{equation}
where $v_\mathrm{abs}$ is the absolute radial velocity and $v_\mathrm{obs}$ the radial velocity directly measured from a spectrum.

\subsection{The Scheme} \label{sec:schemes}
The LAMOST has 16 spectrographs of which each has 250 fibers (4\,000 fibers in total).
Excluding a few tens of sky fibers and a few problematic fibers, each spectrograph typically produces $\lesssim 200$ spectra in an exposure, depending on targeting, the condition of the instrument, the data quality, and the reduction algorithm.
Let $i$ denote the exposure epoch or LMJM, $j$ the spectrograph ID, and $k$ the fiber ID, ideally, we seek for the solution of the RVZP for each fiber, each spectrograph, and exposure by exposure, namely, the $\Delta v_{i,j,k}$.
This scheme is infeasible because the true/reference RVs $v_{i,j,k,\mathrm{abs}}$ of the targets are not always known.

In this work, assuming that the fibers in a spectrograph in one exposure (hereafter, we refer to it as a spectrograph-exposure-unit, or an SEU) share similar RVZPs, the systematic RVZP $\Delta v_{i,j}$ can be determined as long as a homogeneous reference set of RVs can be found for a fraction of fibers in that SEU.
The assumption is quite reasonable given the fact that the wavelength calibration of a multi-fiber spectrograph is done by fitting a 2D grating equation.
And, as we will see in Section \ref{sec:gaia}, the Gaia DR2 RVs \citep{2019A&A...622A.205K} meet our needs for the reference set.

As a constrast, both the LAMOST pipeline and \cite{2019ApJS..244...27W} calculate $\Delta v_j$ assuming the RVZPs for a specific spectrograph do not vary with time, and, therefore, get around the temporal variation of RVZPs.
However, we noticed that there exist some weird absolute RVs (as we will see in the results in Section \ref{sec:results}).

\subsection{The \textit{Gaia} DR2 RVs as the Reference Set} \label{sec:gaia}
Thanks to the ESA \textit{Gaia} mission\citep{2016A&A...595A...1G}, providing us the largest RV dataset that matches the LAMOST MRS survey in velocity precision and magnitude limit.
The spectral resolution of \textit{Gaia}-RVS \citep[$R\sim 11\,500$,][]{2018A&A...616A...5C} is slightly higher than the MRS ($R\sim 7\,500$).
The magnitude limit of the \textit{Gaia} DR2 RV catalog \citep{2019A&A...622A.205K} is at $G_\mathrm{RVS}=12\,\mathrm{mag}$ or $G\sim14\,\mathrm{mag}$ depending on spectral type and line-of-sight interstellar extinction, which is slightly brighter than the LAMOST MRS magnitude limit.
The \textit{Gaia} DR2 contains qualified median radial velocities for 7 224 631 stars derived from the \textit{Gaia}-RVS spectra, with $T_\mathrm{eff}$ in the range [3550, 6900] K excluding large RV variant stars \citep[see][for details]{2019A&A...622A.205K}.
At the faint end, $G_\mathrm{RVS}$ = 11.75 mag, the precisions for $T_\mathrm{eff}$ = 5000 and 6500 K are 1.4 and 3.7 $\kms$, respectively.

Aiming at studying time-domain astrophysical phenomena, e.g, spectroscopic binaries, we proceed to carry out the second-best scheme -- $\Delta v_{i,j}$.
Cross-matching the LAMOST MRS DR7 catalog with the \textit{Gaia} DR2, 1\,582\,948 out of 3\,753\,659 single-exposure spectra (42.1\%) have \textit{Gaia} RVs, and the common objects usually have good SNR in MRS because they are relatively bright in LAMOST MRS.
The number of objects in \textit{Gaia} DR2 is $\sim1\,000$ times larger than that in catalogs of RV standard stars such as \cite{2018AJ....156...90H}.
The challenge arises because not all of the 7 million objects in the \textit{Gaia}-RVS catalog are RV invariant, i.e., quite a number of them are pulsating stars or binary/multiple systems that have periodic/non-periodic RV variations.
In the below, we demonstrate a robust method that can determine the RVZP self-consistently for each spectrograph in each exposure ($\Delta v_{i, j}$) by comparing the observed RVs to the \textit{Gaia} DR2 RVs without identifying RV standard stars.

\subsection{Self-consistent RVZPs} \label{sec:scrvzp}
Assuming that the RV variables are varying with random periods at random phases or non-periodically, and are not the majority of the observed stars, we can regard them as outliers and use a robust method, e.g, the Least Absolute Residual (LAR) regression \citep[or Least Absolute Deviation regression, see][]{2002nrca.book.....P}, to estimate the $\Delta v_{i,j}$ (the common RV bias shared by the objects in an SEU).
From a Bayesian perspective, the LAR regression originates from an exponential likelihood while Least Squares (LSQ) regression comes from a Gaussian likelihood.
Utilizing the LAR technique, extreme values have a lesser influence on the fit compared to the LSQ regression.
Besides, since we aim at time-domain analysis, as long as our RVZPs are temporally self-consistent, the absolute scales are not very important.

Using the indices proposed in Section \ref{sec:schemes}, for each group of pointings, we construct a global cost function $f$ as below,
\begin{equation}
\begin{aligned}
    f \left(\boldsymbol{\Delta v}\right) = \Lambda_1 \sum_i \sum_{j} \sum_k \frac{\mid v_{i,j,k,\mathrm{obs}} + \Delta v_{i,j} - v_{i,j,k,\mathrm{GAIA}} \mid}{\sqrt{\sigma_\mathrm{min}^2+\sigma_{i,j,k,\mathrm{obs}}^2+\sigma_{i,j,k,\mathrm{GAIA}}^2}} \\
       + \Lambda_2 \sum_i \sum_{j} \sum_{k}  \frac{\mid v_{i,j,k,\mathrm{obs}} + \Delta v_{i,j} - \overline{v_{\cdot,\mathrm{obs}}} \mid}{\sqrt{\sigma_\mathrm{min}^2+\sigma_{i,j,k,\mathrm{obs}}^2+\sigma_{\cdot,\mathrm{obs}}^2}}, \\\label{eq:global_cost}
\end{aligned}
\end{equation} 
where $\boldsymbol{\Delta v}$ is the vector of $\{\Delta v_{i,j}\}$ for all relevant SEUs in the group of pointings, $v_{i,j,k,\mathrm{obs}}$ and $\sigma_{i,j,k,\mathrm{obs}}$ are the RV and associated measurement error of the $k$th star in the SEU $\{i,j\}$, $v_{i,j,k,\mathrm{GAIA}}$ and $\sigma_{i,j,k,\mathrm{GAIA}}$ are the \textit{Gaia} RV and associated uncertainty of the $k$th star in SEU $\{i,j\}$, $\sigma_\mathrm{min}$ is the noise floor of the measured RV which indicates the stability of wavelength calibration, i.e., the dispersion of $\Delta v_{i,j,k}$ in SEU $\{i,j\}$, $\Delta v_{i,j}$ is the RVZP correction value of SUE $\{i,j\}$ which is a free variable to be solved, $\overline{v_{\cdot,\mathrm{obs}}}$ and $\sigma_{\cdot,\mathrm{obs}}$ are the median and scatter of of the (RVZP-corrected) measured radial velocities of the star $\{i,j,k\}$ in other SEUs, $\Lambda_1$ and $\Lambda_2$ are the regularization parameters of the two terms.
In this scheme, the first term guarantees that the absolute scale of our RVZP-corrected RVs is close to that of \textit{Gaia} DR2, while the second term makes use of multiple exposures and guarantees that the relative RVZPs are self-consistent.
We set $\Lambda_1=\Lambda_2=1$ so that the final correction of each SEU is determined as an average of the two effects.
Then the vector $\boldsymbol{\Delta v}$ is determined by minimizing the cost function $f$, i.e., $\boldsymbol{\Delta v} = \mathrm{arg}~\mathrm{min} f$.
This algorithm can be implemented by minimizing 
\begin{equation}
\begin{aligned}
    f_{i,j} \left(\Delta v_{i,j} \right) =& \Lambda_1 \sum_k \frac{\mid v_{i,j,k,\mathrm{obs}} + \Delta v_{i,j} - v_{i,j,k,\mathrm{GAIA}} \mid}{\sqrt{\sigma_\mathrm{min}^2+\sigma_{i,j,k,\mathrm{obs}}^2+\sigma_{i,j,k,\mathrm{GAIA}}^2}} \\
       & + \Lambda_2   \sum_{k}  \frac{\mid v_{i,j,k,\mathrm{obs}} + \Delta v_{i,j} - \overline{v_{\cdot,\mathrm{obs}}} \mid}{\sqrt{\sigma_\mathrm{min}^2+\sigma_{i,j,k,\mathrm{obs}}^2+\sigma_{\cdot,\mathrm{obs}}^2}} \\
\label{eq:local_cost}       
\end{aligned}
\end{equation}
for each SEU $\{i,j\}$ sequentially and iteratively, where the $\Delta v_{i,j}$ is the RVZP correction value for SUE $\{i,j\}$.
Therefore, the problem is to solve the vector $\boldsymbol{\Delta v}$ which has an $N_\mathrm{SEU}$ elements where $N_\mathrm{SEU}$ is the number of related SEUs.
We claim that the $\boldsymbol{\Delta v}$ is determined when the $L_\infty$ norm of the difference between the solutions in $l$th and $(l+1)$th iteration is less than a specified value, i.e., $\mathrm{max}(\mid \boldsymbol{\Delta v}_{l}-\boldsymbol{\Delta v}_{l+1} \mid)<\epsilon$, where $\epsilon$ is the tolerance and is set to 0.075 $\kms$. 

The SEUs with RVZP correction values $|\Delta v_{i,j}|>50\,\kms$ or associated uncertainties $\sigma_{\Delta v_{i,j}}>10\,\kms$ (see Section \ref{sec:uncertainty}) are excluded in the iteration.
These results are generally because (1) \textit{Gaia} DR2 objects are $\leq 10$ (2) spectral SNRs are too low (3) bad spectra due to saturation or instrumental problems.
We do not think for these SEUs our scheme and assumptions are valid, so we only keep their initial guesses of $\Delta v_{i,j}$ (see Section \ref{sec:tricks}) and their uncertainties (see Section \ref{sec:uncertainty}) in our catalog (see Section \ref{sec:dataproducts}).

\subsection{Tricks to Accelerate the Algorithm} \label{sec:tricks}
Several tricks are used to accelerate the algorithm.
The first trick is to get a good initial estimation of $\boldsymbol{\Delta v}$.
A good approximation can be made by igoring the second term in Eq. (\ref{eq:local_cost}), so that with \textit{Gaia} DR2 RVs we can roughly estimate $\Delta v_{i,j}$ by minimizing  
\begin{equation}
    f_{i,j,\mathrm{init}} \left(\Delta v_{i,j} \right) = \sum_k \frac{\mid v_{i,j,k,\mathrm{obs}} + \Delta v_{i,j} - v_{i,j,k,\mathrm{GAIA}}\mid}{\sqrt{\sigma_\mathrm{min}^2+\sigma_{i,j,k,\mathrm{obs}}^2+\sigma_{i,j,k,\mathrm{GAIA}}^2}}.\\
\end{equation}
We note that if an SEU has only a few objects common with the \textit{Gaia} catalog, the estimation is risky.
Therefore, we require that an SEU at least have 10 objects common with \textit{Gaia} DR2 to proceed otherwise we calculate the initial guess $\Delta v_{i,j,\mathrm{init}}$ but exclude this SEU in the iteration process.

The second trick is to cut down $N_\mathrm{SEU}$ by separating physically detached SEUs, which fastens the index evaluation in each iteration.
In the Eq. (\ref{eq:global_cost}) and (\ref{eq:local_cost}), the second terms contain cross terms, meaning that when solving the $N_\mathrm{SEU}$ elements, an iteration process is needed to guarantee that the solution of $\boldsymbol{\Delta v}$ is stable.
However, the evaluation of indices is computationally expensive when the $N_\mathrm{SEU}$ grows.
Therefore, before performing the optimization process, physically detached sky areas can be separated, so that the many index evaluation processes can be accelerated by cutting down the array sizes.
We group the pointings of LAMOST MRS DR7 using a friends-of-friends algorithm with a 5-degree linking length which is double of the radius of the field-of-view of LAMOST in case of any possible common stars between them.
Eventually, 137 groups are obtained, as shown in Figure \ref{fig:groups}.
Initial guesses of $\Delta v$ for each SEU is made by optimizing the cost function Eq. (\ref{eq:global_cost}) for each group of plates.

In practice, we find that if the $\epsilon$ is too small, there is the possibility that the $\boldsymbol{\Delta v}$ jumps back and forth between two solutions and does not converge.
Further analysis shows that this is an optimization method related problem (we used the Nelder-Mead solver, changing it to the Powell solver does not solve the problem but the two solutions are different).
We guess that this might due to the numerical problem of the optimization routine.
To avoid such a situation, we then added random processes into the algorithm, i.e., if in $l$th iteration the solution is $\boldsymbol{\Delta v}_{l}$ and after looping over all related SEUs the solution is $\boldsymbol{\Delta v}_{l,\mathrm{opt}}$, we evaluate the $l+1$th solution by $\boldsymbol{\Delta v}_{l+1} = \eta (\boldsymbol{\Delta v}_{l,\mathrm{opt}}-\boldsymbol{\Delta v}_{l})+\boldsymbol{\Delta v}_{l}$, where the $\eta$ is a \textit{learning rate} randomly generated between $\eta_0$ and $\eta_1$.
We set $\eta_0=0.5$, $\eta_1=1.0$, and $\epsilon=0.075$, considering the expectation of $\eta$ is 0.75, the effective tolerance of our solution of $\boldsymbol{\Delta v}$ is 0.10 $\mathrm{km\,s}^{-1}$, which is acceptable when compared to the typical precision of measured radial velocities \citep[e.g., $\sim1.5\,\mathrm{km\,s}^{-1}$ reported by][]{2019ApJS..244...27W}.

Finally, the RVZP corrections for all the 137 groups of pointings converge after several tens of iterations with a Dell Precision R740 workstation with 2 Intel Xeon Platinum 8260 CPUs (2.40GHz), among which the longest solution takes $\sim$ 10 hours.
Compared to the computational cost of RV measurements using CCF for 3.8 million spectra including blue and red arms ($\sim1$ week on the same machine), computing the RVZPs is quite fast.

\subsection{Uncertainty Estimation} \label{sec:uncertainty}
Rigorous uncertainties are very difficult to obtain for our RVZP corrections.
The uncertainties of the RVZP correction values consists of two parts, namely the tolerance in the iteration process and the formal error.
The tolerance is $\epsilon=0.1~\kms$ as mentioned above.
For the latter part, based on the discussion presented in Appendix \ref{app:bias}, we use the 16th and 84th percentiles to construct a fiducial error of our RVZP correction values $\Delta v_{i,j}$ divided by an empirical correction $\xi$ to construct the formal error.
Hence, the total uncertainties of the RVZP correction values are evaluated via
\begin{equation}
	\sigma_{\boldsymbol \Delta v_{i,j}}^2 = \left( \frac{q84_{i,j}-q16_{i,j}}{2\, \xi \sqrt{N_{i,j}}}  \right)^2 + \epsilon^2,
\end{equation}
where ${i,j}$ index the SEUs, $q16$ and $q84$ denote the 16th and 84th percentiles of the residuals of the Gaia DR2 RVs and the RVZP-corrected LAMOST MRS RVs, the $N_{i,j}$ is the number of \textit{Gaia} DR2 objects with RVs, and $\xi$ is the empirical correction factor for small number statistics.

\section{Results and Validation}\label{sec:results}

In total, we have measured RVs from 3\,181\,157/3\,723\,934 single-exposure blue/red arm spectra in LAMOST MRS DR7 with SNR higher than 5. 
For 36\,301/37\,122/37\,122 (B/R/Rm) out of 37\,624 SEUs, we have successfully derived initial values of the RVZPs.
After eliminating the bad SEUs with criteria described at the end of Section \ref{sec:scrvzp}, we estimate the final RVZPs for 33\,073/35\,207/35\,199 (B/R/Rm) SEUs which cover 2\,985\,015/3\,631\,023/3\,629\,895 B/R/Rm RVs.
Roughly, the percentages of coverage are 87.9/93.6/93.6\% for B/R/Rm in terms of SEUs and 93.8/97.5/97.5\% for B/R/Rm in terms of RVs.

\subsection{The Temporal Variation of RVZPs}

In Figure \ref{fig:rvzpcorr2}, we present the $\Delta v_{i,j}$ for each SEU for Sc and ThAr arc lamps versus date.
The Sc lamp was in use until Oct. 2018, after when it is totally replaced by ThAr lamp.
The mean uncertainties of $\Delta v_\mathrm{B}$, $\Delta v_\mathrm{R}$ and $\Delta v_\mathrm{Rm}$ are all $\sim$ 0.38 $\kms$ which are quite good.
The median uncertainty is even 20\% less.
The $\Delta v_\mathrm{B}$ and $\Delta v_\mathrm{R}$ have different patterns while the $\Delta v_\mathrm{R}$ and $\Delta v_\mathrm{Rm}$ are very similar.
In Figure \ref{fig:rvzpcorr_hist} we show the distribution of the $\Delta v_\mathrm{B}$, $\Delta v_\mathrm{R}$ and $\Delta v_\mathrm{Rm}$ of the SEUs solved.
The $\mu$ here is estimated using the median, and $\sigma$ estimated using $(q84-q16)/2$.
The $\mu$s are 0.49 and 6.47 $\kms$ for ThAr and Sc lamp in the blue arm, while in the red arm they are 0.26 and 4.91 $\kms$, respectively.
The $\sigma$s are 1.07 and 1.06 $\kms$ for Thar and Sc lamp in blue arm, and in red arm are 0.85 and 0.68 $\kms$.
Generally, the different systematics for Sc and ThAr lamp-calibrated data, which is consistent with \citep{2019ApJS..244...27W}.
Despite the large systematics, we find that the precision of the Sc lamp calibrated data is no worse than those calibrated by ThAr lamp.
The Rm results are very similar to that of R, except their $\mu$s have a 0.3 $\kms$ difference.
We note that this is reasonable considering that the systematics vary with wavelength as shown in \cite{2020arXiv200806236R}.
In addition, $\sim$ 4\,000 single-exposure spectra calibrated using Ne lamp are also found in DR7 v1.1.
We confirm that the Ne lamp is used to calibrate the LRS spectra and these mistakenly calibrated data will be removed in DR7 v2 (the internationally public version), so we exclude these data in the following analysis.

It is clear that the RVZPs are reasonably stable except after $\sim$ 1 May 2019, which seems to be correlated with the arc lamp exposure flux\footnote{We also get access to the peak flux of Th5231 and Ar6752 of each ThAr lamp exposure since Nov. 2018 which are not included in the formal data release.}.
We plot the $\Delta v$ for each spectrograph for the time interval with available arc lamp intensity in Figure \ref{fig:temporal_variance_B} and \ref{fig:temporal_variance_R}.
Since our mean uncertainty of $\Delta v_{i,j}$ is 0.38 $\kms$, we regard the $\Delta v_{i,j}$ larger than 1 $\kms$ as significant values, including the 7, 12, 13, 15th spectrographs of the blue arm and 7, 9, 15th spectrographs of the red arm.
It is currently not sure whether these shifts are due to that the new ThAr lamps are brought into use at around 1 May 2019 or some other issues induced in the maintenance.
Probably in DR8, with the RVZP data in a longer time baseline we can address the problem.

\subsection{Validating with Other Datasets} \label{sec:valitation}
We also validate our RVs with the stars common in other datesets, namely, the \textit{Gaia} DR2 \citep{2019A&A...622A.205K}, APOGEE DR16 \citep[Column {\ttfamily VHELIO\_AVG},][]{2020AJ....160..120J}, RV standard stars from \cite{2018AJ....156...90H}, GALAH DR3 \citep[Column {\ttfamily rv\_galah},][]{2020arXiv201102505B}, and RAVE DR6 \citep[Column {\ttfamily hrv\_sparv},][]{2020AJ....160...82S}.
The mean $\mu$ and scatter $\sigma$ derived from Gaussian fitting to the residuals are tabulated in Table \ref{tab:precision} and shown as functions of SNR in Figure \ref{fig:compare_rvscale}.
Also shown are the results of the LAMOST pipeline and \cite{2019ApJS..244...27W}.
Note that, in DR7 v1.1, the LAMOST pipeline only provides two RVs measured from the blue arm using ELODIE empirical templates \citep{2004PASP..116..693M} and ATLAS9 synthetic templates \citep{2003IAUS..210P.A20C}, respectively.
In the next version of DR7 (v1.2) and future data releases, the RVs using ELODIE templates will be removed, and the RVs based on ATLAS9 will be provided for both blue and red arms for better performance.
Therefore, we use their calibrated RVs of blue arms based on ATLAS9 templates in our comparison (Column {\ttfamily rv\_ku1}).
The \cite{2019ApJS..244...27W} catalog is a subset of ours (including data taken during the first one year and a half with SNR cut at 10).
The spectra without our measurements (i.e., either $\mathrm{SNR}<5$ or $\Delta v$ is invalid) are excluded in this comparison, so that it is fair to the other two RV sources.
At high SNR end (50 to 100), we found the standard deviations derived from Gaussian fitting for our results can reach 1.00/1.10, 0.84/0.80, 0.69/0.74, 0.72/0.78, and 1.77/1.88 $\kms$ with respect to the \textit{Gaia} DR2, APOGEE DR16, \cite{2018AJ....156...90H}, GALAH DR3, and RAVE DR6 in blue/red arm.
The common stars among LAMOST MRS DR7, \textit{Gaia} DR2 and the other four reference sets are used to calculate the fiducial accuracy of \textit{Gaia} DR2. At the high-SNR end ($50<\mathrm{SNR}<100$), the LAMOST MRS gets close to the performance of \textit{Gaia} DR2 (see Figure \ref{fig:compare_rvscale}).
Note that, in our algorithm, the \textit{Gaia} DR2 RVs are used as a reference set, so the comparison with \textit{Gaia} DR2 is not an independent validation.
These results of the comparisons are quite fascinating.
At high-SNR end, we outperform \cite{2019ApJS..244...27W} and LAMOST pipeline by $\sim$ 20\% and 30\%, respectively, according to the blue arm results compared to APOGEE DR16.
At low-SNR end (10 to 20), this advantage increases to 58\% and 47\%, indicating that our algorithm of RV measurements and RVZP determinations are quite efficient.
Since the \cite{2018AJ....156...90H} sample is from APOGEE DR14, the comparison to \cite{2018AJ....156...90H} has a 0.4 $\kms$ systematic bias which does not exist in our comparison to APOGEE DR16.
This may be due to the update of the APOGEE data release.
The GALAH DR3 has a 0.23 $\kms$ systematic bias compared to \textit{Gaia} DR2.
The comparison with RAVE DR6, whose spectral resolution is the same as the LAMOST MRS but lower than \textit{Gaia}-RVS, APOGEE and GALAH, shows a large scatter but is still reasonable.

\begin{figure*}[ht]
\plotone{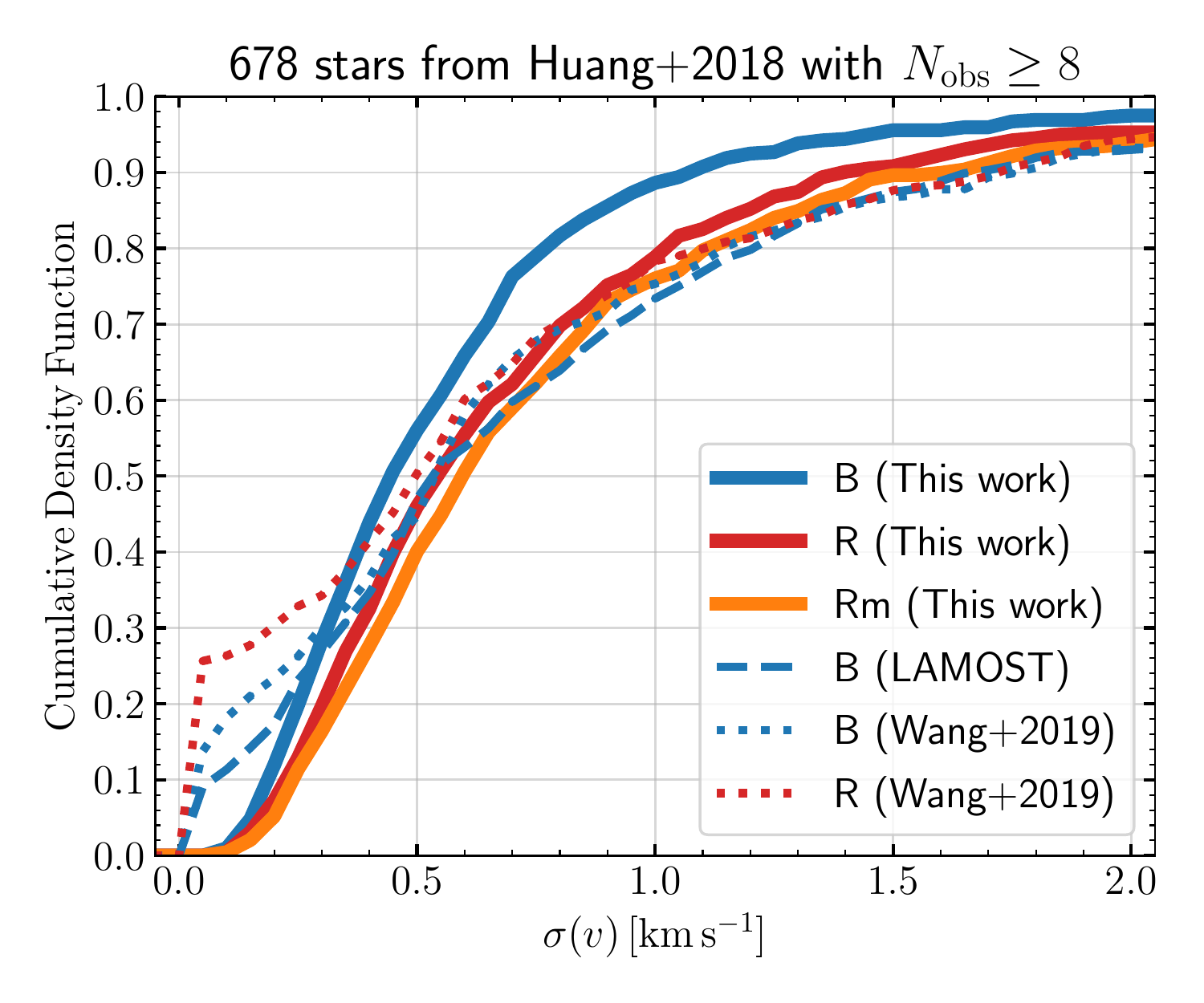}
\caption{The CDFs of the standard deviations of multiple observations for 678 RV standard stars \citep{2018AJ....156...90H} observed in LAMOST MRS DR7. The blue/red/orange color denote the B/R/Rm measurements, and the solid/dashed/dotted lines represent the results of this work/LAMOST/\cite{2019ApJS..244...27W}.}
\label{fig:consistency}
\end{figure*}

\begin{deluxetable*}{c|c|ccc|c|cc|c}
\tabletypesize{\scriptsize}
\tablenum{2}
\tablecaption{Compare the LAMOST MRS RVs to other datasets \label{tab:precision}}
\tablewidth{0pt}
\tablehead{
\colhead{} & \colhead{ } & \multicolumn3c{This Work} & \colhead{LAMOST} & \multicolumn2c{\cite{2019ApJS..244...27W}} & \colhead{\textit{Gaia} DR2} \\
\colhead{$\mathrm{SNR}$} & \colhead{Ref. Dataset} & \colhead{Blue arm} & \colhead{Red arm} & \colhead{$\mathrm{H}\alpha$ masked} & \colhead{Blue arm} & \colhead{Blue arm} & \colhead{Red arm} & \colhead{RVS}
}
\startdata
\multirow{12}{*}{\textit{Gaia} DR2} & {} & $N=166218$ & $N=89032$ & $N=89009$ & $N=166218$ & $N=0$ & $N=0$ & \multirow{10}{*}{$-$} \\
\multirow{12}{*}{} & $5<\mathrm{SNR}<10$ & $\mu=-0.02$ & $\mu=-0.05$ & $\mu=-0.07$ & $\mu=-0.56$ & $-$ & $-$ & \multirow{10}{*}{$-$} \\
\multirow{12}{*}{} & {} & $\sigma=1.68$ & $\sigma=2.38$ & $\sigma=2.47$ & $\sigma=3.78$ & $-$ & $-$ & \multirow{10}{*}{$-$} \\
\cline{2-8}
\multirow{12}{*}{} & {} & $N=301963$ & $N=164068$ & $N=164010$ & $N=301963$ & $N=209446$ & $N=26195$ & \multirow{12}{*}{} \\
\multirow{12}{*}{} & $10<\mathrm{SNR}<20$ & $\mu=-0.04$ & $\mu=-0.10$ & $\mu=-0.10$ & $\mu=-0.53$ & $\mu=-0.55$ & $\mu=-0.59$ & \multirow{12}{*}{} \\
\multirow{12}{*}{} & {} & $\sigma=1.42$ & $\sigma=1.74$ & $\sigma=1.81$ & $\sigma=2.52$ & $\sigma=2.14$ & $\sigma=2.72$ & \multirow{12}{*}{} \\
\cline{2-8}
\multirow{12}{*}{} & {} & $N=572355$ & $N=531044$ & $N=530913$ & $N=572355$ & $N=389277$ & $N=313059$ & \multirow{12}{*}{} \\
\multirow{12}{*}{} & $20<\mathrm{SNR}<50$ & $\mu=-0.03$ & $\mu=-0.07$ & $\mu=-0.08$ & $\mu=-0.44$ & $\mu=-0.46$ & $\mu=-0.66$ & \multirow{12}{*}{} \\
\multirow{12}{*}{} & {} & $\sigma=1.22$ & $\sigma=1.39$ & $\sigma=1.44$ & $\sigma=1.88$ & $\sigma=1.66$ & $\sigma=2.01$ & \multirow{12}{*}{} \\
\cline{2-8}
\multirow{12}{*}{} & {} & $N=270880$ & $N=497328$ & $N=497126$ & $N=270880$ & $N=180183$ & $N=322647$ & \multirow{12}{*}{} \\
\multirow{12}{*}{} & $50<\mathrm{SNR}<100$ & $\mu=0.00$ & $\mu=-0.00$ & $\mu=-0.01$ & $\mu=-0.33$ & $\mu=-0.35$ & $\mu=-0.58$ & \multirow{12}{*}{} \\
\multirow{12}{*}{} & {} & $\sigma=1.00$ & $\sigma=1.10$ & $\sigma=1.14$ & $\sigma=1.45$ & $\sigma=1.32$ & $\sigma=1.55$ & \multirow{12}{*}{} \\
\hline
\multirow{12}{*}{APOGEE DR16} & {} & $N=35352$ & $N=23362$ & $N=23354$ & $N=35352$ & $N=0$ & $N=0$ & \multirow{10}{*}{$N=28155$} \\
\multirow{12}{*}{} & $5<\mathrm{SNR}<10$ & $\mu=0.22$ & $\mu=0.08$ & $\mu=0.06$ & $\mu=-0.39$ & $-$ & $-$ & \multirow{10}{*}{$\mu=0.02$} \\
\multirow{12}{*}{} & {} & $\sigma=1.26$ & $\sigma=1.99$ & $\sigma=2.17$ & $\sigma=3.72$ & $-$ & $-$ & \multirow{10}{*}{$\sigma=0.67$} \\
\cline{2-8}
\multirow{12}{*}{} & {} & $N=49482$ & $N=38971$ & $N=38987$ & $N=49482$ & $N=33297$ & $N=4781$ & \multirow{12}{*}{} \\
\multirow{12}{*}{} & $10<\mathrm{SNR}<20$ & $\mu=0.13$ & $\mu=0.08$ & $\mu=0.02$ & $\mu=-0.41$ & $\mu=-0.46$ & $\mu=-0.47$ & \multirow{12}{*}{} \\
\multirow{12}{*}{} & {} & $\sigma=0.95$ & $\sigma=1.27$ & $\sigma=1.38$ & $\sigma=2.24$ & $\sigma=1.81$ & $\sigma=2.14$ & \multirow{12}{*}{} \\
\cline{2-8}
\multirow{12}{*}{} & {} & $N=80556$ & $N=87168$ & $N=87159$ & $N=80556$ & $N=52872$ & $N=48104$ & \multirow{12}{*}{} \\
\multirow{12}{*}{} & $20<\mathrm{SNR}<50$ & $\mu=0.06$ & $\mu=0.06$ & $\mu=0.03$ & $\mu=-0.41$ & $\mu=-0.43$ & $\mu=-0.49$ & \multirow{12}{*}{} \\
\multirow{12}{*}{} & {} & $\sigma=0.82$ & $\sigma=0.92$ & $\sigma=0.99$ & $\sigma=1.51$ & $\sigma=1.28$ & $\sigma=1.60$ & \multirow{12}{*}{} \\
\cline{2-8}
\multirow{12}{*}{} & {} & $N=41479$ & $N=73363$ & $N=73350$ & $N=41479$ & $N=26724$ & $N=45191$ & \multirow{12}{*}{} \\
\multirow{12}{*}{} & $50<\mathrm{SNR}<100$ & $\mu=-0.04$ & $\mu=0.04$ & $\mu=0.04$ & $\mu=-0.41$ & $\mu=-0.42$ & $\mu=-0.51$ & \multirow{12}{*}{} \\
\multirow{12}{*}{} & {} & $\sigma=0.84$ & $\sigma=0.80$ & $\sigma=0.85$ & $\sigma=1.22$ & $\sigma=1.08$ & $\sigma=1.26$ & \multirow{12}{*}{} \\
\hline
\multirow{12}{*}{\cite{2018AJ....156...90H}} & {} & $N=2753$ & $N=1659$ & $N=1661$ & $N=2753$ & $N=0$ & $N=0$ & \multirow{10}{*}{$N=2070$} \\
\multirow{12}{*}{} & $5<\mathrm{SNR}<10$ & $\mu=0.44$ & $\mu=0.33$ & $\mu=0.36$ & $\mu=0.28$ & $-$ & $-$ & \multirow{10}{*}{$\mu=0.34$} \\
\multirow{12}{*}{} & {} & $\sigma=1.11$ & $\sigma=1.53$ & $\sigma=1.69$ & $\sigma=3.34$ & $-$ & $-$ & \multirow{10}{*}{$\sigma=0.59$} \\
\cline{2-8}
\multirow{12}{*}{} & {} & $N=3404$ & $N=2890$ & $N=2890$ & $N=3404$ & $N=2316$ & $N=161$ & \multirow{12}{*}{} \\
\multirow{12}{*}{} & $10<\mathrm{SNR}<20$ & $\mu=0.43$ & $\mu=0.36$ & $\mu=0.36$ & $\mu=0.01$ & $\mu=-0.18$ & $\mu=0.18$ & \multirow{12}{*}{} \\
\multirow{12}{*}{} & {} & $\sigma=0.84$ & $\sigma=1.13$ & $\sigma=1.18$ & $\sigma=2.02$ & $\sigma=1.54$ & $\sigma=2.07$ & \multirow{12}{*}{} \\
\cline{2-8}
\multirow{12}{*}{} & {} & $N=5050$ & $N=5923$ & $N=5920$ & $N=5050$ & $N=3352$ & $N=2938$ & \multirow{12}{*}{} \\
\multirow{12}{*}{} & $20<\mathrm{SNR}<50$ & $\mu=0.42$ & $\mu=0.39$ & $\mu=0.40$ & $\mu=0.01$ & $\mu=-0.05$ & $\mu=0.10$ & \multirow{12}{*}{} \\
\multirow{12}{*}{} & {} & $\sigma=0.76$ & $\sigma=0.81$ & $\sigma=0.87$ & $\sigma=1.31$ & $\sigma=1.13$ & $\sigma=1.47$ & \multirow{12}{*}{} \\
\cline{2-8}
\multirow{12}{*}{} & {} & $N=2507$ & $N=4917$ & $N=4916$ & $N=2507$ & $N=1711$ & $N=3127$ & \multirow{12}{*}{} \\
\multirow{12}{*}{} & $50<\mathrm{SNR}<100$ & $\mu=0.38$ & $\mu=0.44$ & $\mu=0.47$ & $\mu=0.06$ & $\mu=0.05$ & $\mu=-0.08$ & \multirow{12}{*}{} \\
\multirow{12}{*}{} & {} & $\sigma=0.69$ & $\sigma=0.74$ & $\sigma=0.79$ & $\sigma=0.98$ & $\sigma=0.87$ & $\sigma=1.03$ & \multirow{12}{*}{} \\
\hline
\multirow{12}{*}{GALAH DR3} & {} & $N=30272$ & $N=17317$ & $N=17271$ & $N=30272$ & $N=0$ & $N=0$ & \multirow{10}{*}{$N=8778$} \\
\multirow{12}{*}{} & $5<\mathrm{SNR}<10$ & $\mu=0.32$ & $\mu=0.26$ & $\mu=0.22$ & $\mu=-0.25$ & $-$ & $-$ & \multirow{10}{*}{$\mu=0.23$} \\
\multirow{12}{*}{} & {} & $\sigma=1.37$ & $\sigma=2.18$ & $\sigma=2.37$ & $\sigma=4.06$ & $-$ & $-$ & \multirow{10}{*}{$\sigma=0.77$} \\
\cline{2-8}
\multirow{12}{*}{} & {} & $N=42242$ & $N=32885$ & $N=32768$ & $N=42242$ & $N=29780$ & $N=5135$ & \multirow{12}{*}{} \\
\multirow{12}{*}{} & $10<\mathrm{SNR}<20$ & $\mu=0.27$ & $\mu=0.29$ & $\mu=0.23$ & $\mu=-0.20$ & $\mu=-0.28$ & $\mu=-0.19$ & \multirow{12}{*}{} \\
\multirow{12}{*}{} & {} & $\sigma=1.00$ & $\sigma=1.40$ & $\sigma=1.54$ & $\sigma=2.48$ & $\sigma=1.99$ & $\sigma=2.13$ & \multirow{12}{*}{} \\
\cline{2-8}
\multirow{12}{*}{} & {} & $N=52015$ & $N=63642$ & $N=63593$ & $N=52015$ & $N=37996$ & $N=40401$ & \multirow{12}{*}{} \\
\multirow{12}{*}{} & $20<\mathrm{SNR}<50$ & $\mu=0.26$ & $\mu=0.27$ & $\mu=0.22$ & $\mu=-0.10$ & $\mu=-0.14$ & $\mu=-0.32$ & \multirow{12}{*}{} \\
\multirow{12}{*}{} & {} & $\sigma=0.83$ & $\sigma=0.96$ & $\sigma=1.01$ & $\sigma=1.66$ & $\sigma=1.47$ & $\sigma=1.75$ & \multirow{12}{*}{} \\
\cline{2-8}
\multirow{12}{*}{} & {} & $N=18796$ & $N=37239$ & $N=37239$ & $N=18796$ & $N=13727$ & $N=27296$ & \multirow{12}{*}{} \\
\multirow{12}{*}{} & $50<\mathrm{SNR}<100$ & $\mu=0.24$ & $\mu=0.29$ & $\mu=0.27$ & $\mu=-0.07$ & $\mu=-0.13$ & $\mu=-0.31$ & \multirow{12}{*}{} \\
\multirow{12}{*}{} & {} & $\sigma=0.72$ & $\sigma=0.78$ & $\sigma=0.81$ & $\sigma=1.16$ & $\sigma=1.08$ & $\sigma=1.30$ & \multirow{12}{*}{} \\
\hline
\multirow{12}{*}{RAVE DR6} & {} & $N=593$ & $N=414$ & $N=414$ & $N=593$ & $N=0$ & $N=0$ & \multirow{10}{*}{$N=1674$} \\
\multirow{12}{*}{} & $5<\mathrm{SNR}<10$ & $\mu=0.53$ & $\mu=0.93$ & $\mu=0.77$ & $\mu=0.13$ & $-$ & $-$ & \multirow{10}{*}{$\mu=0.33$} \\
\multirow{12}{*}{} & {} & $\sigma=2.17$ & $\sigma=2.99$ & $\sigma=3.08$ & $\sigma=4.63$ & $-$ & $-$ & \multirow{10}{*}{$\sigma=1.60$} \\
\cline{2-8}
\multirow{12}{*}{} & {} & $N=1273$ & $N=719$ & $N=719$ & $N=1273$ & $N=862$ & $N=146$ & \multirow{12}{*}{} \\
\multirow{12}{*}{} & $10<\mathrm{SNR}<20$ & $\mu=0.47$ & $\mu=0.50$ & $\mu=0.37$ & $\mu=-0.29$ & $\mu=-0.19$ & $\mu=0.14$ & \multirow{12}{*}{} \\
\multirow{12}{*}{} & {} & $\sigma=1.94$ & $\sigma=2.08$ & $\sigma=2.03$ & $\sigma=3.10$ & $\sigma=2.55$ & $\sigma=3.02$ & \multirow{12}{*}{} \\
\cline{2-8}
\multirow{12}{*}{} & {} & $N=3343$ & $N=2273$ & $N=2273$ & $N=3343$ & $N=2539$ & $N=1487$ & \multirow{12}{*}{} \\
\multirow{12}{*}{} & $20<\mathrm{SNR}<50$ & $\mu=0.39$ & $\mu=0.62$ & $\mu=0.55$ & $\mu=-0.09$ & $\mu=-0.16$ & $\mu=0.15$ & \multirow{12}{*}{} \\
\multirow{12}{*}{} & {} & $\sigma=1.81$ & $\sigma=2.15$ & $\sigma=2.17$ & $\sigma=2.34$ & $\sigma=2.14$ & $\sigma=2.57$ & \multirow{12}{*}{} \\
\cline{2-8}
\multirow{12}{*}{} & {} & $N=2332$ & $N=3209$ & $N=3209$ & $N=2332$ & $N=1798$ & $N=2368$ & \multirow{12}{*}{} \\
\multirow{12}{*}{} & $50<\mathrm{SNR}<100$ & $\mu=0.35$ & $\mu=0.41$ & $\mu=0.35$ & $\mu=0.08$ & $\mu=0.05$ & $\mu=-0.25$ & \multirow{12}{*}{} \\
\multirow{12}{*}{} & {} & $\sigma=1.77$ & $\sigma=1.88$ & $\sigma=1.86$ & $\sigma=2.11$ & $\sigma=1.93$ & $\sigma=2.10$ & \multirow{12}{*}{} \\
\hline
\enddata
\tablecomments{In each cell, we present the number of star ($N$), the Gaussian fitted systematic bias ($\mu/ \kms$) and the standard error ($\sigma/ \kms$) with respect to a specific reference set. }
\end{deluxetable*}

\begin{deluxetable*}{c|ccc|c|cc}
\tabletypesize{\normalsize}
\tablenum{3}
\tablecaption{The comparison of self-consistencies between this work, the LAMOST pipeline and \cite{2019ApJS..244...27W}.\label{tab:consistency}}
\tablewidth{0pt}
\tablehead{
\colhead{ } & \multicolumn3c{This Work} & \colhead{LAMOST} & \multicolumn2c{\cite{2019ApJS..244...27W}}  \\
\colhead{Percentile} & \colhead{Blue arm} & \colhead{Red arm} & \colhead{$\mathrm{H}\alpha$ masked} & \colhead{Blue arm} & \colhead{Blue arm} & \colhead{Red arm}
}
\startdata
 {$q=50\%$} & 0.45 (16.5\%) & 0.54 & 0.59 & 0.53 & 0.54 (-0.7\%) & 0.50 \\ 
 {$q=90\%$} & 1.07 (35.5\%) & 1.39 & 1.61 & 1.66 & 1.76 (-5.8\%) & 1.72 \\ 
 {$q=95\%$} & 1.45 (37.5\%) & 1.86 & 2.10 & 2.32 & 2.29 (1.5\%) & 2.08 \\
\hline
\enddata
\tablecomments{Each column shows the RV standard deviation for the 678 standard stars at corresponding levels of the CDF. In the parentheses we show the advantages over the LAMOST blue arm results.}
\end{deluxetable*}

\subsection{The Self-consistency}
Besides the precision, a check on the self-consistency is necessary before using RVs in time-domain research.
As a demonstration, in Figure \ref{fig:rv_comparison_HIP8426401} and \ref{fig:rv_comparison_HIP4312101} we validate the temporal RV variations of the standard stars (whose RVs are assumed invariant) from \cite{2018AJ....156...90H} in two pointings, namely, {\ttfamily planid}=HIP8426401 and HIP4312101, which have 10 and 17 exposures and contain 27 and 41 RV standard stars, respectively.
It is obvious that the RVs from the LAMOST pipeline show large fluctuations, which is as expected from the comparison in Section \ref{sec:valitation}.
It turns out that for many stars, the RVs of multiple exposures from \cite{2019ApJS..244...27W} catalog have exactly the same values.
This is due to the failure of their Gaussian fitting process and then they fall back on best estimation from the 1 $\kms$ RV grid, which is a defect in the algorithm.
By comparing the measured RVs and the RVZP-corrected RVs, we find that the RVZP-corrected RVs are much cleaner, indicating that the RVs do benefit from our algorithm for $\Delta v_{i, j}$ determination.
The LAMOST pipeline and \cite{2019ApJS..244...27W} assume that the RVZP for a spectrograph is static, so that the RVZP corrections are basically a shift of the RVs.

To quantify the performance in self--consistency, we select RV standard stars from \cite{2018AJ....156...90H} with at least 8 exposures have valid RVZP-corrected RVs in our catalog and calculated their standard deviation empirically corrected for the small-number-statistic effect which is discussed in Appendix \ref{app:bias}.
The cumulative distribution function (CDF) of the standard deviations is shown in Figure \ref{fig:consistency}.
The RVs measured from blue arm (B) shows the best consistency while those from red arm (R) and $\mathrm{H}\alpha$-masked red arm (Rm) follow.
By calculating the 50th, 90th and 95th percentiles of the CDFs for the blue arm results, we find our RVs have significant advantages over the LAMOST pipeline, namely 16.5\%, 35.5\% and 37.5\% better, while \cite{2019ApJS..244...27W} is at nearly the same level as LAMOST pipeline.
This reveals the excellent self-consistency of our RVs in the time-domain analysis.
Note that, these estimations of advantages are very conservative due to the facts such as \cite{2019ApJS..244...27W} dataset cut SNR at 10 but LAMOST and our sample cut at 5.
Besides, we do not exclude any of the "exactly the same" RVs from \cite{2019ApJS..244...27W} as readers may find that the CDFs for LAMOST and \cite{2019ApJS..244...27W} jump at around 0 $\kms$ position.

With the self-consistency performance, we select 10\,320 candidate RV standard stars by requiring that in blue arm (B) and red arm (R)
\begin{enumerate}
	\item their numbers of exposures are at least 8,
	\item their absolute RVs have standard deviations (corrected for small number statistics) are less than 1.45 and 1.85 $\kms$ \citep[corresponding to the $95\%$ level of the CDF, or $95\%$ completeness with respect to][]{2018AJ....156...90H},
	\item their time baselines are at least longer than 180 days.
\end{enumerate}
These stars can be useful for RV calibration of low-resolution surveys such the LAMOST LRS survey ($R \sim 1\,800$).

\section{Data products} \label{sec:dataproducts}
The data products of this work include
\begin{enumerate}
	\item a catalog of $\sim$3.8 million measured RVs (but 5 million rows for completeness), associated errors and the information of observations for $\sim 0.8$ million stars (Table \ref{tab:pub_rv}),
	\item a catalog of the $\Delta v_{i,j}$ for B, R, and Rm measurements in each SEU, and their uncertainties corresponding to $\sim$3.6 million RVs (Table \ref{tab:pub_rvzpc}). 
	\item a catalog of 10\,320 candidate RV standard stars with more than 8 exposures and standard deviation less than 1.45/1.86 $\kms$ in blue/red arm over a time baseline longer than 180 days (Table \ref{tab:pub_rvss}). 
\end{enumerate}
The catalogs can be cross-matched using the columns {\ttfamily spid} and {\ttfamily lmjm}.
They will be available with the journal and also at \url{https://github.com/hypergravity/paperdata} once the paper is accepted.

A few tips: the users who want to correct Doppler effects of their spectra \citep[e.g.,][]{2020ApJS..246....9Z} should use RVs without RVZP corrections, while the users who want to use absolute RVs can obtain them from our catalog via Eq (\ref{eq:vabs}).
The uncertainties of the absolute RVs can be evaluated via 
\begin{equation}
	\sigma_\mathrm{abs}^2 = \sigma_{v,\mathrm{obs}}^2 + \sigma_\mathrm{min}^2 + \sigma_{\Delta v}^2 + \sigma_\mathrm{mod}^2,
\end{equation}
where $\sigma_{v,\mathrm{obs}}$ is the measurement error, $\sigma_\mathrm{min}$ is the wavelength calibration error floor which we can infer from the comparison to APOGEE DR16 that it is approximately 0.85 $\kms$ or conservatively 1 $\kms$, $\sigma_{\Delta v}$ is the uncertainty of the RVZP, and $\sigma_\mathrm{mod}$ is the contribution from the sparsity of the spectral templates (0.10 $\kms$ for blue arm and 0.20 $\kms$ for red arm).

\begin{splitdeluxetable*}{ccccccccccccBcccccccccccc}
\tabletypesize{\scriptsize}
\tablewidth{0pt} 
\tablenum{4}
\tablecaption{The 3.8 million RVs obtained from the LAMOST MRS DR7 (v1.1). \label{tab:pub_rv}}
\tablehead{
\colhead{OBSID} & \colhead{LMJM} & \colhead{BJD} & \colhead{PLANID} & \colhead{SPID} & \colhead{FIBERID} & \colhead{RA} & \colhead{DEC} & \colhead{SNRB} & \colhead{SNRR} & \colhead{LAMPB} & \colhead{LAMPR} & \colhead{V\_B} & \colhead{VERR\_B} & \colhead{TEFF\_B} & \colhead{CCFMAX\_B} & \colhead{V\_R} & \colhead{VERR\_R} & \colhead{TEFF\_R} & \colhead{CCFMAX\_R} & \colhead{V\_RM} & \colhead{VERR\_RM} & \colhead{TEFF\_RM} & \colhead{CCFMAX\_RM}\\
\colhead{ } & \colhead{ } & \colhead{ } & \colhead{ } & \colhead{ } & \colhead{ } & \colhead{$\mathrm{deg}$} & \colhead{$\mathrm{deg}$} & \colhead{ } & \colhead{ } & \colhead{ } & \colhead{ } & \colhead{$\kms$} & \colhead{$\kms$} & \colhead{ $\rm K$} & \colhead{ } & \colhead{$\kms$} & \colhead{$\kms$} & \colhead{ $\rm K$} & \colhead{ } & \colhead{$\kms$} & \colhead{$\kms$} & \colhead{ $\rm K$} & \colhead{ }}
\colnumbers
\startdata
588902003 & 83556146 & 2458025.2773235 & HIP507401 & 2 & 3 & 15.3672776 & 4.0094024 & 25.2 & 44.7 & sc & sc & -10.847 & 0.140 & 4976.0 & 0.921 & -8.859 & 0.158 & 5813.7 & 0.829 & -8.847 & 0.176 & 5813.7 & 0.815 \\
588902003 & 83556132 & 2458025.2679947 & HIP507401 & 2 & 3 & 15.3672776 & 4.0094024 & 29.8 & 52.1 & sc & sc & -10.628 & 0.078 & 4976.0 & 0.929 & -8.637 & 0.157 & 4976.0 & 0.845 & -8.859 & 0.191 & 5813.7 & 0.832 \\
588902003 & 83556119 & 2458025.2587006 & HIP507401 & 2 & 3 & 15.3672776 & 4.0094024 & 30.6 & 52.4 & sc & sc & -10.735 & 0.090 & 4976.0 & 0.931 & -8.297 & 0.183 & 4976.0 & 0.850 & -8.599 & 0.144 & 5813.7 & 0.842\\
\enddata
\tablecomments{Column 1 is the observational ID in LAMOST, Column 2 is the local Modified Julian Minite (1440 $\times$ the local Modified Julian Date) of the beginning of exposure which is an 8-bit integer assigned to each exposure, Column 3 is the barycentric Julian Day of the middle of exposure, Column 4 is the plan ID, Column 5 is the spectrograph ID, Column 6 is the fiber ID, Column 7-8 are equatorial coordinates, Column 9-10 are the SNRs in blue and red arms, Column 11-12 are the calibration arc lamp. Column 13-16 are the measured RV, associated measurement error, the effective temperature of the best-match template, the maximum of the CCF, respectively, for the blue arm. Column 17-20 and 21-24 are for the red arm and H$\alpha$-masked red arm. We refer readers to \url{http://dr7.lamost.org/v1.1/doc/mr-data-production-description} for detailed explanations of the Column 1-6.}
\end{splitdeluxetable*}

\begin{splitdeluxetable*}{ccccccccccccBcccccccccBccccccccc}
\tabletypesize{\scriptsize}
\tablewidth{0pt} 
\tablenum{5}
\tablecaption{The RVZP correction values for LAMOST MRS DR7 (v1.1). \label{tab:pub_rvzpc}}
\tablehead{
\colhead{PLANID} & \colhead{LMJM} & \colhead{SPID} 
& \colhead{DV0\_B} & \colhead{N\_B} & \colhead{DV\_B} & \colhead{NF1\_B} & \colhead{NF2\_B} & \colhead{NO\_MED\_B} & \colhead{NO\_MAX\_B} & \colhead{NO\_MIN\_B} & \colhead{DVERR\_B} 
& \colhead{DV0\_R} & \colhead{N\_R} & \colhead{DV\_R} & \colhead{NF1\_R} & \colhead{NF2\_R} & \colhead{NO\_MED\_R} & \colhead{NO\_MAX\_R} & \colhead{NO\_MIN\_R} & \colhead{DVERR\_R} 
& \colhead{DV0\_RM} & \colhead{N\_RM} & \colhead{DV\_RM} & \colhead{NF1\_RM} & \colhead{NF2\_RM} & \colhead{NO\_MED\_RM} & \colhead{NO\_MAX\_RM} & \colhead{NO\_MIN\_RM} & \colhead{DVERR\_RM} \\
\colhead{} & \colhead{} & \colhead{} 
& \colhead{$\kms$} & \colhead{} & \colhead{$\kms$} & \colhead{} & \colhead{} & \colhead{} & \colhead{} & \colhead{} & \colhead{} 
& \colhead{$\kms$} & \colhead{} & \colhead{$\kms$} & \colhead{} & \colhead{} & \colhead{} & \colhead{} & \colhead{} & \colhead{} 
& \colhead{$\kms$} & \colhead{} & \colhead{$\kms$} & \colhead{} & \colhead{} & \colhead{} & \colhead{} & \colhead{} & \colhead{} \\}
\colnumbers
\startdata
HIP9645901 & 83555758 & 2 & 6.412 & 156 & 6.353 & 111 & 82 & 7 & 47 & 3 & 0.188 & 5.012 & 159 & 4.886 & 113 & 88 & 7 & 47 & 3 & 0.183 & 5.176 & 159 & 5.056 & 113 & 88 & 7 & 47 & 3 & 0.190 \\
HIP9645901 & 83555758 & 3 & 8.004 & 170 & 7.998 & 132 & 168 & 3 & 46 & 0 & 0.178 & 3.921 & 170 & 3.905 & 132 & 170 & 3 & 46 & 3 & 0.169 & 3.958 & 170 & 3.954 & 132 & 170 & 3 & 46 & 3 & 0.195 \\
HIP9645901 & 83555758 & 4 & 6.126 & 159 & 6.128 & 127 & 158 & 3 & 7 & 2 & 0.213 & 3.738 & 163 & 3.706 & 129 & 163 & 3 & 7 & 3 & 0.192 & 3.818 & 163 & 3.830 & 129 & 163 & 3 & 7 & 3 & 0.202 \\
\enddata
\tablecomments{Column 1 is the plan ID, Column 2 is the local Modified Julian Minute, Column 3 is the spectrograph ID. Column 4-12 are the initial RVZP correction value ($\Delta v_{i,j,\mathrm{init}}$), number of objects, final RVZP correction value ($\Delta v_{i,j}$), number of objects for the first term of the Eq. (\ref{eq:local_cost}), number of objects for the second term of the Eq. (\ref{eq:local_cost}), the median number of exposures, the maximum number of exposures, the minimum number of exposures, the estimated uncertainty, respectively, for the blue arm. Column 13-21 and 22-30 are for the red arm and H$\alpha$-masked red arm, respectively.}
\end{splitdeluxetable*}

\begin{deluxetable*}{cccccccccc}
\tabletypesize{\scriptsize}
\tablenum{6}
\tablecaption{The selected candidates of RV standard stars from LAMOST MRS DR7 (v1.1). \label{tab:pub_rvss}}
\tablehead{
\colhead{RA} & \colhead{DEC} & \colhead{RVMED\_B} & \colhead{RVSTD\_B} & \colhead{NEXP\_B} & \colhead{TBL\_B} & \colhead{RVMED\_R} & \colhead{RVSTD\_R} & \colhead{NEXP\_R} & \colhead{TBL\_R}\\
\colhead{deg} & \colhead{deg} & \colhead{$\kms$} & \colhead{$\kms$} & \colhead{} & \colhead{day} & \colhead{$\kms$} & \colhead{$\kms$} & \colhead{} & \colhead{day}\\
}
\colnumbers
\startdata
15.367275 & 4.009401 & -4.644 & 0.350 & 47 & 427.896 & -4.215 & 0.358 & 47 & 427.896 \\
15.306525 & 3.796851 & 15.830 & 1.027 & 30 & 452.830 & 16.281 & 1.063 & 33 & 452.830 \\
15.541034 & 3.710481 & 20.450 & 0.884 & 29 & 427.896 & 20.973 & 0.894 & 36 & 427.896\\
\enddata
\tablecomments{Column 1 and 2 are equatorial coordinates, Column 3-6 are the median RV, standard deviation, number of exposures and the time baseline in the blue arm, respectively, and Column 7-10 are for the red arm.}
\end{deluxetable*}

\section{Summary} \label{sec:conclusion}
In this paper, we measure the RVs from LAMOST Medium-Resolution Survey (MRS) DR7 stellar spectra and determine the RVZPs with the help of \textit{Gaia} DR2 RVs, aiming at making the absolute RVs self-consistent and proper for time-domain analysis.
More specifically,
\begin{enumerate}
	\item we have measured RVs of $\sim$ 3.8 million single-exposure spectra for more than 0.8 million stars obtained from the LAMOST MRS DR7, including the blue arm, red arm (with and without H$\alpha$),
	\item we determine the RVZPs exposure by exposure (for 3.6 million spectra) by comparing the measured RVs to the \textit{Gaia} DR2 and multiple MRS exposures using a robust method to a mean precision of 0.38 $\kms$,  
	\item we find the RVZP vary significantly for some spectrographs before/after 1 May 2019, which confirmed the necessity of our algorithm to determine the RVZPs,
	\item we find good agreements in the comparisons of our absolute RVs with APOGEE DR16, RV standard stars \citep{2018AJ....156...90H}, GALAH DR3, and RAVE DR6, and the precision at $50<\mathrm{SNR}<100$ can reach 1.00/1.10, 0.84/0.80, 0.69/0.74, 0.72/0.78, and 1.77/1.88 $\kms$ in the blue/red arm, respectively,
	\item we show that our absolute RVs have 16.5, 35.5, and 37.5\% better self-consistency at 50\%, 90\%, and 95\% levels of the CDF of standard deviations, respectively, which benefits subsequent time-domain analysis.
	\item we select a set of \replaced{9\,708}{10\,320} candidate RV standard stars whose standard deviations of RVs are less than 1.45 and 1.86 $\kms$ in the blue arm and red arm, respectively, over a time baseline of at least 180 days.
\end{enumerate} 
The algorithms and results presented in this work will be used in the subsequent works on spectroscopic binaries (Xiong et al. in prep., Zhang et al. in prep.). 

The LAMOST MRS DR7 v1.2 and v1.3 have been released.
We confirm that, in DR7 v1.2/1.3 the spectra are the same as v1.1 but catalogs and parameters will have some changes\footnote{\url{http://dr7.lamost.org/v1.2/doc/dr7_update}}\footnote{\url{http://dr7.lamost.org/v1.3/doc/dr7_update}}.
Therefore, our results can be cross-matched with the v1.2/1.3 catalogs directly.
And we will release a new version of RVs on github once DR8 is released.
On the other hand, since \textit{Gaia} eDR3 is the same as in DR2 but with moderate filtering, our absolute RVs should be consistent with \textit{Gaia} eDR3.
In future LAMOST MRS data releases, we will update our RVs using the most recent \textit{Gaia} RVs as reference set.
 
\acknowledgments
BZ thanks Qin Lai, Feng Luo, Dr. Rui Wang, Dr. Hai-Bo Yuan, Prof. Jian-Jun Chen, Prof. Hao-Tong Zhang, Prof. A-Li Luo, Prof. Jian-Rong Shi for constructive discussion and generous help.
BZ also thanks the support from the LAMOST FELLOWSHIP fund.

JNF acknowledges the support from the National Natural Science Foundation of China (NSFC) through the grants 11833002, 12090042 and 12090040.

This work is supported by National Key R\&D Program of China No. 2019YFA0405500. C.L. thanks the National Natural Science Foundation of China (NSFC) with grant No. 11835057.

This work is supported by Chinese Space Station Telescope (CSST) pre-research projects of \textit{Key Problems in Binaries and Chemical Evolution of the Milky Way and its Nearby Galaxies}, the National Natural Science Foundation of China (No. U1931209).

W.Z. is supported by the Fundamental Research Funds for the Central Universities.

The authors also thank the reviewer of this paper for the patience and constructive comments during these days.

The LAMOST FELLOWSHIP is supported by Special Funding for Advanced Users, budgeted and administrated by Center for Astronomical Mega-Science, Chinese Academy of Sciences (CAMS). This work is supported by Cultivation Project for LAMOST Scientific Payoff and Research Achievement of CAMS-CAS.


Guoshoujing Telescope (the Large Sky Area Multi-Object Fiber Spectroscopic
Telescope LAMOST) is a National Major Scientific Project built by the Chinese
Academy of Sciences. Funding for the project has been provided by the National
Development and Reform Commission. LAMOST is operated and managed by the
National Astronomical Observatories, Chinese Academy of Sciences.

%

\vspace{5mm}
\facilities{LAMOST}


\software{
laspec\citep{laspec},
regli\citep{regli},
berliner\citep{berliner},
scikit-learn\citep{2012arXiv1201.0490P},
astropy\citep{2018AJ....156..123A}}



\appendix
\section{Cross-correlation Function (CCF)} \label{app:covariance}
In this section, we explain our definitions of mean, variance, covariance, and CCF.
Let $\boldsymbol{X}$ denote a vector containing $N$ elements $\{X_i\}$ (e.g., a continuum-normalized spectrum with $N$ pixels), the mean is defined as
\begin{equation}
	\overline{\boldsymbol{X}} = \frac{1}{N}\sum_i X_i,
\end{equation}
and the variance as
\begin{equation}
	\mathrm{Var}(\boldsymbol{X}) = \frac{1}{N} \sum_i (X_i  - \overline{\boldsymbol{X}})^2,
\end{equation}
and the covariance of two vectors $\boldsymbol{X}$ and $\boldsymbol{Y}$ as
\begin{equation}
	\mathrm{Cov}(\boldsymbol{X},\boldsymbol{Y}) = \frac{1}{N} \sum_i (X_i  - \overline{\boldsymbol{X}})(Y_i  - \overline{\boldsymbol{Y}}).
\end{equation}
A normalize cross-correlation function can be calculated with standardized $f$ and $g$, namely 
\begin{equation}
	\mathrm{CCF}(v|\boldsymbol{F},\boldsymbol{G}) = \frac{\mathrm{Cov}(\boldsymbol{F}, \boldsymbol{G}(v))}{ \sqrt{\mathrm{Var}(\boldsymbol{F})\mathrm{Var}(\boldsymbol{G}(v))}},
\end{equation}
where $\boldsymbol{F}$ is one vector and $\boldsymbol{G}(v)$ is the other vector but shifted by radial velocity $v$.
When utilizing this CCF to estimate stellar RVs, the $\boldsymbol{G}(v)$ is usually a spectral template whose signal-to-noise ratio is infinite and covers the wavelength range of $\boldsymbol{F}$.
Therefore, the shift could be implemented with an interpolation.
The CCF in this form is essentially the linear correlation coefficient and varies between $-1$ and $1$.

\begin{figure*}[ht]
\plottwo{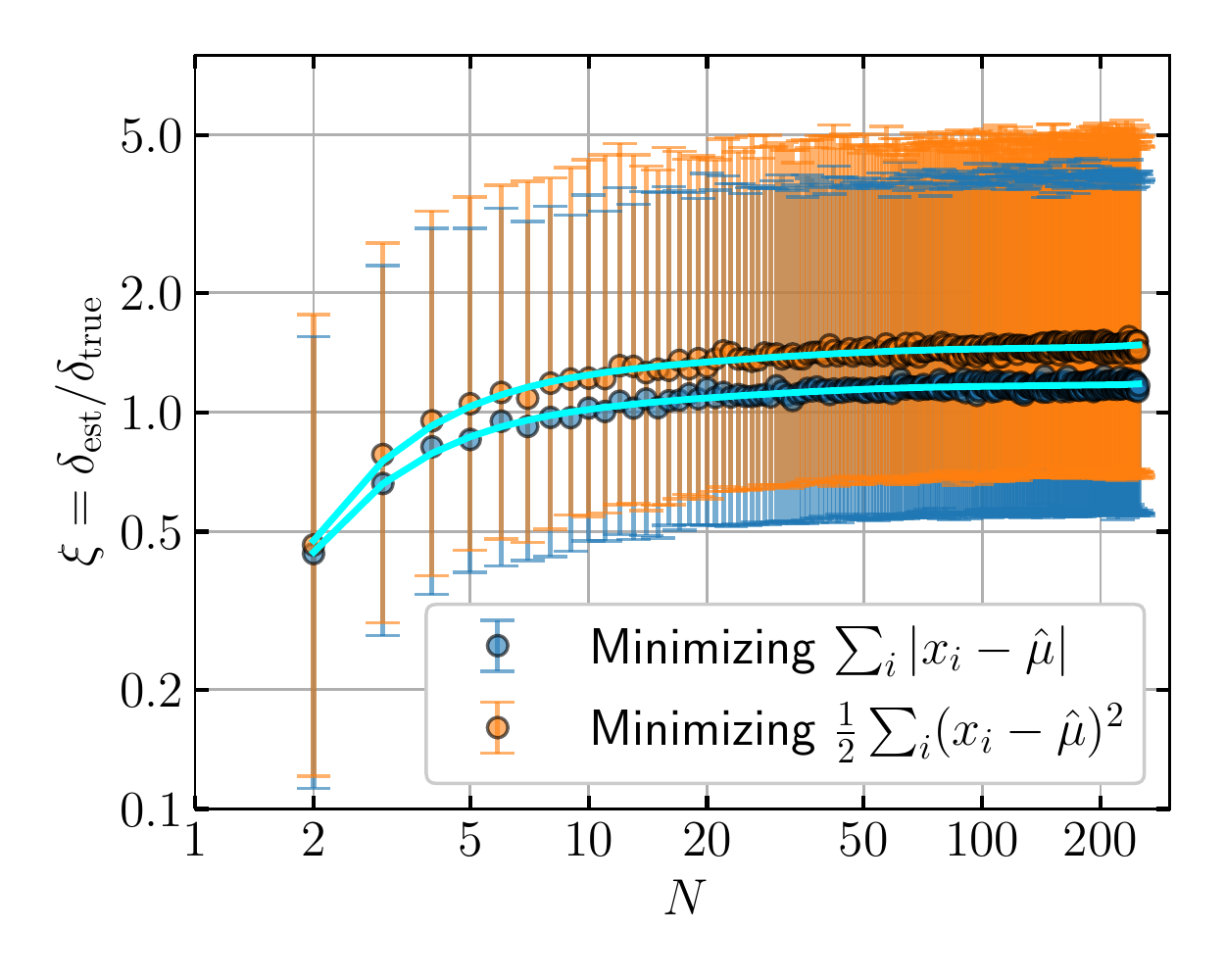}{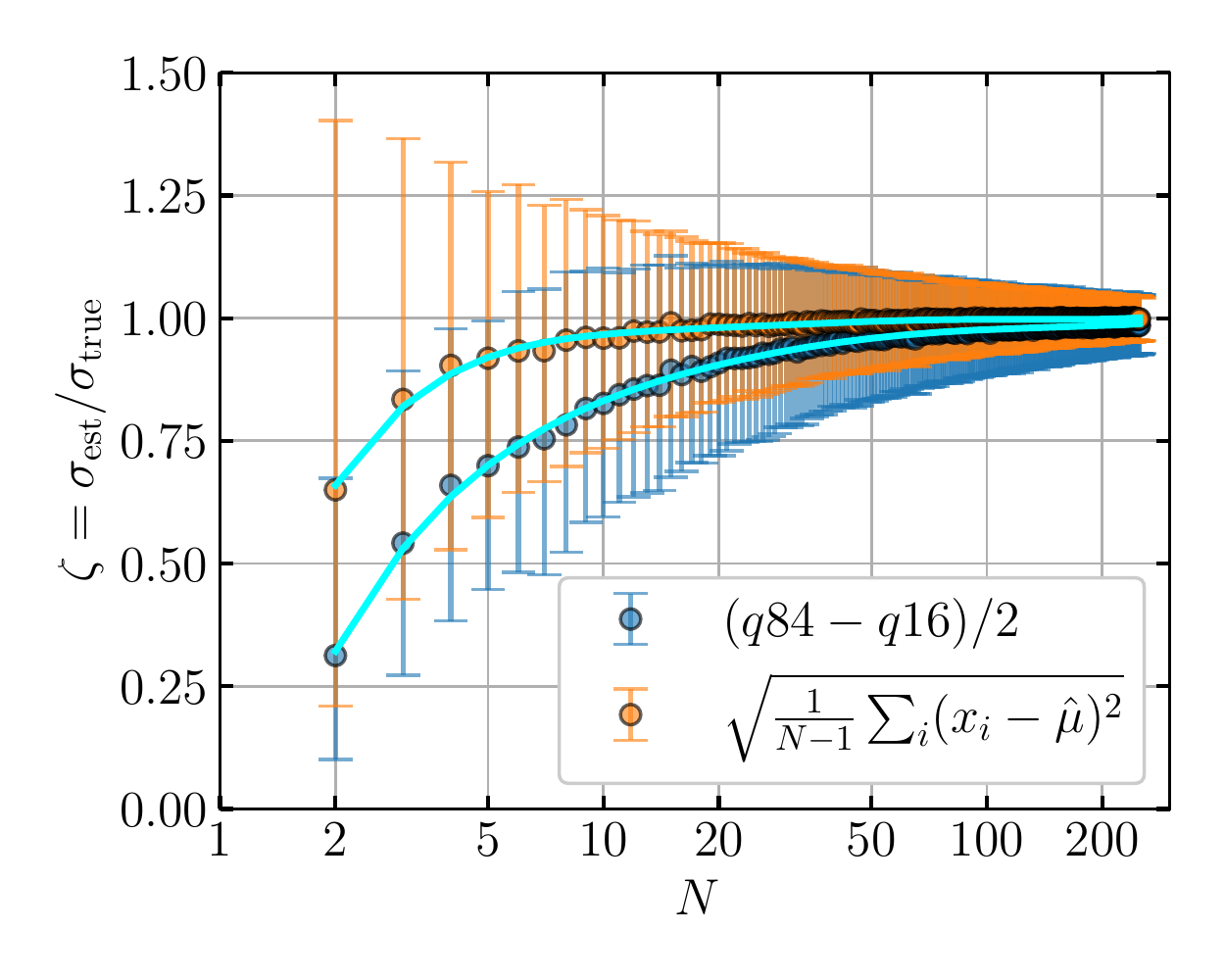}
\caption{The empirical correction factor for error of mean ($\xi$) and for standard error ($\zeta$) of Gaussian distributions in small number statistics. \label{fig:xi}}
\end{figure*}

\section{Bias in small number statistics} \label{app:bias}
The estimators that characterize dispersion are often underestimated when the number of samples is small.
For example, if we only have 3 or 5 measurements of a physical quantity, the standard deviation could be underestimated.
In this section, we propose an empirical correction of this bias for the error of mean and standard deviation assuming Gaussian distribution $P(x| \mu, \sigma)$ where $\mu$ is its position and $\sigma$ is its standard error.

\subsection{The Deviation of Mean}
We can minimize an $L_1$-norm or $L_2$-norm cost function, namely, $\sum_i |x_i-\hat{\mu}|$ and $\sum_i (x_i-\hat{\mu})^2/2$, respectively, to get an estimate of the mean $\hat{\mu}$.
The true deviation is by definition
\begin{equation}
	\delta_\mathrm{true} = \mid \hat{\mu}-\mu \mid.
\end{equation}
However, in practice we do not know $\mu$ when we tackle such a problem.
A fiducial deviation associated with $\hat{\mu}$ can be constructed using the 84 and 16 percentiles \citep[or interquantiles, see][]{1993stp..book.....L, 2014sdmm.book.....I}, i.e., 
\begin{equation}
	\delta_\mathrm{est} = \frac{\{x_i\}_{q84} - \{x_i\}_{q16}}{2\sqrt{N}},
\end{equation}
where $N$  is the sample size.
To obtain an empirical relation between $\delta_\mathrm{est}$ and $\delta_\mathrm{true}$, we assume the following form, 
\begin{equation}
	\delta_\mathrm{true} = \delta_\mathrm{est} / \xi (N),
\end{equation}
where the $\xi(N)$ is the empirical correction factor and is a function of $N$.
Then, we draw mock data from a standard Gaussian distribution with the {\tt numpy.random} module.
In each experiment, we draw $N$ samples and calculated $\delta_\mathrm{est}$ and $\delta_\mathrm{true}$, and derive $\xi$.
We repeat this experiment for 3000 times for each N ranges from 2 to 250 which is enough for our purpose, and show the 16, 50 and 84th percentiles of the results for each $N$  in the left panel of Figure \ref{fig:xi}.
It is obvious that this fiducial deviation is underestimation of $\delta_\mathrm{true}$.
The relation between medians of $\log{\xi}$ and $\log{N}$ is fitted with a 5th order polynomial function, whose coefficients are tabulated in Table \ref{tab:xi}.
With this relation, we can scale the fiducial deviation to a standard that is less affected by the sample size $N$.

\subsection{The Standard Error}
For a Gaussian distribution $P(x| \mu, \sigma)$, we can use sample-based standard deviation and the 16 and 84th percentiles to estimate the true standard error $\sigma$, i.e., 
\begin{equation}\label{eq:sigma_std}
	\sigma_\mathrm{est} = \sqrt{\frac{\sum_i (x_i-\hat{\mu})^2}{N-1}},
\end{equation}
and 
\begin{equation}\label{eq:sigma_q}
	\sigma_\mathrm{est} = \frac{\{x_i\}_{q84} - \{x_i\}_{q16}}{2},
\end{equation}
respectively.
We define $\zeta$ by
\begin{equation}
	\sigma_\mathrm{true} = \sigma_\mathrm{est}/\zeta (N),
\end{equation}
The results clearly show that both method underestimate the standard error.
Similar to the procedures in the previous test, we fit the relation with a 5th order polynomial to the medians of $\zeta$ and $\log_{10} N$, the best-fit polynomials are shown in the right panel of Figure \ref{fig:xi} and coefficients are tabulated in Table~\ref{tab:zeta}.
Compared to Eq. (\ref{eq:sigma_std}), Eq. (\ref{eq:sigma_q}) is more robust to outliers but suffers from more significant underestimation when $N$ is small.

\begin{deluxetable*}{c|ccccccc}[h]
\tabletypesize{\scriptsize}
\tablenum{B.1}
\tablecaption{The best fit coefficients of the 5th order polynomials for the empirical relationship between $\log_{10}{\xi}$ and $\log_{10}{N}$. \label{tab:xi}}
\tablewidth{0pt}
\tablehead{
\colhead{Cost function} & \colhead{$\beta_5$} & \colhead{$\beta_4$} & \colhead{$\beta_3$} & \colhead{$\beta_2$} & \colhead{$\beta_1$} & \colhead{$\beta_0$}}
\startdata
$\sum_i |x_i-\hat{\mu}|$ & 0.07349721 & -0.60647022 & 1.97806105 & -3.22994084 & 2.72007585 & -0.92812989 \\ 
$\sum_i (x_i-\hat{\mu})^2/2$ & 0.08434813 & -0.69694429 & 2.28047526 & -3.73821453 &  3.15612997 & -0.99158461\\
\enddata
\tablecomments{Our definition of polynomial is $\mathrm{Poly}(x|\beta)=\sum_i \beta_i x^i$.}
\end{deluxetable*}

\begin{deluxetable*}{c|ccccccc}[h]
\tabletypesize{\scriptsize}
\tablenum{B.2}
\tablecaption{The best-fit coefficients of the 5th order polynomials for the empirical relationship between $\zeta$ and $\log_{10}{N}$. \label{tab:zeta}}
\tablewidth{0pt}
\tablehead{
\colhead{Estimator} & \colhead{$\beta_5$} & \colhead{$\beta_4$} & \colhead{$\beta_3$} & \colhead{$\beta_2$} & \colhead{$\beta_1$} & \colhead{$\beta_0$}}
\startdata
$(q84-q16)/2$ & 0.05712779 & -0.47050592 & 1.56888875 & -2.75649024 &  2.71900493 & -0.28637106 \\ 
$\sqrt{\frac{1}{N-1}\sum (x_i-\hat{\mu})^2}$ & 0.08344418 & -0.68630504 & 2.21084202 & -3.50529186 &  2.77854093 & 0.08576048\\
\enddata
\end{deluxetable*}

\section{Several related Python packages}
Three packages are developed in this work.
\begin{enumerate}
	\item laspec \citep{laspec}: A toolkit for LAMOST MRS/LRS spectra, including modules for file IO, spectral convolution, continuum normalization, removal of cosmic rays, cross-correlation function and the empirical correction evaluation (Appendix \ref{app:bias}). 
	\item regli \citep{regli}: REgular Grid Linear Interpolator, a multi-dimensional linear interpolator based on gridded data. It is faster than the {\tt scipy.interpolate.LinearNDInterpolator} in the python standard library in our performance test.
	\item berliner \citep{berliner}: A toolkit for manipulating the MIST \citep{ 2016ApJS..222....8D} and PARSEC \citep{2012MNRAS.427..127B} stellar evolutionary tracks and isochrones, including a python interface to download PARSEC isochrones from the CMD 3.4 website \url{http://stev.oapd.inaf.it/cgi-bin/cmd}.
\end{enumerate}
The source code and some tutorials of these packages can be found at \url{https://github.com/hypergravity}.
Readers who are interested in LAMOST MRS spectra might find them useful for their research.


\bibliography{rvzp.bib}{}

\begin{thebibliography}{}
\expandafter\ifx\csname natexlab\endcsname\relax\def\natexlab#1{#1}\fi
\providecommand{\url}[1]{\href{#1}{#1}}
\providecommand{\dodoi}[1]{doi:~\href{http://doi.org/#1}{\nolinkurl{#1}}}
\providecommand{\doeprint}[1]{\href{http://ascl.net/#1}{\nolinkurl{http://ascl.net/#1}}}
\providecommand{\doarXiv}[1]{\href{https://arxiv.org/abs/#1}{\nolinkurl{https://arxiv.org/abs/#1}}}

\bibitem[{{Allende Prieto} {et~al.}(2018){Allende Prieto}, {Koesterke},
  {Hubeny}, {Bautista}, {Barklem}, \& {Nahar}}]{2018A&A...618A..25A}
{Allende Prieto}, C., {Koesterke}, L., {Hubeny}, I., {et~al.} 2018, \aap, 618,
  A25, \dodoi{10.1051/0004-6361/201732484}

\bibitem[{{Astropy Collaboration} {et~al.}(2018){Astropy Collaboration},
  {Price-Whelan}, {Sip{\H{o}}cz}, {G{\"u}nther}, {Lim}, {Crawford}, {Conseil},
  {Shupe}, {Craig}, {Dencheva}, {Ginsburg}, {VanderPlas}, {Bradley},
  {P{\'e}rez-Su{\'a}rez}, {de Val-Borro}, {Aldcroft}, {Cruz}, {Robitaille},
  {Tollerud}, {Ardelean}, {Babej}, {Bach}, {Bachetti}, {Bakanov}, {Bamford},
  {Barentsen}, {Barmby}, {Baumbach}, {Berry}, {Biscani}, {Boquien}, {Bostroem},
  {Bouma}, {Brammer}, {Bray}, {Breytenbach}, {Buddelmeijer}, {Burke},
  {Calderone}, {Cano Rodr{\'\i}guez}, {Cara}, {Cardoso}, {Cheedella}, {Copin},
  {Corrales}, {Crichton}, {D'Avella}, {Deil}, {Depagne}, {Dietrich}, {Donath},
  {Droettboom}, {Earl}, {Erben}, {Fabbro}, {Ferreira}, {Finethy}, {Fox},
  {Garrison}, {Gibbons}, {Goldstein}, {Gommers}, {Greco}, {Greenfield},
  {Groener}, {Grollier}, {Hagen}, {Hirst}, {Homeier}, {Horton}, {Hosseinzadeh},
  {Hu}, {Hunkeler}, {Ivezi{\'c}}, {Jain}, {Jenness}, {Kanarek}, {Kendrew},
  {Kern}, {Kerzendorf}, {Khvalko}, {King}, {Kirkby}, {Kulkarni}, {Kumar},
  {Lee}, {Lenz}, {Littlefair}, {Ma}, {Macleod}, {Mastropietro}, {McCully},
  {Montagnac}, {Morris}, {Mueller}, {Mumford}, {Muna}, {Murphy}, {Nelson},
  {Nguyen}, {Ninan}, {N{\"o}the}, {Ogaz}, {Oh}, {Parejko}, {Parley}, {Pascual},
  {Patil}, {Patil}, {Plunkett}, {Prochaska}, {Rastogi}, {Reddy Janga},
  {Sabater}, {Sakurikar}, {Seifert}, {Sherbert}, {Sherwood-Taylor}, {Shih},
  {Sick}, {Silbiger}, {Singanamalla}, {Singer}, {Sladen}, {Sooley},
  {Sornarajah}, {Streicher}, {Teuben}, {Thomas}, {Tremblay}, {Turner},
  {Terr{\'o}n}, {van Kerkwijk}, {de la Vega}, {Watkins}, {Weaver}, {Whitmore},
  {Woillez}, {Zabalza}, \& {Astropy Contributors}}]{2018AJ....156..123A}
{Astropy Collaboration}, {Price-Whelan}, A.~M., {Sip{\H{o}}cz}, B.~M., {et~al.}
  2018, \aj, 156, 123, \dodoi{10.3847/1538-3881/aabc4f}

\bibitem[{{Bird} {et~al.}(2020){Bird}, {Xue}, {Liu}, {Shen}, {Flynn}, \&
  {Yang}}]{2020arXiv200505980B}
{Bird}, S.~A., {Xue}, X.-X., {Liu}, C., {et~al.} 2020, arXiv e-prints,
  arXiv:2005.05980.
\newblock \doarXiv{2005.05980}

\bibitem[{{Bressan} {et~al.}(2012){Bressan}, {Marigo}, {Girardi}, {Salasnich},
  {Dal Cero}, {Rubele}, \& {Nanni}}]{2012MNRAS.427..127B}
{Bressan}, A., {Marigo}, P., {Girardi}, L., {et~al.} 2012, \mnras, 427, 127,
  \dodoi{10.1111/j.1365-2966.2012.21948.x}

\bibitem[{{Buder} {et~al.}(2020){Buder}, {Sharma}, {Kos}, {Amarsi},
  {Nordlander}, {Lind}, {Martell}, {Asplund}, {Bland-Hawthorn}, {Casey}, {De
  Silva}, {D'Orazi}, {Freeman}, {Hayden}, {Lewis}, {Lin}, {Schlesinger},
  {Simpson}, {Stello}, {Zucker}, {Zwitter}, {Beeson}, {Buck}, {Casagrande},
  {Clark}, {Cotar}, {Da Costa}, {de Grijs}, {Feuillet}, {Horner}, {Kafle},
  {Khanna}, {Kobayashi}, {Liu}, {Montet}, {Nandakumar}, {Nataf}, {Ness},
  {Spina}, {Tepper-Garcia}, {Ting}, {Traven}, {Vogrincic}, {Wittenmyer},
  {Wyse}, {Zerjal}, \& {the GALAH collaboration}}]{2020arXiv201102505B}
{Buder}, S., {Sharma}, S., {Kos}, J., {et~al.} 2020, arXiv e-prints,
  arXiv:2011.02505.
\newblock \doarXiv{2011.02505}

\bibitem[{{Castelli} \& {Kurucz}(2003)}]{2003IAUS..210P.A20C}
{Castelli}, F., \& {Kurucz}, R.~L. 2003, in Modelling of Stellar Atmospheres,
  ed. N.~{Piskunov}, W.~W. {Weiss}, \& D.~F. {Gray}, Vol. 210, A20.
\newblock \doarXiv{astro-ph/0405087}

\bibitem[{{Cropper} {et~al.}(2018){Cropper}, {Katz}, {Sartoretti}, {Prusti},
  {de Bruijne}, {Chassat}, {Charvet}, {Boyadjian}, {Perryman}, {Sarri}, {Gare},
  {Erdmann}, {Munari}, {Zwitter}, {Wilkinson}, {Arenou}, {Vallenari},
  {G{\'o}mez}, {Panuzzo}, {Seabroke}, {Allende Prieto}, {Benson}, {Marchal},
  {Huckle}, {Smith}, {Dolding}, {Jan{\ss}en}, {Viala}, {Blomme}, {Baker},
  {Boudreault}, {Crifo}, {Soubiran}, {Fr{\'e}mat}, {Jasniewicz}, {Guerrier},
  {Guy}, {Turon}, {Jean-Antoine-Piccolo}, {Th{\'e}venin}, {David}, {Gosset}, \&
  {Damerdji}}]{2018A&A...616A...5C}
{Cropper}, M., {Katz}, D., {Sartoretti}, P., {et~al.} 2018, \aap, 616, A5,
  \dodoi{10.1051/0004-6361/201832763}

\bibitem[{{Cui} {et~al.}(2012){Cui}, {Zhao}, {Chu}, {Li}, {Li}, {Zhang}, {Su},
  {Yao}, {Wang}, {Xing}, {Li}, {Zhu}, {Wang}, {Gu}, {Luo}, {Xu}, {Zhang},
  {Liu}, {Zhang}, {Yang}, {Cao}, {Chen}, {Chen}, {Chen}, {Chen}, {Chu}, {Feng},
  {Gong}, {Hou}, {Hu}, {Hu}, {Hu}, {Jia}, {Jiang}, {Jiang}, {Jiang}, {Jin},
  {Li}, {Li}, {Li}, {Liu}, {Liu}, {Lu}, {Mao}, {Men}, {Qi}, {Qi}, {Shi},
  {Tang}, {Tao}, {Wang}, {Wang}, {Wang}, {Wang}, {Wang}, {Wang}, {Wang},
  {Wang}, {Wang}, {Wang}, {Wang}, {Wang}, {Xu}, {Xu}, {Yang}, {Yu}, {Yuan},
  {Yuan}, {Zhai}, {Zhang}, {Zhang}, {Zhang}, {Zhao}, {Zhou}, {Zhou}, {Zhu}, \&
  {Zou}}]{2012RAA....12.1197C}
{Cui}, X.-Q., {Zhao}, Y.-H., {Chu}, Y.-Q., {et~al.} 2012, Research in Astronomy
  and Astrophysics, 12, 1197, \dodoi{10.1088/1674-4527/12/9/003}

\bibitem[{{De Silva} {et~al.}(2015){De Silva}, {Freeman}, {Bland-Hawthorn},
  {Martell}, {de Boer}, {Asplund}, {Keller}, {Sharma}, {Zucker}, {Zwitter},
  {Anguiano}, {Bacigalupo}, {Bayliss}, {Beavis}, {Bergemann}, {Campbell},
  {Cannon}, {Carollo}, {Casagrande}, {Casey}, {Da Costa}, {D'Orazi}, {Dotter},
  {Duong}, {Heger}, {Ireland}, {Kafle}, {Kos}, {Lattanzio}, {Lewis}, {Lin},
  {Lind}, {Munari}, {Nataf}, {O'Toole}, {Parker}, {Reid}, {Schlesinger},
  {Sheinis}, {Simpson}, {Stello}, {Ting}, {Traven}, {Watson}, {Wittenmyer},
  {Yong}, \& {{\v{Z}}erjal}}]{2015MNRAS.449.2604D}
{De Silva}, G.~M., {Freeman}, K.~C., {Bland-Hawthorn}, J., {et~al.} 2015,
  \mnras, 449, 2604, \dodoi{10.1093/mnras/stv327}

\bibitem[{{Deng} {et~al.}(2012){Deng}, {Newberg}, {Liu}, {Carlin}, {Beers},
  {Chen}, {Chen}, {Christlieb}, {Grillmair}, {Guhathakurta}, {Han}, {Hou},
  {Lee}, {L{\'e}pine}, {Li}, {Liu}, {Pan}, {Sellwood}, {Wang}, {Wang}, {Yang},
  {Yanny}, {Zhang}, {Zhang}, {Zheng}, \& {Zhu}}]{2012RAA....12..735D}
{Deng}, L.-C., {Newberg}, H.~J., {Liu}, C., {et~al.} 2012, Research in
  Astronomy and Astrophysics, 12, 735, \dodoi{10.1088/1674-4527/12/7/003}

\bibitem[{{Dotter}(2016)}]{2016ApJS..222....8D}
{Dotter}, A. 2016, \apjs, 222, 8, \dodoi{10.3847/0067-0049/222/1/8}

\bibitem[{{El-Badry} {et~al.}(2018){El-Badry}, {Ting}, {Rix}, {Quataert},
  {Weisz}, {Cargile}, {Conroy}, {Hogg}, {Bergemann}, \&
  {Liu}}]{2018MNRAS.476..528E}
{El-Badry}, K., {Ting}, Y.-S., {Rix}, H.-W., {et~al.} 2018, \mnras, 476, 528,
  \dodoi{10.1093/mnras/sty240}

\bibitem[{{Fu} {et~al.}(2020){Fu}, {Cat}, {Zong}, {Frasca}, {Gray}, {Ren},
  {Molenda-{\.Z}akowicz}, {Corbally}, {Catanzaro}, {Shi}, {Luo}, \&
  {Zhang}}]{2020RAA....20..167F}
{Fu}, J.-N., {Cat}, P.~D., {Zong}, W., {et~al.} 2020, Research in Astronomy and
  Astrophysics, 20, 167, \dodoi{10.1088/1674-4527/20/10/167}

\bibitem[{{Gaia Collaboration} {et~al.}(2016){Gaia Collaboration}, {Prusti},
  {de Bruijne}, {Brown}, {Vallenari}, {Babusiaux}, {Bailer-Jones}, {Bastian},
  {Biermann}, {Evans}, {Eyer}, {Jansen}, {Jordi}, {Klioner}, {Lammers},
  {Lindegren}, {Luri}, {Mignard}, {Milligan}, {Panem}, {Poinsignon},
  {Pourbaix}, {Randich}, {Sarri}, {Sartoretti}, {Siddiqui}, {Soubiran},
  {Valette}, {van Leeuwen}, {Walton}, {Aerts}, {Arenou}, {Cropper}, {Drimmel},
  {H{\o}g}, {Katz}, {Lattanzi}, {O'Mullane}, {Grebel}, {Holland}, {Huc},
  {Passot}, {Bramante}, {Cacciari}, {Casta{\~n}eda}, {Chaoul}, {Cheek}, {De
  Angeli}, {Fabricius}, {Guerra}, {Hern{\'a}ndez}, {Jean-Antoine-Piccolo},
  {Masana}, {Messineo}, {Mowlavi}, {Nienartowicz}, {Ord{\'o}{\~n}ez-Blanco},
  {Panuzzo}, {Portell}, {Richards}, {Riello}, {Seabroke}, {Tanga},
  {Th{\'e}venin}, {Torra}, {Els}, {Gracia-Abril}, {Comoretto},
  {Garcia-Reinaldos}, {Lock}, {Mercier}, {Altmann}, {Andrae}, {Astraatmadja},
  {Bellas-Velidis}, {Benson}, {Berthier}, {Blomme}, {Busso}, {Carry},
  {Cellino}, {Clementini}, {Cowell}, {Creevey}, {Cuypers}, {Davidson}, {De
  Ridder}, {de Torres}, {Delchambre}, {Dell'Oro}, {Ducourant}, {Fr{\'e}mat},
  {Garc{\'\i}a-Torres}, {Gosset}, {Halbwachs}, {Hambly}, {Harrison}, {Hauser},
  {Hestroffer}, {Hodgkin}, {Huckle}, {Hutton}, {Jasniewicz}, {Jordan},
  {Kontizas}, {Korn}, {Lanzafame}, {Manteiga}, {Moitinho}, {Muinonen},
  {Osinde}, {Pancino}, {Pauwels}, {Petit}, {Recio-Blanco}, {Robin}, {Sarro},
  {Siopis}, {Smith}, {Smith}, {Sozzetti}, {Thuillot}, {van Reeven}, {Viala},
  {Abbas}, {Abreu Aramburu}, {Accart}, {Aguado}, {Allan}, {Allasia},
  {Altavilla}, {{\'A}lvarez}, {Alves}, {Anderson}, {Andrei}, {Anglada Varela},
  {Antiche}, {Antoja}, {Ant{\'o}n}, {Arcay}, {Atzei}, {Ayache}, {Bach},
  {Baker}, {Balaguer-N{\'u}{\~n}ez}, {Barache}, {Barata}, {Barbier}, {Barblan},
  {Baroni}, {Barrado y Navascu{\'e}s}, {Barros}, {Barstow}, {Becciani},
  {Bellazzini}, {Bellei}, {Bello Garc{\'\i}a}, {Belokurov}, {Bendjoya},
  {Berihuete}, {Bianchi}, {Bienaym{\'e}}, {Billebaud}, {Blagorodnova},
  {Blanco-Cuaresma}, {Boch}, {Bombrun}, {Borrachero}, {Bouquillon}, {Bourda},
  {Bouy}, {Bragaglia}, {Breddels}, {Brouillet}, {Br{\"u}semeister},
  {Bucciarelli}, {Budnik}, {Burgess}, {Burgon}, {Burlacu}, {Busonero}, {Buzzi},
  {Caffau}, {Cambras}, {Campbell}, {Cancelliere}, {Cantat-Gaudin}, {Carlucci},
  {Carrasco}, {Castellani}, {Charlot}, {Charnas}, {Charvet}, {Chassat},
  {Chiavassa}, {Clotet}, {Cocozza}, {Collins}, {Collins}, {Costigan}, {Crifo},
  {Cross}, {Crosta}, {Crowley}, {Dafonte}, {Damerdji}, {Dapergolas}, {David},
  {David}, {De Cat}, {de Felice}, {de Laverny}, {De Luise}, {De March}, {de
  Martino}, {de Souza}, {Debosscher}, {del Pozo}, {Delbo}, {Delgado},
  {Delgado}, {di Marco}, {Di Matteo}, {Diakite}, {Distefano}, {Dolding}, {Dos
  Anjos}, {Drazinos}, {Dur{\'a}n}, {Dzigan}, {Ecale}, {Edvardsson}, {Enke},
  {Erdmann}, {Escolar}, {Espina}, {Evans}, {Eynard Bontemps}, {Fabre},
  {Fabrizio}, {Faigler}, {Falc{\~a}o}, {Farr{\`a}s Casas}, {Faye}, {Federici},
  {Fedorets}, {Fern{\'a}ndez-Hern{\'a}ndez}, {Fernique}, {Fienga}, {Figueras},
  {Filippi}, {Findeisen}, {Fonti}, {Fouesneau}, {Fraile}, {Fraser}, {Fuchs},
  {Furnell}, {Gai}, {Galleti}, {Galluccio}, {Garabato}, {Garc{\'\i}a-Sedano},
  {Gar{\'e}}, {Garofalo}, {Garralda}, {Gavras}, {Gerssen}, {Geyer}, {Gilmore},
  {Girona}, {Giuffrida}, {Gomes}, {Gonz{\'a}lez-Marcos},
  {Gonz{\'a}lez-N{\'u}{\~n}ez}, {Gonz{\'a}lez-Vidal}, {Granvik}, {Guerrier},
  {Guillout}, {Guiraud}, {G{\'u}rpide}, {Guti{\'e}rrez-S{\'a}nchez}, {Guy},
  {Haigron}, {Hatzidimitriou}, {Haywood}, {Heiter}, {Helmi}, {Hobbs},
  {Hofmann}, {Holl}, {Holland}, {Hunt}, {Hypki}, {Icardi}, {Irwin}, {Jevardat
  de Fombelle}, {Jofr{\'e}}, {Jonker}, {Jorissen}, {Julbe}, {Karampelas},
  {Kochoska}, {Kohley}, {Kolenberg}, {Kontizas}, {Koposov}, {Kordopatis},
  {Koubsky}, {Kowalczyk}, {Krone-Martins}, {Kudryashova}, {Kull}, {Bachchan},
  {Lacoste-Seris}, {Lanza}, {Lavigne}, {Le Poncin-Lafitte}, {Lebreton},
  {Lebzelter}, {Leccia}, {Leclerc}, {Lecoeur-Taibi}, {Lemaitre}, {Lenhardt},
  {Leroux}, {Liao}, {Licata}, {Lindstr{\o}m}, {Lister}, {Livanou}, {Lobel},
  {L{\"o}ffler}, {L{\'o}pez}, {Lopez-Lozano}, {Lorenz}, {Loureiro},
  {MacDonald}, {Magalh{\~a}es Fernandes}, {Managau}, {Mann}, {Mantelet},
  {Marchal}, {Marchant}, {Marconi}, {Marie}, {Marinoni}, {Marrese},
  {Marschalk{\'o}}, {Marshall}, {Mart{\'\i}n-Fleitas}, {Martino}, {Mary},
  {Matijevi{\v{c}}}, {Mazeh}, {McMillan}, {Messina}, {Mestre}, {Michalik},
  {Millar}, {Miranda}, {Molina}, {Molinaro}, {Molinaro}, {Moln{\'a}r},
  {Moniez}, {Montegriffo}, {Monteiro}, {Mor}, {Mora}, {Morbidelli}, {Morel},
  {Morgenthaler}, {Morley}, {Morris}, {Mulone}, {Muraveva}, {Musella},
  {Narbonne}, {Nelemans}, {Nicastro}, {Noval}, {Ord{\'e}novic},
  {Ordieres-Mer{\'e}}, {Osborne}, {Pagani}, {Pagano}, {Pailler}, {Palacin},
  {Palaversa}, {Parsons}, {Paulsen}, {Pecoraro}, {Pedrosa}, {Pentik{\"a}inen},
  {Pereira}, {Pichon}, {Piersimoni}, {Pineau}, {Plachy}, {Plum}, {Poujoulet},
  {Pr{\v{s}}a}, {Pulone}, {Ragaini}, {Rago}, {Rambaux}, {Ramos-Lerate},
  {Ranalli}, {Rauw}, {Read}, {Regibo}, {Renk}, {Reyl{\'e}}, {Ribeiro},
  {Rimoldini}, {Ripepi}, {Riva}, {Rixon}, {Roelens}, {Romero-G{\'o}mez},
  {Rowell}, {Royer}, {Rudolph}, {Ruiz-Dern}, {Sadowski}, {Sagrist{\`a}
  Sell{\'e}s}, {Sahlmann}, {Salgado}, {Salguero}, {Sarasso}, {Savietto},
  {Schnorhk}, {Schultheis}, {Sciacca}, {Segol}, {Segovia}, {Segransan},
  {Serpell}, {Shih}, {Smareglia}, {Smart}, {Smith}, {Solano}, {Solitro},
  {Sordo}, {Soria Nieto}, {Souchay}, {Spagna}, {Spoto}, {Stampa}, {Steele},
  {Steidelm{\"u}ller}, {Stephenson}, {Stoev}, {Suess}, {S{\"u}veges}, {Surdej},
  {Szabados}, {Szegedi-Elek}, {Tapiador}, {Taris}, {Tauran}, {Taylor},
  {Teixeira}, {Terrett}, {Tingley}, {Trager}, {Turon}, {Ulla}, {Utrilla},
  {Valentini}, {van Elteren}, {Van Hemelryck}, {van Leeuwen}, {Varadi},
  {Vecchiato}, {Veljanoski}, {Via}, {Vicente}, {Vogt}, {Voss}, {Votruba},
  {Voutsinas}, {Walmsley}, {Weiler}, {Weingrill}, {Werner}, {Wevers},
  {Whitehead}, {Wyrzykowski}, {Yoldas}, {{\v{Z}}erjal}, {Zucker}, {Zurbach},
  {Zwitter}, {Alecu}, {Allen}, {Allende Prieto}, {Amorim},
  {Anglada-Escud{\'e}}, {Arsenijevic}, {Azaz}, {Balm}, {Beck}, {Bernstein},
  {Bigot}, {Bijaoui}, {Blasco}, {Bonfigli}, {Bono}, {Boudreault}, {Bressan},
  {Brown}, {Brunet}, {Bunclark}, {Buonanno}, {Butkevich}, {Carret}, {Carrion},
  {Chemin}, {Ch{\'e}reau}, {Corcione}, {Darmigny}, {de Boer}, {de Teodoro}, {de
  Zeeuw}, {Delle Luche}, {Domingues}, {Dubath}, {Fodor}, {Fr{\'e}zouls},
  {Fries}, {Fustes}, {Fyfe}, {Gallardo}, {Gallegos}, {Gardiol}, {Gebran},
  {Gomboc}, {G{\'o}mez}, {Grux}, {Gueguen}, {Heyrovsky}, {Hoar}, {Iannicola},
  {Isasi Parache}, {Janotto}, {Joliet}, {Jonckheere}, {Keil}, {Kim},
  {Klagyivik}, {Klar}, {Knude}, {Kochukhov}, {Kolka}, {Kos}, {Kutka}, {Lainey},
  {LeBouquin}, {Liu}, {Loreggia}, {Makarov}, {Marseille}, {Martayan},
  {Martinez-Rubi}, {Massart}, {Meynadier}, {Mignot}, {Munari}, {Nguyen},
  {Nordlander}, {Ocvirk}, {O'Flaherty}, {Olias Sanz}, {Ortiz}, {Osorio},
  {Oszkiewicz}, {Ouzounis}, {Palmer}, {Park}, {Pasquato}, {Peltzer}, {Peralta},
  {P{\'e}turaud}, {Pieniluoma}, {Pigozzi}, {Poels}, {Prat}, {Prod'homme},
  {Raison}, {Rebordao}, {Risquez}, {Rocca-Volmerange}, {Rosen}, {Ruiz-Fuertes},
  {Russo}, {Sembay}, {Serraller Vizcaino}, {Short}, {Siebert}, {Silva},
  {Sinachopoulos}, {Slezak}, {Soffel}, {Sosnowska}, {Strai{\v{z}}ys}, {ter
  Linden}, {Terrell}, {Theil}, {Tiede}, {Troisi}, {Tsalmantza}, {Tur},
  {Vaccari}, {Vachier}, {Valles}, {Van Hamme}, {Veltz}, {Virtanen}, {Wallut},
  {Wichmann}, {Wilkinson}, {Ziaeepour}, \& {Zschocke}}]{2016A&A...595A...1G}
{Gaia Collaboration}, {Prusti}, T., {de Bruijne}, J.~H.~J., {et~al.} 2016,
  \aap, 595, A1, \dodoi{10.1051/0004-6361/201629272}

\bibitem[{{Gaia Collaboration} {et~al.}(2018){Gaia Collaboration}, {Brown},
  {Vallenari}, {Prusti}, {de Bruijne}, {Babusiaux}, {Bailer-Jones}, {Biermann},
  {Evans}, {Eyer}, {Jansen}, {Jordi}, {Klioner}, {Lammers}, {Lindegren},
  {Luri}, {Mignard}, {Panem}, {Pourbaix}, {Randich}, {Sartoretti}, {Siddiqui},
  {Soubiran}, {van Leeuwen}, {Walton}, {Arenou}, {Bastian}, {Cropper},
  {Drimmel}, {Katz}, {Lattanzi}, {Bakker}, {Cacciari}, {Casta{\~n}eda},
  {Chaoul}, {Cheek}, {De Angeli}, {Fabricius}, {Guerra}, {Holl}, {Masana},
  {Messineo}, {Mowlavi}, {Nienartowicz}, {Panuzzo}, {Portell}, {Riello},
  {Seabroke}, {Tanga}, {Th{\'e}venin}, {Gracia-Abril}, {Comoretto},
  {Garcia-Reinaldos}, {Teyssier}, {Altmann}, {Andrae}, {Audard},
  {Bellas-Velidis}, {Benson}, {Berthier}, {Blomme}, {Burgess}, {Busso},
  {Carry}, {Cellino}, {Clementini}, {Clotet}, {Creevey}, {Davidson}, {De
  Ridder}, {Delchambre}, {Dell'Oro}, {Ducourant},
  {Fern{\'a}ndez-Hern{\'a}ndez}, {Fouesneau}, {Fr{\'e}mat}, {Galluccio},
  {Garc{\'\i}a-Torres}, {Gonz{\'a}lez-N{\'u}{\~n}ez}, {Gonz{\'a}lez-Vidal},
  {Gosset}, {Guy}, {Halbwachs}, {Hambly}, {Harrison}, {Hern{\'a}ndez},
  {Hestroffer}, {Hodgkin}, {Hutton}, {Jasniewicz}, {Jean-Antoine-Piccolo},
  {Jordan}, {Korn}, {Krone-Martins}, {Lanzafame}, {Lebzelter}, {L{\"o}ffler},
  {Manteiga}, {Marrese}, {Mart{\'\i}n-Fleitas}, {Moitinho}, {Mora}, {Muinonen},
  {Osinde}, {Pancino}, {Pauwels}, {Petit}, {Recio-Blanco}, {Richards},
  {Rimoldini}, {Robin}, {Sarro}, {Siopis}, {Smith}, {Sozzetti}, {S{\"u}veges},
  {Torra}, {van Reeven}, {Abbas}, {Abreu Aramburu}, {Accart}, {Aerts},
  {Altavilla}, {{\'A}lvarez}, {Alvarez}, {Alves}, {Anderson}, {Andrei},
  {Anglada Varela}, {Antiche}, {Antoja}, {Arcay}, {Astraatmadja}, {Bach},
  {Baker}, {Balaguer-N{\'u}{\~n}ez}, {Balm}, {Barache}, {Barata}, {Barbato},
  {Barblan}, {Barklem}, {Barrado}, {Barros}, {Barstow}, {Bartholom{\'e}
  Mu{\~n}oz}, {Bassilana}, {Becciani}, {Bellazzini}, {Berihuete}, {Bertone},
  {Bianchi}, {Bienaym{\'e}}, {Blanco-Cuaresma}, {Boch}, {Boeche}, {Bombrun},
  {Borrachero}, {Bossini}, {Bouquillon}, {Bourda}, {Bragaglia}, {Bramante},
  {Breddels}, {Bressan}, {Brouillet}, {Br{\"u}semeister}, {Brugaletta},
  {Bucciarelli}, {Burlacu}, {Busonero}, {Butkevich}, {Buzzi}, {Caffau},
  {Cancelliere}, {Cannizzaro}, {Cantat-Gaudin}, {Carballo}, {Carlucci},
  {Carrasco}, {Casamiquela}, {Castellani}, {Castro-Ginard}, {Charlot},
  {Chemin}, {Chiavassa}, {Cocozza}, {Costigan}, {Cowell}, {Crifo}, {Crosta},
  {Crowley}, {Cuypers}, {Dafonte}, {Damerdji}, {Dapergolas}, {David}, {David},
  {de Laverny}, {De Luise}, {De March}, {de Martino}, {de Souza}, {de Torres},
  {Debosscher}, {del Pozo}, {Delbo}, {Delgado}, {Delgado}, {Di Matteo},
  {Diakite}, {Diener}, {Distefano}, {Dolding}, {Drazinos}, {Dur{\'a}n},
  {Edvardsson}, {Enke}, {Eriksson}, {Esquej}, {Eynard Bontemps}, {Fabre},
  {Fabrizio}, {Faigler}, {Falc{\~a}o}, {Farr{\`a}s Casas}, {Federici},
  {Fedorets}, {Fernique}, {Figueras}, {Filippi}, {Findeisen}, {Fonti},
  {Fraile}, {Fraser}, {Fr{\'e}zouls}, {Gai}, {Galleti}, {Garabato},
  {Garc{\'\i}a-Sedano}, {Garofalo}, {Garralda}, {Gavel}, {Gavras}, {Gerssen},
  {Geyer}, {Giacobbe}, {Gilmore}, {Girona}, {Giuffrida}, {Glass}, {Gomes},
  {Granvik}, {Gueguen}, {Guerrier}, {Guiraud}, {Guti{\'e}rrez-S{\'a}nchez},
  {Haigron}, {Hatzidimitriou}, {Hauser}, {Haywood}, {Heiter}, {Helmi}, {Heu},
  {Hilger}, {Hobbs}, {Hofmann}, {Holland}, {Huckle}, {Hypki}, {Icardi},
  {Jan{\ss}en}, {Jevardat de Fombelle}, {Jonker}, {Juh{\'a}sz}, {Julbe},
  {Karampelas}, {Kewley}, {Klar}, {Kochoska}, {Kohley}, {Kolenberg},
  {Kontizas}, {Kontizas}, {Koposov}, {Kordopatis}, {Kostrzewa-Rutkowska},
  {Koubsky}, {Lambert}, {Lanza}, {Lasne}, {Lavigne}, {Le Fustec}, {Le
  Poncin-Lafitte}, {Lebreton}, {Leccia}, {Leclerc}, {Lecoeur-Taibi},
  {Lenhardt}, {Leroux}, {Liao}, {Licata}, {Lindstr{\o}m}, {Lister}, {Livanou},
  {Lobel}, {L{\'o}pez}, {Managau}, {Mann}, {Mantelet}, {Marchal}, {Marchant},
  {Marconi}, {Marinoni}, {Marschalk{\'o}}, {Marshall}, {Martino}, {Marton},
  {Mary}, {Massari}, {Matijevi{\v{c}}}, {Mazeh}, {McMillan}, {Messina},
  {Michalik}, {Millar}, {Molina}, {Molinaro}, {Moln{\'a}r}, {Montegriffo},
  {Mor}, {Morbidelli}, {Morel}, {Morris}, {Mulone}, {Muraveva}, {Musella},
  {Nelemans}, {Nicastro}, {Noval}, {O'Mullane}, {Ord{\'e}novic},
  {Ord{\'o}{\~n}ez-Blanco}, {Osborne}, {Pagani}, {Pagano}, {Pailler},
  {Palacin}, {Palaversa}, {Panahi}, {Pawlak}, {Piersimoni}, {Pineau}, {Plachy},
  {Plum}, {Poggio}, {Poujoulet}, {Pr{\v{s}}a}, {Pulone}, {Racero}, {Ragaini},
  {Rambaux}, {Ramos-Lerate}, {Regibo}, {Reyl{\'e}}, {Riclet}, {Ripepi}, {Riva},
  {Rivard}, {Rixon}, {Roegiers}, {Roelens}, {Romero-G{\'o}mez}, {Rowell},
  {Royer}, {Ruiz-Dern}, {Sadowski}, {Sagrist{\`a} Sell{\'e}s}, {Sahlmann},
  {Salgado}, {Salguero}, {Sanna}, {Santana-Ros}, {Sarasso}, {Savietto},
  {Schultheis}, {Sciacca}, {Segol}, {Segovia}, {S{\'e}gransan}, {Shih},
  {Siltala}, {Silva}, {Smart}, {Smith}, {Solano}, {Solitro}, {Sordo}, {Soria
  Nieto}, {Souchay}, {Spagna}, {Spoto}, {Stampa}, {Steele},
  {Steidelm{\"u}ller}, {Stephenson}, {Stoev}, {Suess}, {Surdej}, {Szabados},
  {Szegedi-Elek}, {Tapiador}, {Taris}, {Tauran}, {Taylor}, {Teixeira},
  {Terrett}, {Teyssandier}, {Thuillot}, {Titarenko}, {Torra Clotet}, {Turon},
  {Ulla}, {Utrilla}, {Uzzi}, {Vaillant}, {Valentini}, {Valette}, {van Elteren},
  {Van Hemelryck}, {van Leeuwen}, {Vaschetto}, {Vecchiato}, {Veljanoski},
  {Viala}, {Vicente}, {Vogt}, {von Essen}, {Voss}, {Votruba}, {Voutsinas},
  {Walmsley}, {Weiler}, {Wertz}, {Wevers}, {Wyrzykowski}, {Yoldas},
  {{\v{Z}}erjal}, {Ziaeepour}, {Zorec}, {Zschocke}, {Zucker}, {Zurbach}, \&
  {Zwitter}}]{2018A&A...616A...1G}
{Gaia Collaboration}, {Brown}, A.~G.~A., {Vallenari}, A., {et~al.} 2018, \aap,
  616, A1, \dodoi{10.1051/0004-6361/201833051}

\bibitem[{{Gao} {et~al.}(2014){Gao}, {Liu}, {Zhang}, {Justham}, {Deng}, \&
  {Yang}}]{2014ApJ...788L..37G}
{Gao}, S., {Liu}, C., {Zhang}, X., {et~al.} 2014, \apjl, 788, L37,
  \dodoi{10.1088/2041-8205/788/2/L37}

\bibitem[{{Gao} {et~al.}(2017){Gao}, {Zhao}, {Yang}, \&
  {Gao}}]{2017MNRAS.469L..68G}
{Gao}, S., {Zhao}, H., {Yang}, H., \& {Gao}, R. 2017, \mnras, 469, L68,
  \dodoi{10.1093/mnrasl/slx048}

\bibitem[{{Gilmore} {et~al.}(2012){Gilmore}, {Randich}, {Asplund}, {Binney},
  {Bonifacio}, {Drew}, {Feltzing}, {Ferguson}, {Jeffries}, {Micela},
  {Negueruela}, {Prusti}, {Rix}, {Vallenari}, {Alfaro}, {Allende-Prieto},
  {Babusiaux}, {Bensby}, {Blomme}, {Bragaglia}, {Flaccomio}, {Fran{\c{c}}ois},
  {Irwin}, {Koposov}, {Korn}, {Lanzafame}, {Pancino}, {Paunzen},
  {Recio-Blanco}, {Sacco}, {Smiljanic}, {Van Eck}, {Walton}, {Aden}, {Aerts},
  {Affer}, {Alcala}, {Altavilla}, {Alves}, {Antoja}, {Arenou}, {Argiroffi},
  {Asensio Ramos}, {Bailer-Jones}, {Balaguer-Nunez}, {Bayo}, {Barbuy},
  {Barisevicius}, {Barrado y Navascues}, {Battistini}, {Bellas Velidis},
  {Bellazzini}, {Belokurov}, {Bergemann}, {Bertelli}, {Biazzo}, {Bienayme},
  {Bland-Hawthorn}, {Boeche}, {Bonito}, {Boudreault}, {Bouvier}, {Brandao},
  {Brown}, {de Bruijne}, {Burleigh}, {Caballero}, {Caffau}, {Calura},
  {Capuzzo-Dolcetta}, {Caramazza}, {Carraro}, {Casagrande}, {Casewell},
  {Chapman}, {Chiappini}, {Chorniy}, {Christlieb}, {Cignoni}, {Cocozza},
  {Colless}, {Collet}, {Collins}, {Correnti}, {Covino}, {Crnojevic}, {Cropper},
  {Cunha}, {Damiani}, {David}, {Delgado}, {Duffau}, {Edvardsson}, {Eldridge},
  {Enke}, {Eriksson}, {Evans}, {Eyer}, {Famaey}, {Fellhauer}, {Ferreras},
  {Figueras}, {Fiorentino}, {Flynn}, {Folha}, {Franciosini}, {Frasca},
  {Freeman}, {Fremat}, {Friel}, {Gaensicke}, {Gameiro}, {Garzon}, {Geier},
  {Geisler}, {Gerhard}, {Gibson}, {Gomboc}, {Gomez}, {Gonzalez-Fernandez},
  {Gonzalez Hernandez}, {Gosset}, {Grebel}, {Greimel}, {Groenewegen},
  {Grundahl}, {Guarcello}, {Gustafsson}, {Hadrava}, {Hatzidimitriou}, {Hambly},
  {Hammersley}, {Hansen}, {Haywood}, {Heber}, {Heiter}, {Held}, {Helmi},
  {Hensler}, {Herrero}, {Hill}, {Hodgkin}, {Huelamo}, {Huxor}, {Ibata},
  {Jackson}, {de Jong}, {Jonker}, {Jordan}, {Jordi}, {Jorissen}, {Katz},
  {Kawata}, {Keller}, {Kharchenko}, {Klement}, {Klutsch}, {Knude}, {Koch},
  {Kochukhov}, {Kontizas}, {Koubsky}, {Lallement}, {de Laverny}, {van Leeuwen},
  {Lemasle}, {Lewis}, {Lind}, {Lindstrom}, {Lobel}, {Lopez Santiago}, {Lucas},
  {Ludwig}, {Lueftinger}, {Magrini}, {Maiz Apellaniz}, {Maldonado}, {Marconi},
  {Marino}, {Martayan}, {Martinez-Valpuesta}, {Matijevic}, {McMahon},
  {Messina}, {Meyer}, {Miglio}, {Mikolaitis}, {Minchev}, {Minniti}, {Moitinho},
  {Momany}, {Monaco}, {Montalto}, {Monteiro}, {Monier}, {Montes}, {Mora},
  {Moraux}, {Morel}, {Mowlavi}, {Mucciarelli}, {Munari}, {Napiwotzki},
  {Nardetto}, {Naylor}, {Naze}, {Nelemans}, {Okamoto}, {Ortolani}, {Pace},
  {Palla}, {Palous}, {Parker}, {Penarrubia}, {Pillitteri}, {Piotto}, {Posbic},
  {Prisinzano}, {Puzeras}, {Quirrenbach}, {Ragaini}, {Read}, {Read}, {Reyle},
  {De Ridder}, {Robichon}, {Robin}, {Roeser}, {Romano}, {Royer}, {Ruchti},
  {Ruzicka}, {Ryan}, {Ryde}, {Santos}, {Sanz Forcada}, {Sarro Baro},
  {Sbordone}, {Schilbach}, {Schmeja}, {Schnurr}, {Schoenrich}, {Scholz},
  {Seabroke}, {Sharma}, {De Silva}, {Smith}, {Solano}, {Sordo}, {Soubiran},
  {Sousa}, {Spagna}, {Steffen}, {Steinmetz}, {Stelzer}, {Stempels},
  {Tabernero}, {Tautvaisiene}, {Thevenin}, {Torra}, {Tosi}, {Tolstoy}, {Turon},
  {Walker}, {Wambsganss}, {Worley}, {Venn}, {Vink}, {Wyse}, {Zaggia},
  {Zeilinger}, {Zoccali}, {Zorec}, {Zucker}, {Zwitter}, \& {Gaia-ESO Survey
  Team}}]{2012Msngr.147...25G}
{Gilmore}, G., {Randich}, S., {Asplund}, M., {et~al.} 2012, The Messenger, 147,
  25

\bibitem[{{Gu} {et~al.}(2019){Gu}, {Mu}, {Fu}, {Zheng}, {Yi}, {Bai}, {Wang},
  {Zhang}, {Lei}, {Bai}, {Wu}, {Wang}, \& {Liu}}]{2019ApJ...872L..20G}
{Gu}, W.-M., {Mu}, H.-J., {Fu}, J.-B., {et~al.} 2019, \apjl, 872, L20,
  \dodoi{10.3847/2041-8213/ab04f0}

\bibitem[{{Huang} {et~al.}(2018){Huang}, {Liu}, {Chen}, {Zhang}, {Yuan},
  {Xiang}, {Wang}, \& {Tian}}]{2018AJ....156...90H}
{Huang}, Y., {Liu}, X.~W., {Chen}, B.~Q., {et~al.} 2018, \aj, 156, 90,
  \dodoi{10.3847/1538-3881/aacda5}

\bibitem[{{Ivezi{\'c}} {et~al.}(2014){Ivezi{\'c}}, {Connelly}, {VanderPlas}, \&
  {Gray}}]{2014sdmm.book.....I}
{Ivezi{\'c}}, {\v{Z}}., {Connelly}, A.~J., {VanderPlas}, J.~T., \& {Gray}, A.
  2014, {Statistics, Data Mining, and Machine Learning in Astronomy}

\bibitem[{{J{\"o}nsson} {et~al.}(2020){J{\"o}nsson}, {Holtzman}, {Allende
  Prieto}, {Cunha}, {Garc{\'\i}a-Hern{\'a}ndez}, {Hasselquist}, {Masseron},
  {Osorio}, {Shetrone}, {Smith}, {Stringfellow}, {Bizyaev}, {Edvardsson},
  {Majewski}, {M{\'e}sz{\'a}ros}, {Souto}, {Zamora}, {Beaton}, {Bovy}, {Donor},
  {Pinsonneault}, {Poovelil}, \& {Sobeck}}]{2020AJ....160..120J}
{J{\"o}nsson}, H., {Holtzman}, J.~A., {Allende Prieto}, C., {et~al.} 2020, \aj,
  160, 120, \dodoi{10.3847/1538-3881/aba592}

\bibitem[{{Katz} {et~al.}(2004){Katz}, {Munari}, {Cropper}, {Zwitter},
  {Th{\'e}venin}, {David}, {Viala}, {Crifo}, {Gomboc}, {Royer}, {Arenou},
  {Marrese}, {Sordo}, {Wilkinson}, {Vallenari}, {Turon}, {Helmi}, {Bono},
  {Perryman}, {G{\'o}mez}, {Tomasella}, {Boschi}, {Morin}, {Haywood},
  {Soubiran}, {Castelli}, {Bijaoui}, {Bertelli}, {Prsa}, {Mignot}, {Sellier},
  {Baylac}, {Lebreton}, {Jauregi}, {Siviero}, {Bingham}, {Chemla}, {Coker},
  {Dibbens}, {Hancock}, {Holland}, {Horville}, {Huet}, {Laporte}, {Melse},
  {Say{\`e}de}, {Stevenson}, {Vola}, {Walton}, \&
  {Winter}}]{2004MNRAS.354.1223K}
{Katz}, D., {Munari}, U., {Cropper}, M., {et~al.} 2004, \mnras, 354, 1223,
  \dodoi{10.1111/j.1365-2966.2004.08282.x}

\bibitem[{{Katz} {et~al.}(2019){Katz}, {Sartoretti}, {Cropper}, {Panuzzo},
  {Seabroke}, {Viala}, {Benson}, {Blomme}, {Jasniewicz}, {Jean-Antoine},
  {Huckle}, {Smith}, {Baker}, {Crifo}, {Damerdji}, {David}, {Dolding},
  {Fr{\'e}mat}, {Gosset}, {Guerrier}, {Guy}, {Haigron}, {Jan{\ss}en},
  {Marchal}, {Plum}, {Soubiran}, {Th{\'e}venin}, {Ajaj}, {Allende Prieto},
  {Babusiaux}, {Boudreault}, {Chemin}, {Delle Luche}, {Fabre}, {Gueguen},
  {Hambly}, {Lasne}, {Meynadier}, {Pailler}, {Panem}, {Royer}, {Tauran},
  {Zurbach}, {Zwitter}, {Arenou}, {Bossini}, {Gerssen}, {G{\'o}mez},
  {Lemaitre}, {Leclerc}, {Morel}, {Munari}, {Turon}, {Vallenari}, \&
  {{\v{Z}}erjal}}]{2019A&A...622A.205K}
{Katz}, D., {Sartoretti}, P., {Cropper}, M., {et~al.} 2019, \aap, 622, A205,
  \dodoi{10.1051/0004-6361/201833273}

\bibitem[{{Kurucz}(1979)}]{1979ApJS...40....1K}
{Kurucz}, R.~L. 1979, \apjs, 40, 1, \dodoi{10.1086/190589}

\bibitem[{{Liu}(2019)}]{2019MNRAS.490..550L}
{Liu}, C. 2019, \mnras, 490, 550, \dodoi{10.1093/mnras/stz2274}

\bibitem[{{Liu} {et~al.}(2020){Liu}, {Fu}, {Shi}, {Wu}, {Han}, {Chen}, {Dong},
  {Zhao}, {Chen}, {Zhang}, {Bai}, {Chen}, {Cui}, {Du}, {Hsia}, {Jiang}, {Hou},
  {Hou}, {Li}, {Li}, {Li}, {Liu}, {Liu}, {Luo}, {Ren}, {Tian}, {Tian}, {Wang},
  {Wu}, {Xie}, {Yan}, {Yang}, {Yu}, {Zhang}, {Zhang}, {Zhang}, {Zhang}, {Zhao},
  {Zhong}, {Zong}, \& {Zuo}}]{2020arXiv200507210L}
{Liu}, C., {Fu}, J., {Shi}, J., {et~al.} 2020, arXiv e-prints,
  arXiv:2005.07210.
\newblock \doarXiv{2005.07210}

\bibitem[{{Liu} {et~al.}(2019{\natexlab{a}}){Liu}, {Zhang}, {Howard}, {Bai},
  {Lu}, {Soria}, {Justham}, {Li}, {Zheng}, {Wang}, {Belczynski}, {Casares},
  {Zhang}, {Yuan}, {Dong}, {Lei}, {Isaacson}, {Wang}, {Bai}, {Shao}, {Gao},
  {Wang}, {Niu}, {Cui}, {Zheng}, {Mu}, {Zhang}, {Wang}, {Heger}, {Qi}, {Liao},
  {Lattanzi}, {Gu}, {Wang}, {Wu}, {Shao}, {Shen}, {Wang}, {Bregman}, {Di
  Stefano}, {Liu}, {Han}, {Zhang}, {Wang}, {Ren}, {Zhang}, {Zhang}, {Wang},
  {Cabrera-Lavers}, {Corradi}, {Rebolo}, {Zhao}, {Zhao}, {Chu}, \&
  {Cui}}]{2019Natur.575..618L}
{Liu}, J., {Zhang}, H., {Howard}, A.~W., {et~al.} 2019{\natexlab{a}}, \nat,
  575, 618, \dodoi{10.1038/s41586-019-1766-2}

\bibitem[{{Liu} {et~al.}(2019{\natexlab{b}}){Liu}, {Fu}, {Zong}, {Shi}, {Luo},
  {Zhang}, {Cui}, {Hou}, {Pan}, {Shan}, {Chen}, {Bai}, {Chen}, {Du}, {Hou},
  {Liu}, {Tian}, {Wang}, {Wang}, {Wu}, {Wu}, {Yan}, \&
  {Zuo}}]{2019RAA....19...75L}
{Liu}, N., {Fu}, J.-N., {Zong}, W., {et~al.} 2019{\natexlab{b}}, Research in
  Astronomy and Astrophysics, 19, 075, \dodoi{10.1088/1674-4527/19/5/75}

\bibitem[{{Luo} {et~al.}(2015){Luo}, {Zhao}, {Zhao}, {Deng}, {Liu}, {Jing},
  {Wang}, {Zhang}, {Shi}, {Cui}, {Chu}, {Li}, {Bai}, {Wu}, {Cai}, {Cao}, {Cao},
  {Carlin}, {Chen}, {Chen}, {Chen}, {Chen}, {Chen}, {Chen}, {Chen},
  {Christlieb}, {Chu}, {Cui}, {Dong}, {Du}, {Fan}, {Feng}, {Fu}, {Gao}, {Gong},
  {Gu}, {Guo}, {Han}, {He}, {Hou}, {Hou}, {Hou}, {Hu}, {Hu}, {Hu}, {Huo},
  {Jia}, {Jiang}, {Jiang}, {Jiang}, {Jin}, {Kong}, {Kong}, {Lei}, {Li}, {Li},
  {Li}, {Li}, {Li}, {Li}, {Li}, {Li}, {Li}, {Li}, {Li}, {Li}, {Liang}, {Lin},
  {Liu}, {Liu}, {Liu}, {Liu}, {Lu}, {Luo}, {Mao}, {Newberg}, {Ni}, {Qi}, {Qi},
  {Shen}, {Shi}, {Song}, {Song}, {Su}, {Su}, {Tang}, {Tao}, {Tian}, {Wang},
  {Wang}, {Wang}, {Wang}, {Wang}, {Wang}, {Wang}, {Wang}, {Wang}, {Wang},
  {Wang}, {Wang}, {Wang}, {Wang}, {Wang}, {Wang}, {Wang}, {Wang}, {Wang},
  {Wang}, {Wei}, {Wei}, {Wu}, {Wu}, {Wu}, {Wu}, {Xing}, {Xu}, {Xu}, {Xu},
  {Yan}, {Yang}, {Yang}, {Yang}, {Yang}, {Yao}, {Yu}, {Yuan}, {Yuan}, {Yuan},
  {Yuan}, {Zhai}, {Zhang}, {Zhang}, {Zhang}, {Zhang}, {Zhang}, {Zhang},
  {Zhang}, {Zhang}, {Zhao}, {Zhou}, {Zhou}, {Zhu}, {Zhu}, {Zou}, \&
  {Zuo}}]{2015RAA....15.1095L}
{Luo}, A.~L., {Zhao}, Y.-H., {Zhao}, G., {et~al.} 2015, Research in Astronomy
  and Astrophysics, 15, 1095, \dodoi{10.1088/1674-4527/15/8/002}

\bibitem[{{Lupton}(1993)}]{1993stp..book.....L}
{Lupton}, R. 1993, {Statistics in theory and practice}

\bibitem[{{Majewski} {et~al.}(2017){Majewski}, {Schiavon}, {Frinchaboy},
  {Allende Prieto}, {Barkhouser}, {Bizyaev}, {Blank}, {Brunner}, {Burton},
  {Carrera}, {Chojnowski}, {Cunha}, {Epstein}, {Fitzgerald}, {Garc{\'\i}a
  P{\'e}rez}, {Hearty}, {Henderson}, {Holtzman}, {Johnson}, {Lam}, {Lawler},
  {Maseman}, {M{\'e}sz{\'a}ros}, {Nelson}, {Nguyen}, {Nidever}, {Pinsonneault},
  {Shetrone}, {Smee}, {Smith}, {Stolberg}, {Skrutskie}, {Walker}, {Wilson},
  {Zasowski}, {Anders}, {Basu}, {Beland}, {Blanton}, {Bovy}, {Brownstein},
  {Carlberg}, {Chaplin}, {Chiappini}, {Eisenstein}, {Elsworth}, {Feuillet},
  {Fleming}, {Galbraith-Frew}, {Garc{\'\i}a}, {Garc{\'\i}a-Hern{\'a}ndez},
  {Gillespie}, {Girardi}, {Gunn}, {Hasselquist}, {Hayden}, {Hekker}, {Ivans},
  {Kinemuchi}, {Klaene}, {Mahadevan}, {Mathur}, {Mosser}, {Muna}, {Munn},
  {Nichol}, {O'Connell}, {Parejko}, {Robin}, {Rocha-Pinto}, {Schultheis},
  {Serenelli}, {Shane}, {Silva Aguirre}, {Sobeck}, {Thompson}, {Troup},
  {Weinberg}, \& {Zamora}}]{2017AJ....154...94M}
{Majewski}, S.~R., {Schiavon}, R.~P., {Frinchaboy}, P.~M., {et~al.} 2017, \aj,
  154, 94, \dodoi{10.3847/1538-3881/aa784d}

\bibitem[{{M{\'e}sz{\'a}ros} {et~al.}(2012){M{\'e}sz{\'a}ros}, {Allende
  Prieto}, {Edvardsson}, {Castelli}, {Garc{\'\i}a P{\'e}rez}, {Gustafsson},
  {Majewski}, {Plez}, {Schiavon}, {Shetrone}, \& {de
  Vicente}}]{2012AJ....144..120M}
{M{\'e}sz{\'a}ros}, S., {Allende Prieto}, C., {Edvardsson}, B., {et~al.} 2012,
  \aj, 144, 120, \dodoi{10.1088/0004-6256/144/4/120}

\bibitem[{{Morton}(2000)}]{2000ApJS..130..403M}
{Morton}, D.~C. 2000, \apjs, 130, 403, \dodoi{10.1086/317349}

\bibitem[{{Moultaka} {et~al.}(2004){Moultaka}, {Ilovaisky}, {Prugniel}, \&
  {Soubiran}}]{2004PASP..116..693M}
{Moultaka}, J., {Ilovaisky}, S.~A., {Prugniel}, P., \& {Soubiran}, C. 2004,
  \pasp, 116, 693, \dodoi{10.1086/422177}

\bibitem[{Nelder \& Mead(1965)}]{Nelder1965ASM}
Nelder, J., \& Mead, R. 1965, Comput. J., 7, 308

\bibitem[{{Nidever} {et~al.}(2015){Nidever}, {Holtzman}, {Allende Prieto},
  {Beland}, {Bender}, {Bizyaev}, {Burton}, {Desphande}, {Fleming}, {Garc{\'\i}a
  P{\'e}rez}, {Hearty}, {Majewski}, {M{\'e}sz{\'a}ros}, {Muna}, {Nguyen},
  {Schiavon}, {Shetrone}, {Skrutskie}, {Sobeck}, \&
  {Wilson}}]{2015AJ....150..173N}
{Nidever}, D.~L., {Holtzman}, J.~A., {Allende Prieto}, C., {et~al.} 2015, \aj,
  150, 173, \dodoi{10.1088/0004-6256/150/6/173}

\bibitem[{{Pedregosa} {et~al.}(2012){Pedregosa}, {Varoquaux}, {Gramfort},
  {Michel}, {Thirion}, {Grisel}, {Blondel}, {M{\"u}ller}, {Nothman}, {Louppe},
  {Prettenhofer}, {Weiss}, {Dubourg}, {Vanderplas}, {Passos}, {Cournapeau},
  {Brucher}, {Perrot}, \& {Duchesnay}}]{2012arXiv1201.0490P}
{Pedregosa}, F., {Varoquaux}, G., {Gramfort}, A., {et~al.} 2012, arXiv
  e-prints, arXiv:1201.0490.
\newblock \doarXiv{1201.0490}

\bibitem[{{Press} {et~al.}(2002){Press}, {Teukolsky}, {Vetterling}, \&
  {Flannery}}]{2002nrca.book.....P}
{Press}, W.~H., {Teukolsky}, S.~A., {Vetterling}, W.~T., \& {Flannery}, B.~P.
  2002, {Numerical recipes in C++ : the art of scientific computing}

\bibitem[{{Ren} {et~al.}(2020){Ren}, {Wu}, {Wu}, {Zhang}, {Chen}, {Hsia},
  {Yang}, {Liu}, {Shi}, {Wu}, {Zhu}, {Li}, {Bai}, {Tian}, \&
  {Hou}}]{2020arXiv200806236R}
{Ren}, J.-J., {Wu}, H., {Wu}, C.-J., {et~al.} 2020, arXiv e-prints,
  arXiv:2008.06236.
\newblock \doarXiv{2008.06236}

\bibitem[{{Steinmetz} {et~al.}(2006){Steinmetz}, {Zwitter}, {Siebert},
  {Watson}, {Freeman}, {Munari}, {Campbell}, {Williams}, {Seabroke}, {Wyse},
  {Parker}, {Bienaym{\'e}}, {Roeser}, {Gibson}, {Gilmore}, {Grebel}, {Helmi},
  {Navarro}, {Burton}, {Cass}, {Dawe}, {Fiegert}, {Hartley}, {Russell},
  {Saunders}, {Enke}, {Bailin}, {Binney}, {Bland-Hawthorn}, {Boeche}, {Dehnen},
  {Eisenstein}, {Evans}, {Fiorucci}, {Fulbright}, {Gerhard}, {Jauregi}, {Kelz},
  {Mijovi{\'c}}, {Minchev}, {Parmentier}, {Pe{\~n}arrubia}, {Quillen}, {Read},
  {Ruchti}, {Scholz}, {Siviero}, {Smith}, {Sordo}, {Veltz}, {Vidrih}, {von
  Berlepsch}, {Boyle}, \& {Schilbach}}]{2006AJ....132.1645S}
{Steinmetz}, M., {Zwitter}, T., {Siebert}, A., {et~al.} 2006, \aj, 132, 1645,
  \dodoi{10.1086/506564}

\bibitem[{{Steinmetz} {et~al.}(2020{\natexlab{a}}){Steinmetz}, {Guiglion},
  {McMillan}, {Matijevi{\v{c}}}, {Enke}, {Kordopatis}, {Zwitter}, {Valentini},
  {Chiappini}, {Casagrande}, {Wojno}, {Anguiano}, {Bienaym{\'e}}, {Bijaoui},
  {Binney}, {Burton}, {Cass}, {de Laverny}, {Fiegert}, {Freeman}, {Fulbright},
  {Gibson}, {Gilmore}, {Grebel}, {Helmi}, {Kunder}, {Munari}, {Navarro},
  {Parker}, {Ruchti}, {Recio-Blanco}, {Reid}, {Seabroke}, {Siviero}, {Siebert},
  {Stupar}, {Watson}, {Williams}, {Wyse}, {Anders}, {Antoja}, {Birko},
  {Bland-Hawthorn}, {Bossini}, {Garc{\'\i}a}, {Carrillo}, {Chaplin},
  {Elsworth}, {Famaey}, {Gerhard}, {Jofre}, {Just}, {Mathur}, {Miglio},
  {Minchev}, {Monari}, {Mosser}, {Ritter}, {Rodrigues}, {Scholz}, {Sharma},
  {Sysoliatina}, \& {RAVE Collaboration}}]{2020AJ....160...83S}
{Steinmetz}, M., {Guiglion}, G., {McMillan}, P.~J., {et~al.}
  2020{\natexlab{a}}, \aj, 160, 83, \dodoi{10.3847/1538-3881/ab9ab8}

\bibitem[{{Steinmetz} {et~al.}(2020{\natexlab{b}}){Steinmetz},
  {Matijevi{\v{c}}}, {Enke}, {Zwitter}, {Guiglion}, {McMillan}, {Kordopatis},
  {Valentini}, {Chiappini}, {Casagrande}, {Wojno}, {Anguiano}, {Bienaym{\'e}},
  {Bijaoui}, {Binney}, {Burton}, {Cass}, {de Laverny}, {Fiegert}, {Freeman},
  {Fulbright}, {Gibson}, {Gilmore}, {Grebel}, {Helmi}, {Kunder}, {Munari},
  {Navarro}, {Parker}, {Ruchti}, {Recio-Blanco}, {Reid}, {Seabroke}, {Siviero},
  {Siebert}, {Stupar}, {Watson}, {Williams}, {Wyse}, {Anders}, {Antoja},
  {Birko}, {Bland-Hawthorn}, {Bossini}, {Garc{\'\i}a}, {Carrillo}, {Chaplin},
  {Elsworth}, {Famaey}, {Gerhard}, {Jofre}, {Just}, {Mathur}, {Miglio},
  {Minchev}, {Monari}, {Mosser}, {Ritter}, {Rodrigues}, {Scholz}, {Sharma},
  {Sysoliatina}, \& {RAVE Collaboration}}]{2020AJ....160...82S}
{Steinmetz}, M., {Matijevi{\v{c}}}, G., {Enke}, H., {et~al.}
  2020{\natexlab{b}}, \aj, 160, 82, \dodoi{10.3847/1538-3881/ab9ab9}

\bibitem[{{Tian} {et~al.}(2020){Tian}, {Liu}, {Wang}, {Xu}, {Yang}, {Zhang}, \&
  {Xue}}]{2020ApJ...899..110T}
{Tian}, H., {Liu}, C., {Wang}, Y., {et~al.} 2020, \apj, 899, 110,
  \dodoi{10.3847/1538-4357/aba1ec}

\bibitem[{{Tonry} \& {Davis}(1979)}]{1979AJ.....84.1511T}
{Tonry}, J., \& {Davis}, M. 1979, \aj, 84, 1511, \dodoi{10.1086/112569}

\bibitem[{{Wang} {et~al.}(2019){Wang}, {Luo}, {Chen}, {Bai}, {Chen}, {Chen},
  {Dong}, {Du}, {Fu}, {Han}, {Hou}, {Hou}, {Hou}, {Jiang}, {Kong}, {Li}, {Liu},
  {Liu}, {Qin}, {Shi}, {Tian}, {Wu}, {Wu}, {Xie}, {Zhang}, {Zhang}, {Zhao},
  {Zhao}, {Zhong}, {Zong}, \& {Zuo}}]{2019ApJS..244...27W}
{Wang}, R., {Luo}, A.~L., {Chen}, J.~J., {et~al.} 2019, \apjs, 244, 27,
  \dodoi{10.3847/1538-4365/ab3cc0}

\bibitem[{{Wang} {et~al.}(2010){Wang}, {Qin}, \& {Ye}}]{2010ExA....28..195W}
{Wang}, S., {Qin}, H., \& {Ye}, Z. 2010, Experimental Astronomy, 28, 195,
  \dodoi{10.1007/s10686-010-9198-x}

\bibitem[{{Wu} {et~al.}(2020){Wu}, {Wu}, {Zhang}, {Ren}, {Chen}, {Hsia}, {Wu},
  {Zhu}, {Li}, {Hou}, {Wang}, {Yu}, \& {LAMOST MRS
  Collaboration}}]{2020arXiv200705240W}
{Wu}, C.-J., {Wu}, H., {Zhang}, W., {et~al.} 2020, arXiv e-prints,
  arXiv:2007.05240.
\newblock \doarXiv{2007.05240}

\bibitem[{{Xu} {et~al.}(2020){Xu}, {Liu}, {Tian}, {Newberg}, {Laporte},
  {Zhang}, {Wang}, {Fu}, {Li}, \& {Deng}}]{2020ApJ...905....6X}
{Xu}, Y., {Liu}, C., {Tian}, H., {et~al.} 2020, \apj, 905, 6,
  \dodoi{10.3847/1538-4357/abc2cb}

\bibitem[{{Yang} {et~al.}(2019){Yang}, {Xue}, {Li}, {Liu}, {Zhang}, {Rix},
  {Zhang}, {Zhao}, {Tian}, {Zhong}, {Xing}, {Wu}, {Li}, {Carlin}, \&
  {Chang}}]{2019ApJ...886..154Y}
{Yang}, C., {Xue}, X.-X., {Li}, J., {et~al.} 2019, \apj, 886, 154,
  \dodoi{10.3847/1538-4357/ab48e2}

\bibitem[{{Yang} {et~al.}(2020){Yang}, {Long}, {Shan}, {Zhang}, {Guo}, {Bai},
  {Bai}, {Cui}, {Wang}, \& {Liu}}]{2020ApJS..249...31Y}
{Yang}, F., {Long}, R.~J., {Shan}, S.-S., {et~al.} 2020, \apjs, 249, 31,
  \dodoi{10.3847/1538-4365/ab9b77}

\bibitem[{{Yanny} {et~al.}(2009){Yanny}, {Rockosi}, {Newberg}, {Knapp},
  {Adelman-McCarthy}, {Alcorn}, {Allam}, {Allende Prieto}, {An}, {Anderson},
  {Anderson}, {Bailer-Jones}, {Bastian}, {Beers}, {Bell}, {Belokurov},
  {Bizyaev}, {Blythe}, {Bochanski}, {Boroski}, {Brinchmann}, {Brinkmann},
  {Brewington}, {Carey}, {Cudworth}, {Evans}, {Evans}, {Gates}, {G{\"a}nsicke},
  {Gillespie}, {Gilmore}, {Nebot Gomez-Moran}, {Grebel}, {Greenwell}, {Gunn},
  {Jordan}, {Jordan}, {Harding}, {Harris}, {Hendry}, {Holder}, {Ivans},
  {Ivezi{\v{c}}}, {Jester}, {Johnson}, {Kent}, {Kleinman}, {Kniazev},
  {Krzesinski}, {Kron}, {Kuropatkin}, {Lebedeva}, {Lee}, {French Leger},
  {L{\'e}pine}, {Levine}, {Lin}, {Long}, {Loomis}, {Lupton}, {Malanushenko},
  {Malanushenko}, {Margon}, {Martinez-Delgado}, {McGehee}, {Monet}, {Morrison},
  {Munn}, {Neilsen}, {Nitta}, {Norris}, {Oravetz}, {Owen}, {Padmanabhan},
  {Pan}, {Peterson}, {Pier}, {Platson}, {Re Fiorentin}, {Richards}, {Rix},
  {Schlegel}, {Schneider}, {Schreiber}, {Schwope}, {Sibley}, {Simmons},
  {Snedden}, {Allyn Smith}, {Stark}, {Stauffer}, {Steinmetz}, {Stoughton},
  {SubbaRao}, {Szalay}, {Szkody}, {Thakar}, {Sivarani}, {Tucker}, {Uomoto},
  {Vanden Berk}, {Vidrih}, {Wadadekar}, {Watters}, {Wilhelm}, {Wyse}, {Yarger},
  \& {Zucker}}]{2009AJ....137.4377Y}
{Yanny}, B., {Rockosi}, C., {Newberg}, H.~J., {et~al.} 2009, \aj, 137, 4377,
  \dodoi{10.1088/0004-6256/137/5/4377}

\bibitem[{Zhang(2020{\natexlab{a}})}]{laspec}
Zhang, B. 2020{\natexlab{a}}, hypergravity/laspec: A toolkit for LAMOST
  spectra.,  Zenodo, \dodoi{10.5281/ZENODO.4381155}

\bibitem[{Zhang(2020{\natexlab{b}})}]{regli}
---. 2020{\natexlab{b}}, hypergravity/regli: REgular Grid Linear Interpolator,
  Zenodo, \dodoi{10.5281/ZENODO.4381160}

\bibitem[{Zhang(2020{\natexlab{c}})}]{berliner}
---. 2020{\natexlab{c}}, hypergravity/berliner: A toolkit for stellar tracks
  and isochrones.,  Zenodo, \dodoi{10.5281/ZENODO.4381163}

\bibitem[{{Zhang} {et~al.}(2020{\natexlab{a}}){Zhang}, {Liu}, \&
  {Deng}}]{2020ApJS..246....9Z}
{Zhang}, B., {Liu}, C., \& {Deng}, L.-C. 2020{\natexlab{a}}, \apjs, 246, 9,
  \dodoi{10.3847/1538-4365/ab55ef}

\bibitem[{{Zhang} {et~al.}(2020{\natexlab{b}}){Zhang}, {Liu}, {Li}, {Deng},
  {Yan}, \& {Shi}}]{2020RAA....20...51Z}
{Zhang}, B., {Liu}, C., {Li}, C.-Q., {et~al.} 2020{\natexlab{b}}, Research in
  Astronomy and Astrophysics, 20, 051, \dodoi{10.1088/1674-4527/20/4/51}

\bibitem[{{Zhao} {et~al.}(2012){Zhao}, {Zhao}, {Chu}, {Jing}, \&
  {Deng}}]{2012RAA....12..723Z}
{Zhao}, G., {Zhao}, Y.-H., {Chu}, Y.-Q., {Jing}, Y.-P., \& {Deng}, L.-C. 2012,
  Research in Astronomy and Astrophysics, 12, 723,
  \dodoi{10.1088/1674-4527/12/7/002}

\bibitem[{{Zong} {et~al.}(2020){Zong}, {Fu}, {De Cat}, {Wang}, {Shi}, {Luo},
  {Zhang}, {Frasca}, {Molenda-{\.Z}akowicz}, {Gray}, {Corbally}, {Catanzaro},
  {Cang}, {Wang}, {Chen}, {Hou}, {Liu}, {Niu}, {Pan}, {Tian}, {Yan}, {Zhang},
  \& {Zuo}}]{2020ApJS..251...15Z}
{Zong}, W., {Fu}, J.-N., {De Cat}, P., {et~al.} 2020, \apjs, 251, 15,
  \dodoi{10.3847/1538-4365/abbb2d}

\end{thebibliography}
\bibliographystyle{aasjournal}

\end{document}